\documentclass[12pt,a4paper]{article}
\usepackage[dvipdfmx]{graphicx}
\usepackage{color}
\usepackage{citesort}
\usepackage{latexsym}
\usepackage{amsmath}
\usepackage{amssymb}
\usepackage{setspace}

\begin{document}

\title{Existence of outsiders as a characteristic of online communication networks}
	
\author{Taro Takaguchi$^{1,2}$, Takanori Maehara$^{1,2}$, Masashi Toyoda$^3$, \\and Ken-ichi Kawarabayashi$^{1,2}$\\
\\
\\
${}^{1}$ 
National Institute of Informatics,\\
2-1-2 Hitotsubashi, Chiyoda-ku, Tokyo 101-8430, Japan
\\
\\
${}^{2}$
JST, ERATO, Kawarabayashi Large Graph Project,\\
2-1-2 Hitotsubashi, Chiyoda-ku, Tokyo 101-8430, Japan
\\
\\
${}^{3}$
Institute of Industrial Science,
The University of Tokyo,\\
4-6-1 Komaba, Meguro-ku, Tokyo 153-8505, Japan
\\
\\ 
k\_keniti@nii.ac.jp
}

\begin{singlespace}
\maketitle
\end{singlespace}

\newpage

\begin{abstract}
Online social networking services (SNSs) involve communication activities between large number of individuals over the public Internet and their crawled records are often regarded as proxies of real (i.e., offline) interaction structure. However, structure observed in these records might differ from real counterparts because individuals may behave differently online and non-human accounts may even participate. To understand the difference between online and real social networks, we investigate an empirical communication network between users on Twitter, which is perhaps one of the largest SNSs. We define a network of user pairs that send reciprocal messages. Based on the mixing pattern observed in this network, we argue that this network differs from conventional understandings in the sense that there is a small number of distinctive users that we call outsiders. Outsiders do not belong to any user groups but they are connected with different groups, while not being well connected with each other.  We identify outsiders by maximizing the degree assortativity coefficient of the network via node removal, thereby confirming that local structural properties of outsiders identified are consistent with our hypothesis. Our findings suggest that the existence of outsiders should be considered when using Twitter communication networks for social network analysis.
\end{abstract}

\newpage

Online social networking services (SNSs) facilitate the instantaneous and inexpensive exchange of information.
These SNSs usually provide daily updates of the status of users~\cite{Kwak2010,Golder2011,Sasahara2013} but they can also circulate urgent information during natural disasters, accidents~\cite{Sakaki2010,Sano2013,Szell2014}, or political movements~\cite{Gonzalez-Bailon2011}.
SNSs are increasingly becoming a crucial form of infrastructure for social interactions.
From the perspective of social network research, the communication records stored by these websites provide unprecedented opportunities to analyze the large-scale structure of social networks.

However, the networks observed in SNSs might differ from real (i.e., offline) social networks~\cite{Ahn2007,Grabowicz2012,Arnaboldi2013},
because individuals may behave differently and non-human accounts, such as news media or companies, may also participate in interactions,
which is why SNSs are sometimes called social media~\cite{Kwak2010}.
Therefore, quantifying the differences between SNSs as communication networks and offline social networks is considered to be a fundamental question if we want to use these networks as proxies for real social networks.

In this study, we investigate a network defined by conversations between user accounts on Twitter~\cite{twitterURL}, which is one of the largest SNSs in the world.
Social networks are usually assumed to exhibit a positive correlation in node degree (i.e., the number of connections) between adjacent nodes (i.e., users)~\cite{Newman2002,Newman2003,Newman2003a},
but we find that the Twitter conversation network has a more complex mixing pattern.
However, when a small set of specific nodes is removed from the network, the resultant network exhibits a strongly positive degree correlation.
Therefore, we hypothesize that the network roughly comprises two types of nodes: users who belong to tightly-connected groups, the members of which have similar degree values; and users located outside these groups who are connected to different groups, but who are not well connected to each other.
We postulate that the removed nodes correspond to the users of the latter type, who we refer to as outsiders.
We verify the outsider hypothesis by investigating the local structural properties of nodes in the network.
We also find that outsiders are different from the remaining users in terms of their other activities on Twitter. 
The same hypothesis is tested in data sets derived from other online communication networks,
and the results suggest that the existence of outsiders is a unique property of online conversation networks that mediate fairly private communication.

The existence of the outsiders, notable in interaction networks based on microblogging services such as Twitter, 
might reflect the characteristic communication patterns of these services, such as the exchanges of messages in a rather casual manner with small costs
and the incentives of sending messages to the general public with the aim of self-promotion and commercial campaigns.
Thus, our finding of outsiders in the Twitter conversation network may be an important first step when trying to understand the impact of these communication patterns on the formation of user network structure in online communication services.

\section*{Results}
\subsection*{Basic statistics of the Mention network}
We construct a network between Twitter users in which links are drawn between pairs of nodes that share reciprocal interactions via Mentions (see Methods for further details).
Mention is a function of Twitter that allows a user to send a tweet (i.e., a short message) directly to other users by naming their unique IDs.
Mention can be regarded as a method for mutual communication between users rather than broadcasting information to the public.
To contract the network, we employ the relationship defined by exchanges of Mentions, instead of the so-called follower-friend relationship~\cite{Java2007,Boyd2010,Cha2010},
because Mentions indicate more active and closer communication between users~\cite{Sousa2010,Goncalves2011,Bliss2012}.   
We collected all of the tweets posted by a set of users for one week in 2011 and extracted all the Mention tweets from them.
Next, we connect two nodes with an undirected and unweighted link if the corresponding two users exchange Mentions with each other at least once in both directions during the observation period.
We focus on the largest connected component of the obtained network obtained, which comprises $N=330,114$ nodes ($66\%$ of all the users who have sent Mentions) and $M=927,352$ links, which we refer to as the Mention network.
We denote the Mention network by $G=(V, E)$, where the node set $V$ and link set $E$ are equal in size to $N$ and $M$, respectively.

\begin{figure}
\centering
\includegraphics[width=0.45\hsize]{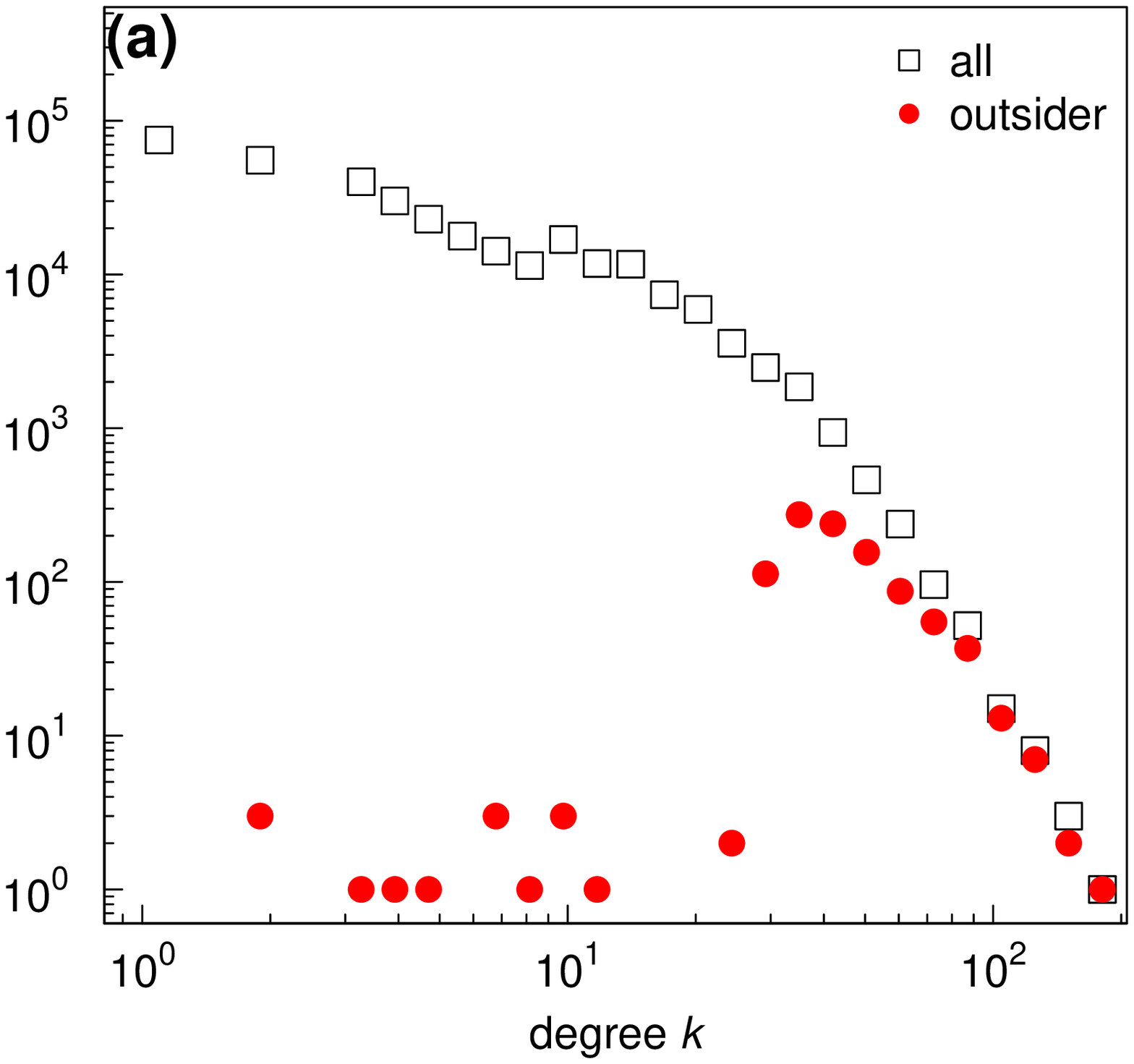}
\includegraphics[width=0.45\hsize]{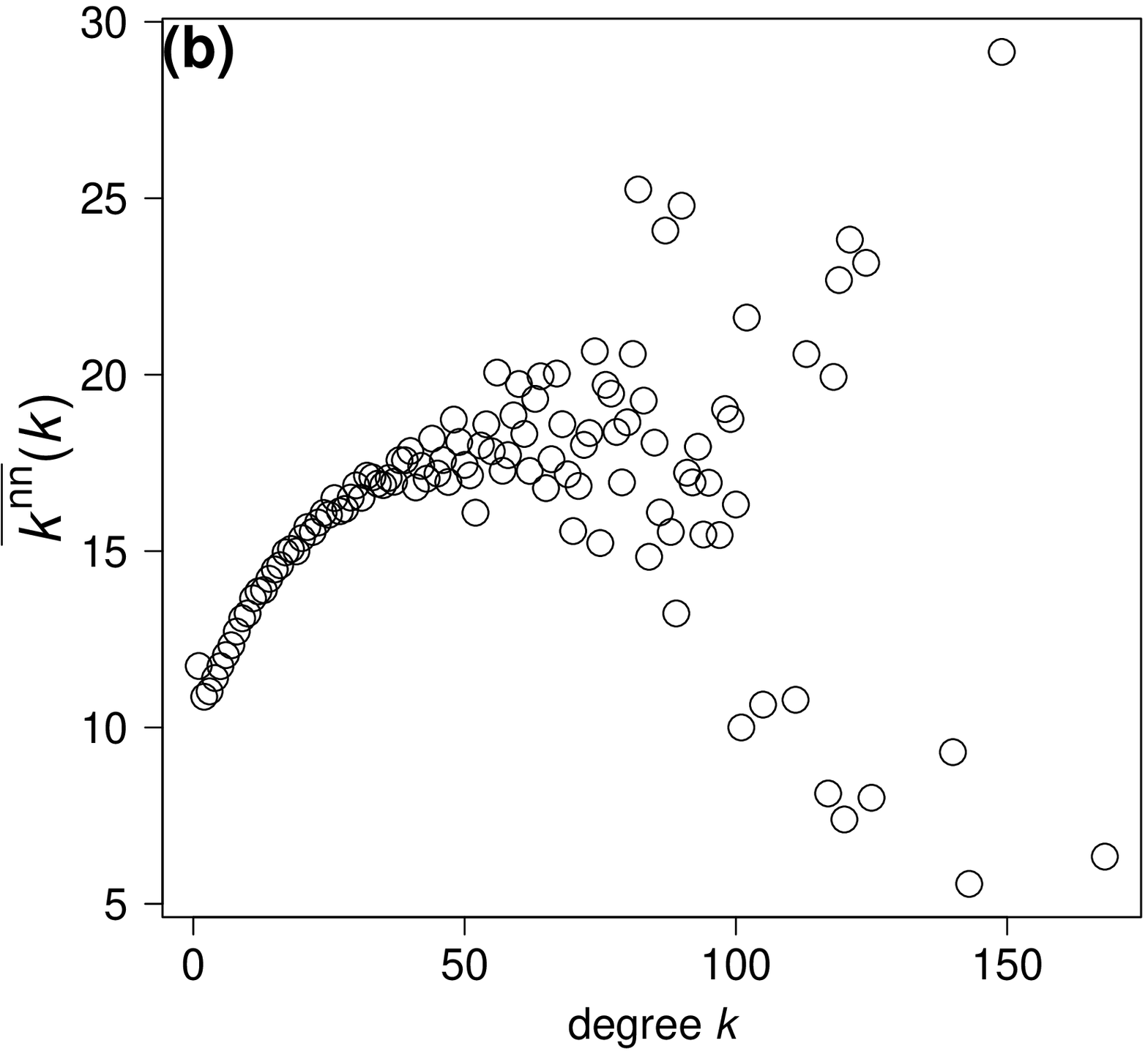}
\caption{(a) Histograms showing the node degree for the Mention network for all the nodes (squares) and for the outsiders (circles).
(b) Average degree of nodes adjacent to the nodes with degree~$k$, denoted by $\overline{k^{\rm nn}}(k)$, as a function of $k$.}
\label{fig:deg_dist}
\end{figure}

We analyze the basic statistics of the Mention network.
First, the histogram of degree, denoted by $k$, for all the nodes is shown in Fig.~\ref{fig:deg_dist}(a),
where the tail of the distribution is not as heavy as that for a power-law function.
This is probably because a user can only manage mutual communication with a limited number of other users at once,
as reported in previous studies~\cite{Goncalves2011}.

Next, we examine two features that social networks are widely believed to possess~\cite{Newman2002,Newman2003,Newman2003a}: a high clustering coefficient and degree assortative mixing.
The average clustering coefficient~\cite{Watts1998} is defined by $C \equiv (1/N) \sum_i C_i$, where $C_i = 2 \times [{\rm number \ of \ triangles \ between} \ i {\rm 's \ neighbors}] / k_i(k_i-1)$ and $k_i$ is the degree of node~$i$ $(1 \leq i \leq N)$.
The $C$ value ranges in $0 \leq C \leq 1$ and a large $C$ indicates an abundance of triangles in the network.
Degree assortativity indicates the correlation of degree between adjacent node pairs over all the links~\cite{Newman2002,Newman2003}. 
If nodes with similar values of $k$ tend to be connected with links, the network is said to be positively correlated and assortative.
Otherwise, if two nodes with a small and large $k$ tend to be connected, the network is negatively correlated and disassortative.
A standard measure used to quantify the extent of degree assortative mixing is the degree assortativity coefficient~\cite{Newman2002,Newman2003} which is defined by
\begin{equation}
r \equiv \frac{\langle k_i k_j \rangle_E - \langle (k_i + k_j)/2 \rangle_E^2} {\langle (k_i^2 + k_j^2)/2 \rangle_E - \langle (k_i + k_j)/2 \rangle_E^2},
\label{eq:assortativity}
\end{equation}
where $k_i$ and $k_j$ are the degree of nodes $i$ and $j$, respectively, and $\langle \cdot \rangle_E$ represents the average over all of links $(i, j) \in E$.
By the definition, the $r$ value is within $-1 \leq r \leq 1$ and the sign of $r$ is used as a discriminator of assortative and disassortative mixings.
Large values of $C$ and $r$ can be outcomes of homophily~\cite{Newman2002,Newman2003,Newman2003a}, i.e., a large $C$ value implies that adjacent node pairs tend to share a high proportion of common neighbors and a large $r$ value implies that nodes tend to be connected to other nodes that have a similar $k$.

For the Mention network, we observe $C = 0.132$ and $r=0.135$.
These $C$ and $r$ values are significantly larger than the values expected for random graphs with the same degree sequence
(for $1,000$ networks generated by the configuration model~\cite{Molloy1995,Newman2010}, $C$ and $r$ lie within $[0.00005, 0.0001]$ and $[-0.007, -0.001]$, respectively).
In addition, this $r$ value is larger than those reported for other networks based on online SNSs~\cite{Hu2009} which often take negative values,
but it is close to those reported for real social networks such as professional collaboration networks~\cite{Newman2003}.
In fact, as noted in previous studies~\cite{Serrano2007,Alderson2007,Whitney2008,Menche2010},
it is sometimes difficult to decide whether a network exhibits assortative or disassortative mixing patterns based on the $r$ value alone, especially when the network possesses a heterogeneous degree distribution.
Therefore, to understand the degree correlation more precisely, we investigate the average degree of the nodes adjacent to nodes with degree $k$, denoted by $\overline{k^{\rm nn}}(k)$~\cite{Pastor-Satorras2001,Vazquez2002} (where the superscript ``nn'' denotes the nearest neighbor), which is defined by
\begin{equation}
\overline{k^{\rm nn}}(k) \equiv \frac{1}{N_k} \sum_{i: k_i = k} \left( \frac{1}{k} \sum_{j \in {\cal N}_i} k_j \right),
\label{eq:knn}
\end{equation}
where $N_k$ is the number of nodes with degree~$k$ and ${\cal N}_i$ is the set of nodes adjacent to node~$i$.
An increase in $\overline{k^{\rm nn}}(k)$ with $k$ indicates an assortative mixing, whereas a decrease indicates disassortative mixing.
In Fig.~\ref{fig:deg_dist}(b), $\overline{k^{\rm nn}}(k)$ is plotted as a function of $k$.
We can see that there is an almost monotonic increase in $\overline{k^{\rm nn}}(k)$ when $1 \leq k \lesssim 40$,
which indicates the existence of an assortative mixing pattern among the nodes with a small $k$ value.
By contrast, the nodes with large $k$ value tend to hinder assortative mixing because they are connected to the nodes with small $k$ value.
The $r$ value of the Mention network is not very different from those of other online social networks~\cite{Hu2009},
but the mixing pattern shown in Fig.~\ref{fig:deg_dist}(b) is a unique property which differs from those observed in previous studies~\cite{Mislove2007,Ahn2007,Chun2008,Hu2009}.

The findings shown in Fig.~\ref{fig:deg_dist}(b) suggest that there are two types of nodes in the Mention network and that the Mention network can exhibit an assortative mixing pattern if we remove nodes of a specific type.
Based on this idea, we propose the network model illustrated in Fig.~\ref{fig:network_image} to explain the characteristic structure observed in the Mention network.
In this model, a large proportion of nodes belong to groups where nodes are tightly connected to each other and have a similar $k$ value.
However, the connections between the node groups are sparse and there are few links bridging different groups.
In contrast to the majority of nodes, a small number of nodes mediate contacts between different groups.
These nodes have three properties: (i) they tend to have a large $k$ value; (ii) they are independent of the node groups; and (iii) they have low connectivity with each other.
We refer to these nodes as outsiders because they are located outside of any of the groups.
It should be noted that this outsider model is consistent with the degree correlation observed in the Mention network (see Fig~\ref{fig:deg_dist}(b)).
In the following section, we show that this simple model is completely consistent with the explicit structure of the Mention network.

\subsection*{Increase in the degree assortativity coefficient after node removal}
\begin{figure}
\centering
\includegraphics[width=0.5\hsize]{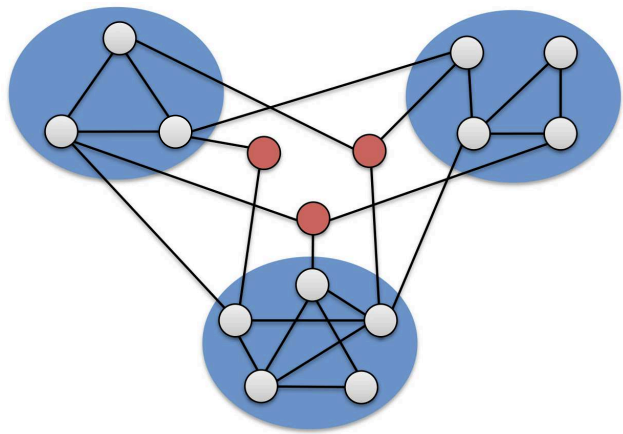}
\caption{Schematic image of the outsider network model. The larger circles filled with blue represent node groups and the smaller circles filled with red represent outsiders.}
\label{fig:network_image}
\end{figure}

First, we show that the removal of a small number of specific nodes increases the $r$ value of the resultant network,
i.e., it makes the resultant network have an assortative mixing pattern.
This supports the validity of the outsider model (shown as a schematic in Fig.~\ref{fig:network_image}) at the global network level.

We assume that outsiders would undermine assortative mixing if they exist and we identify them by removing nodes one by one from the Mention network in order to maximize the $r$ value of the resultant network (see Methods for further details).
We refer to this removal as assortativity-preference scheme.
We also perform node removal based on degree-preference and random schemes for comparison.
In degree-preference scheme, we choose the node with the largest $k$ and remove it from the network.
Next, we recalculate the $k$ value of all the remaining nodes and repeat the removal process. 
In random scheme, we remove the nodes in a uniformly random order.

\begin{figure}
\centering
\includegraphics[width=0.48\hsize]{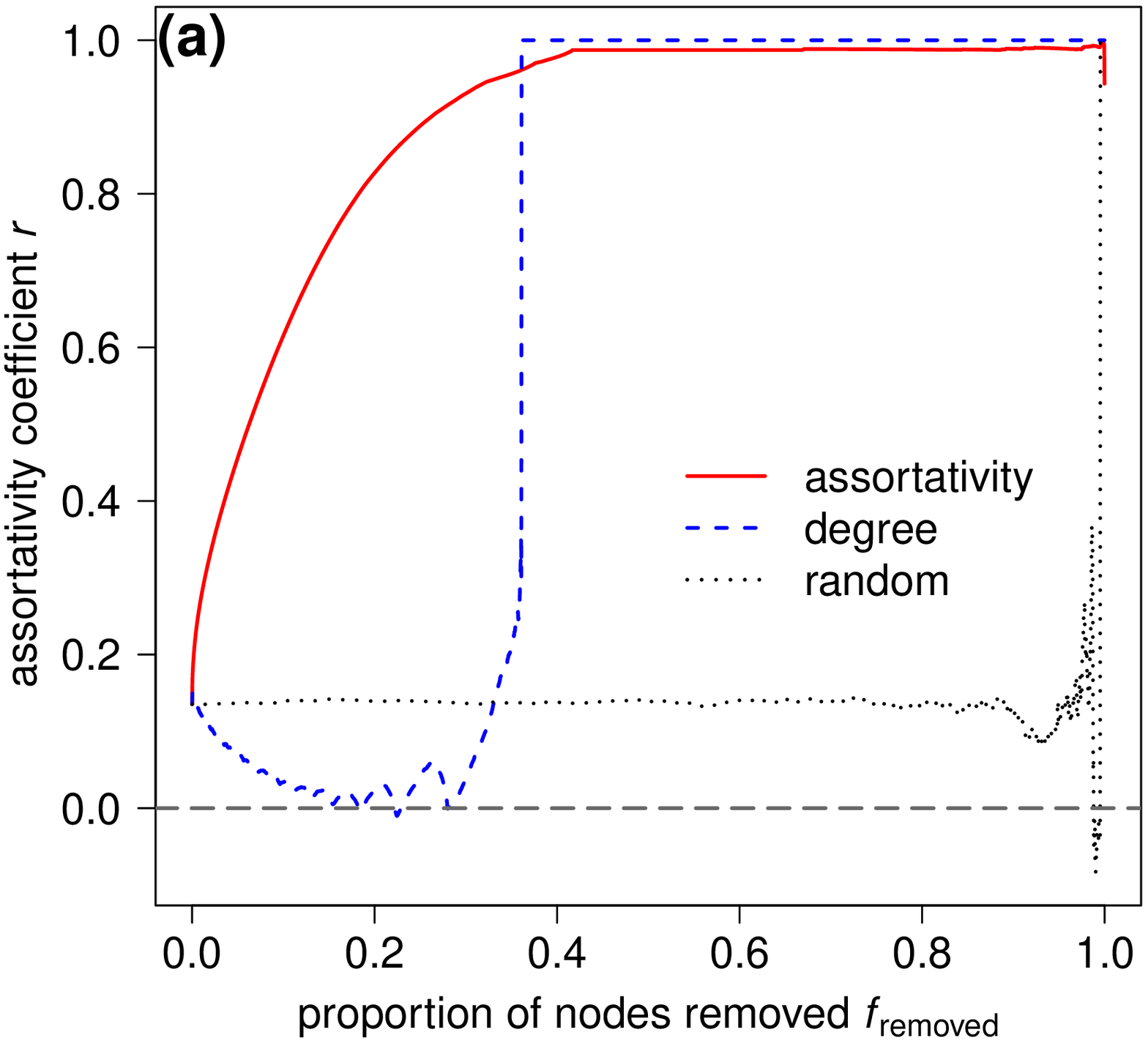}
\includegraphics[width=0.48\hsize]{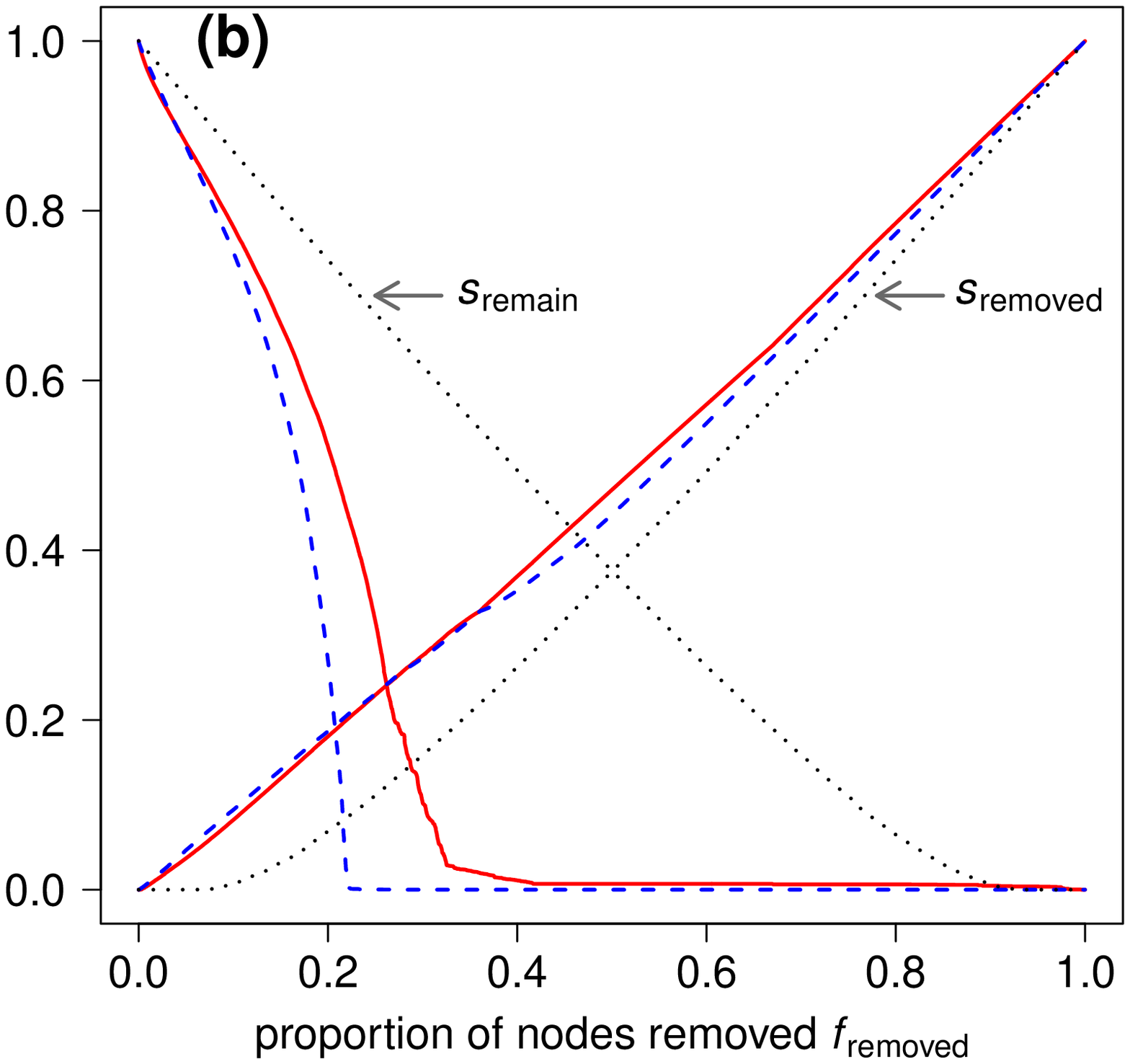}\\
\includegraphics[width=0.48\hsize]{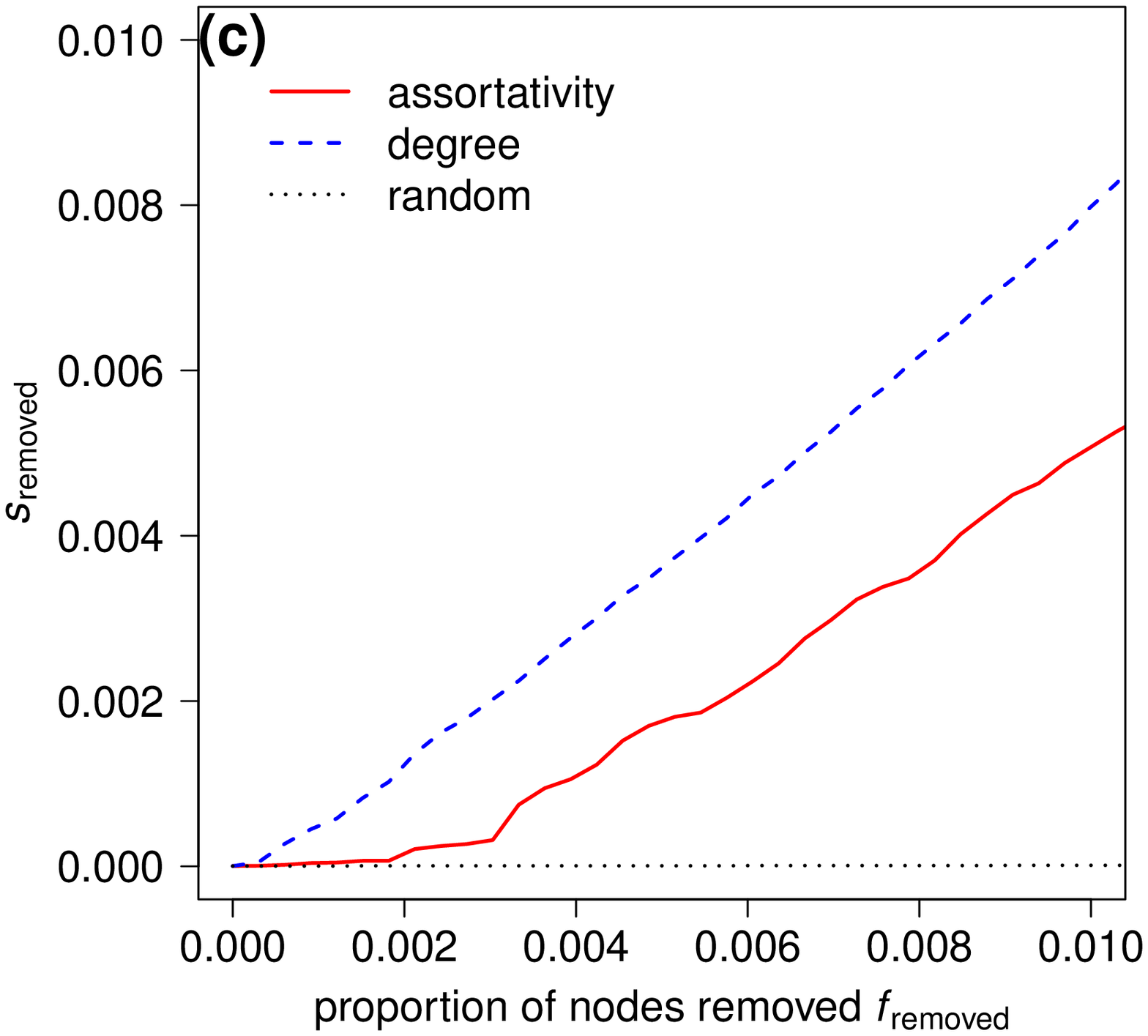}
\caption{(a) Degree assortativity coefficient $r$, (b) the sizes of the largest connected component of the remaining nodes $s_{\rm remain}$ and the removed nodes $s_{\rm removed}$, as a function of the proportion of nodes removed $f_{\rm removed}$. The nodes are removed according to assortativity-preference (solid lines), degree-preference (dashed lines), and random (dotted lines) schemes. Panel~(c) shows an enlargement of $s_{\rm removed}$ for $[0, 0.01]$ from Panel~(b).}
\label{fig:node_removal}
\end{figure}

The resulting $r$ values for the three removal schemes are shown in Fig.~\ref{fig:node_removal}(a) as a function of the proportion of nodes removed, which is denoted by $f_{\rm removed}$.
As expected, $r$ increases monotonically towards unity (i.e., the largest possible value) in assortativity-preference scheme.
However, an increase in $r$ is not observed in degree-preference and random schemes.
Indeed, in degree-preference scheme even decreases, $r$ decreases towards zero when $0 \leq f_{\rm removed} \lesssim 0.3$.
This result also implies that the order of node selection differ for assortativity- and degree-preference schemes.

The difference between the nodes chosen by the two schemes is clearer when we consider the changes in the size of the largest connected components of the remaining and removed nodes, which are denoted by $s_{\rm remain}$ and $s_{\rm removed}$, respectively (Figs.~\ref{fig:node_removal}(b) and \ref{fig:node_removal}(c)).
In the remaining networks, $s_{\rm remain}$ tends towards zero in degree-preference scheme, with a smaller value of $f_{\rm removed}$ than assortativity-preference scheme. 
The fact that $s_{\rm remain}$ maintains a large value even with a relatively large $f_{\rm removed}$ in assortativity-preference scheme is consistent with the outsider model because different groups are connected by a small number of links even in the absence of outsiders. 
The nodes removed first in assortativity-preference scheme, i.e., outsiders according to our definition, are less connected to each other,
which is consistent with the low connectivity between outsiders in the outsider model.
In Fig.~\ref{fig:node_removal}(c), we can see that $s_{\rm removed}$ is larger in degree-preference scheme than assortativity-preference scheme,
especially when $f_{\rm removed}$ is very small.
It should be noted that the large $s_{\rm removed}$ in degree-preference might be evidence of the so-called rich-club phenomenon~\cite{Zhou2004,Colizza2006} where nodes with a large $k$ value tend to be connected to each other.

\subsection*{Local structural properties of outsiders}
In this section, we evaluate the outsider model more deeply by investigating the structural properties at each node level. 
Before the analysis, we determine the number of outsiders explicitly.
We assume that outsiders are the nodes removed first in assortativity-preference scheme, 
but no unique criteria could determine the number of outsiders present in the Mention network.
Nevertheless, we define the number of outsiders as $N_{\rm outsider} = 1,000$, which are the first $1,000$ nodes removed in assortativity-preference scheme.
We set $N_{\rm outsider} = 1,000$ based on the observations shown in Fig.~\ref{fig:node_removal}.
At $f_{\rm removed} = 1,000/330,114 \sim 0.003$, the degree assortativity coefficient $r$ ($0.216$) is sufficiently larger than the original value ($0.135$),
while the remaining network retains its connectivity ($s_{\rm remain} = 0.985$) and the connectivity between outsiders is low ($s_{\rm removed}=0.0003$) .

First, we check the degree distribution of outsiders in the original Mention network.
As indicated by the filled circles in Fig.~\ref{fig:deg_dist}(a), outsiders tend to have a larger $k$ value than other nodes,
i.e., the average $k$ for all outsiders $\langle k \rangle_{\rm outsider} = 46.25$ but for all nodes $\langle k \rangle = 5.618$. 
However, it should be noted that a node with a large $k$ is not always an outsider and that nodes with a small $k$ could also be outsiders.

\begin{figure}
\centering
\includegraphics[width=0.48\hsize]{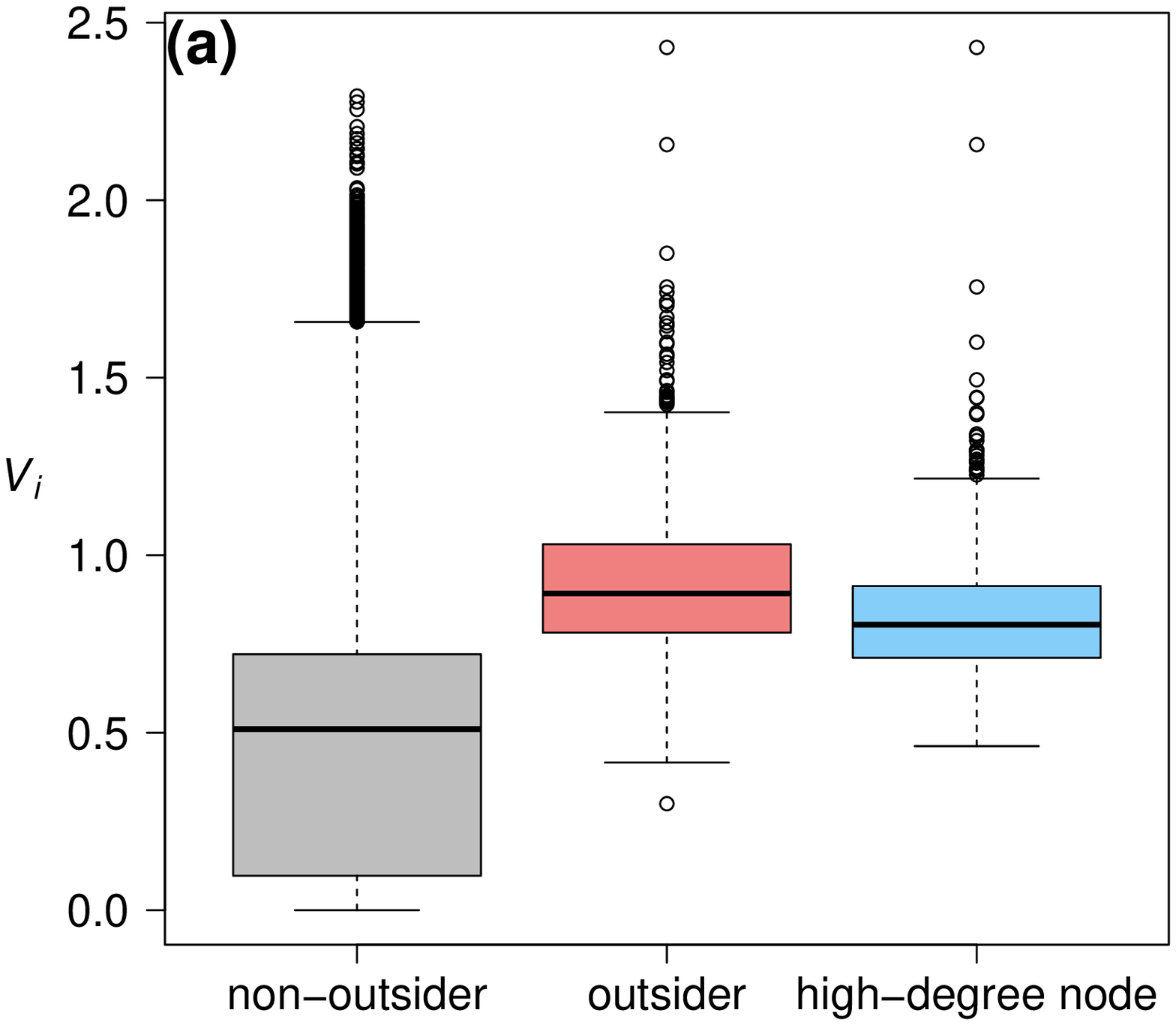}
\includegraphics[width=0.48\hsize]{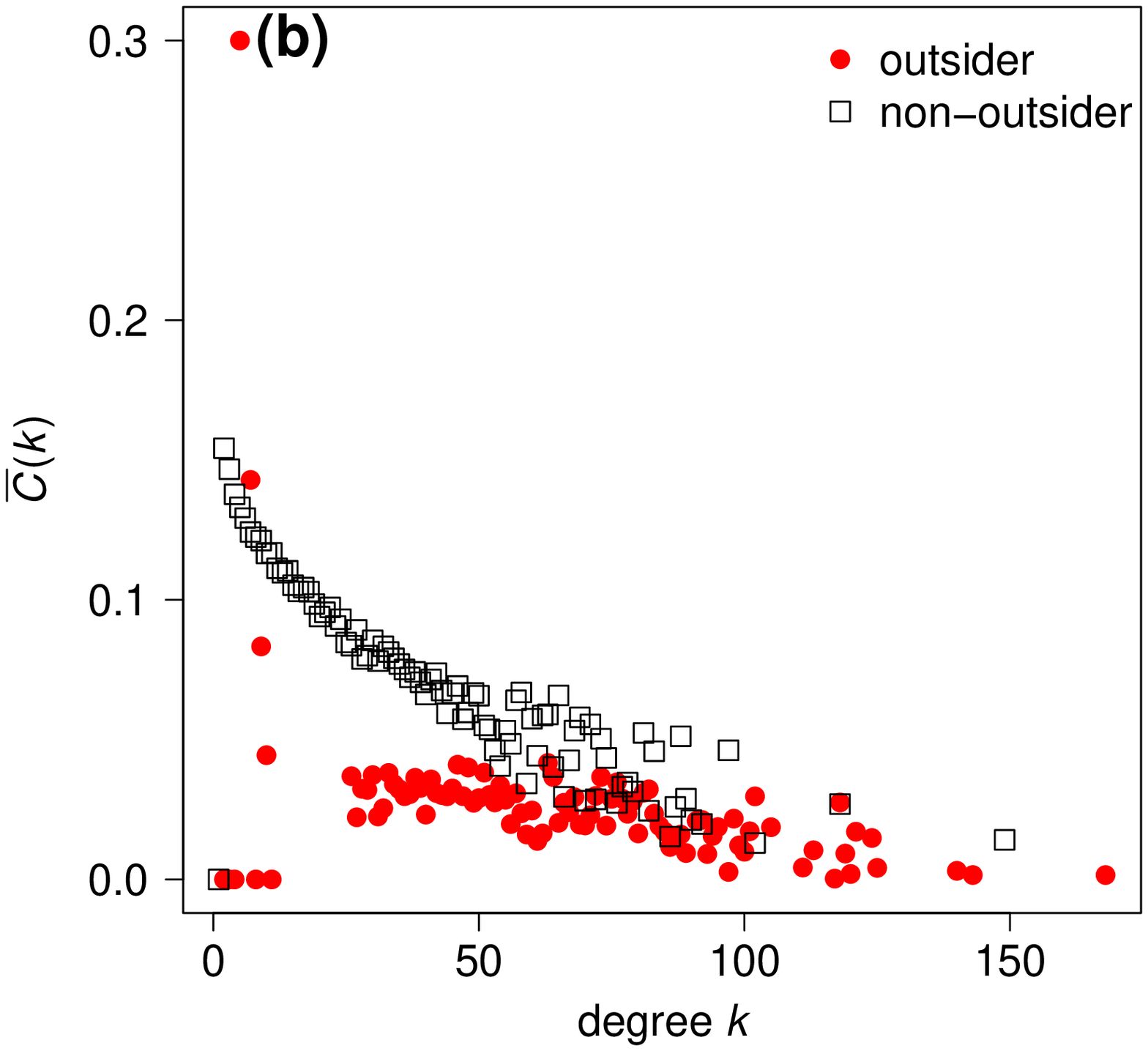}
\includegraphics[width=0.48\hsize]{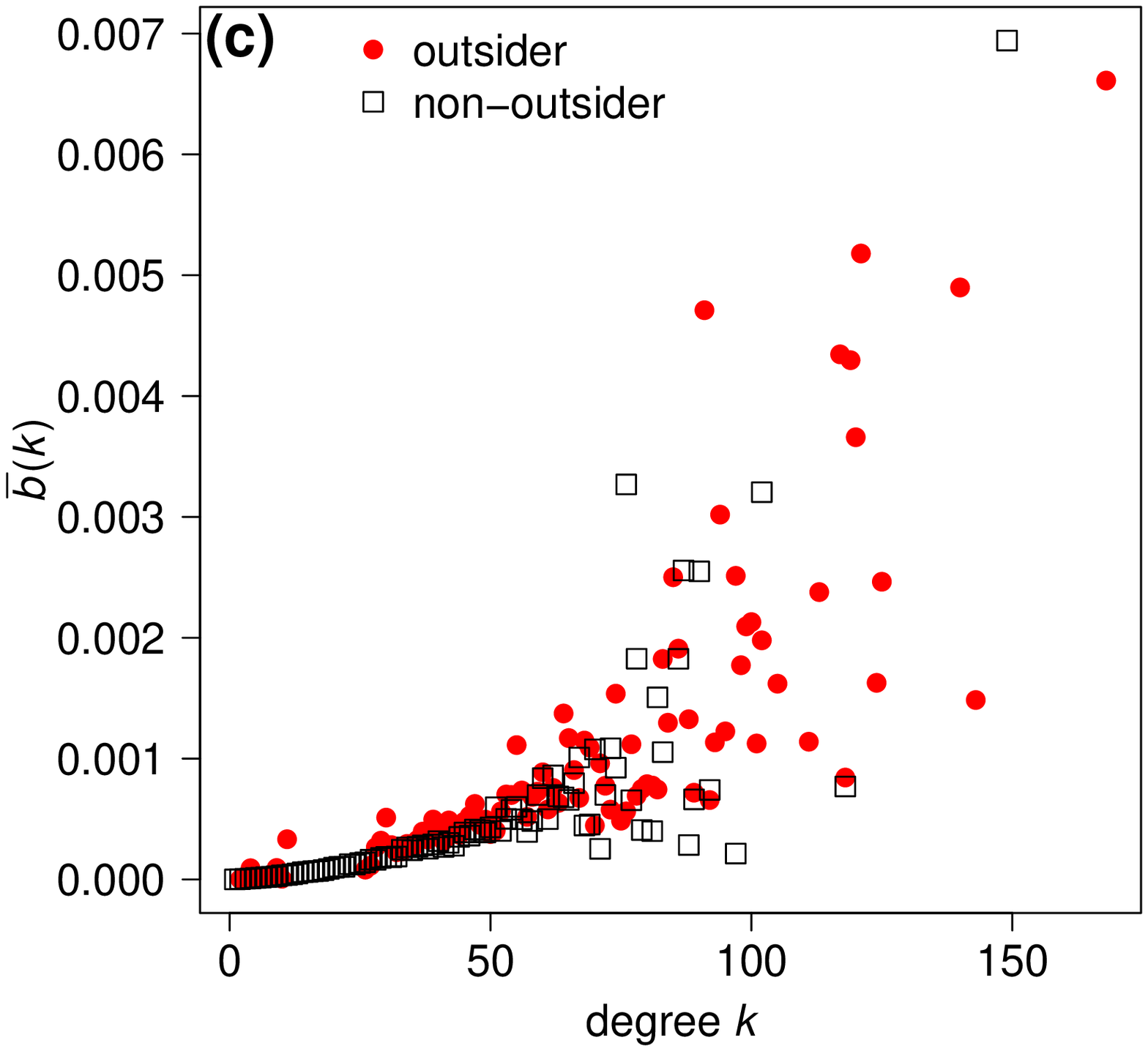}
\caption{(a) Box plots showing the diversity of degree for the neighbor nodes of node~$i$, denoted by $V_i$, for non-outsiders, outsiders, and high-degree nodes.
(b) The local clustering coefficient $C_i$ and (c) node betweenness centrality $b_i$ for outsiders (circles) and non-outsiders (squares).
Both $C_i$ and $b_i$ are the averages of the nodes with degree $k$ and plotted as a function of $k$.}
\label{fig:local_structure}
\end{figure}

Second, we examine the diversity of the $k$ values for the nodes adjacent to outsiders.
If outsiders bridge different groups where nodes have similar $k$ values, the diversity of $k$ for the neighboring nodes would be larger for those of outsiders compared with those of the other nodes.
To quantify this property, we measure the coefficient of variation of $k$ for the neighbors of node~$i$ ($1 \leq i \leq N$), which is given by
\begin{equation}
V_i \equiv
\frac{\sqrt{\mathstrut \frac{1}{k_i} \sum_{j \in {\cal N}_i} k_j^2  - \left( \frac{1}{k_i} \sum_{j \in {\cal N}_i} k_j \right)^2}}{\frac{1}{k_i} \sum_{j \in {\cal N}_i} k_j}
\end{equation}
Thus, $V_i$ is the ratio of the standard deviation relative to the average of $k_j$ over all of $i$'s neighbors $j$.
We use the coefficient of variation instead of the standard deviation or variance because we want to sift out the diversity of $k_j$ rescaled with the average.
In Fig.~\ref{fig:local_structure}(a), the distribution of $V_i$ is shown for outsiders and for all the remaining of nodes (labeled as non-outsiders) using box plots.
As expected, the average value of $V_i$ is significantly larger for outsiders than non-outsiders ($p < 2.2 \times 10^{-16}$ according to Welch's $t$~test),
while some non-outsider nodes have large $V_i$ values.
We also confirm that the large $V_i$ of outsiders is not due simply to their large $k$.
In Fig.~\ref{fig:local_structure}(a), the distribution of $V_i$ is also shown for $1,000$ nodes with the largest $k$ (labeled as high-degree nodes) using a box plot.
The average value of $V_i$ is significantly larger for outsiders than that for high-degree nodes ($p < 2.2 \times 10^{-16}$ according to Welch's $t$~test).
 
Third, we examine the local clustering coefficient $C_i$ of outsiders.
If outsiders bridge different groups that are less connected to each other,
outsiders should be involved in fewer triangles than non-outsiders.
This property is measured using the local clustering coefficient $C_i$ (for its definition, see Basic statistics of the Mention network),
where a lack of triangles associated with a node indicates a small $C_i$ value.
In Fig.~\ref{fig:local_structure}(b),  $\overline{C}(k) \equiv \left( 1/ N_k \right) \sum_{i: k_i = k} C_i$ is plotted as a function of $k_i$ for outsiders and non-outsiders.
We compare the average of $C_i$ over nodes with the same $k_i$ because $C_i$ decreases with $k_i$ by definition.
As expected, $\overline{C}(k)$ is smaller for outsiders than non-outsiders with the same $k$.
Some outsiders have a large $C_i$ with a small $k$, but they are due simply to small $k$ value.
For example, an outsider with $k_i=5$ has $C_i = 0.3$, thereby indicating that there are three links between its five neighbors.  

Finally, we examine the node betweenness centrality~\cite{Freeman1977} of outsiders.
The betweenness centrality of node~$i$, which is denoted by $b_i$, is defined by
\begin{equation}
b_i \equiv \sum_{\stackrel{j \neq \ell}{j, \ell \neq i}} \frac{\sigma_{j\ell}(i)}{\sigma_{j\ell}},
\end{equation}
where $\sigma_{j\ell}$ is the number of different shortest paths between nodes $j$ and $\ell$, and  $\sigma_{j\ell}(i)$ is the number that pass through node~$i$.
Based on the outsider model, $b_i$ for outsiders should not be very large
because there are links between different groups and the paths mediated by outsiders tend not to be the shortest path between nodes in different groups.
In Fig.~\ref{fig:local_structure}(c),  $\overline{b}(k) \equiv \left( 1/ N_k \right) \sum_{i: k_i = k} b_i$ is plotted as a function of $k_i$ for outsiders and non-outsiders.
Again, we used the average $b_i$ over outsiders and non-outsiders with the same $k_i$ because the $b_i$ value tends to increase with $k_i$.
As expected, the $\overline{b}(k)$ values of outsiders are indistinguishable from those of non-outsiders with the same $k$.   

In summary, the outsider model correctly predicts the three structural properties of outsiders: larger $V_i$, smaller $C_i$, and similar $b_i$ compared with the remaining nodes.
Therefore, we confirm that the outsider model captures the structural characteristics of the Mention network. 

Furthermore, we briefly show that the links between outsiders and the remaining nodes are not completely random,
which implies that outsiders might be connected to the nodes of different groups in a specific manner.
To demonstrate this, we perform the same analyses of the Mention network after randomizing the links connected to outsiders (see Supplementary Information for further details).
If outsiders are connected to the other nodes in a random manner in the original Mention network,
the results of the analysis of the randomized network should be similar to those shown in Fig.~\ref{fig:local_structure}. 
However, we obtain different results for the randomized network (see Sec.~S1 and Fig.~S1 in Supplementary Information), which suggests that there is a correlation among the links connected to outsiders.

\subsection*{Characteristics of outsiders in terms of other activities} 
\begin{figure}
\centering
\includegraphics[width=0.48\hsize]{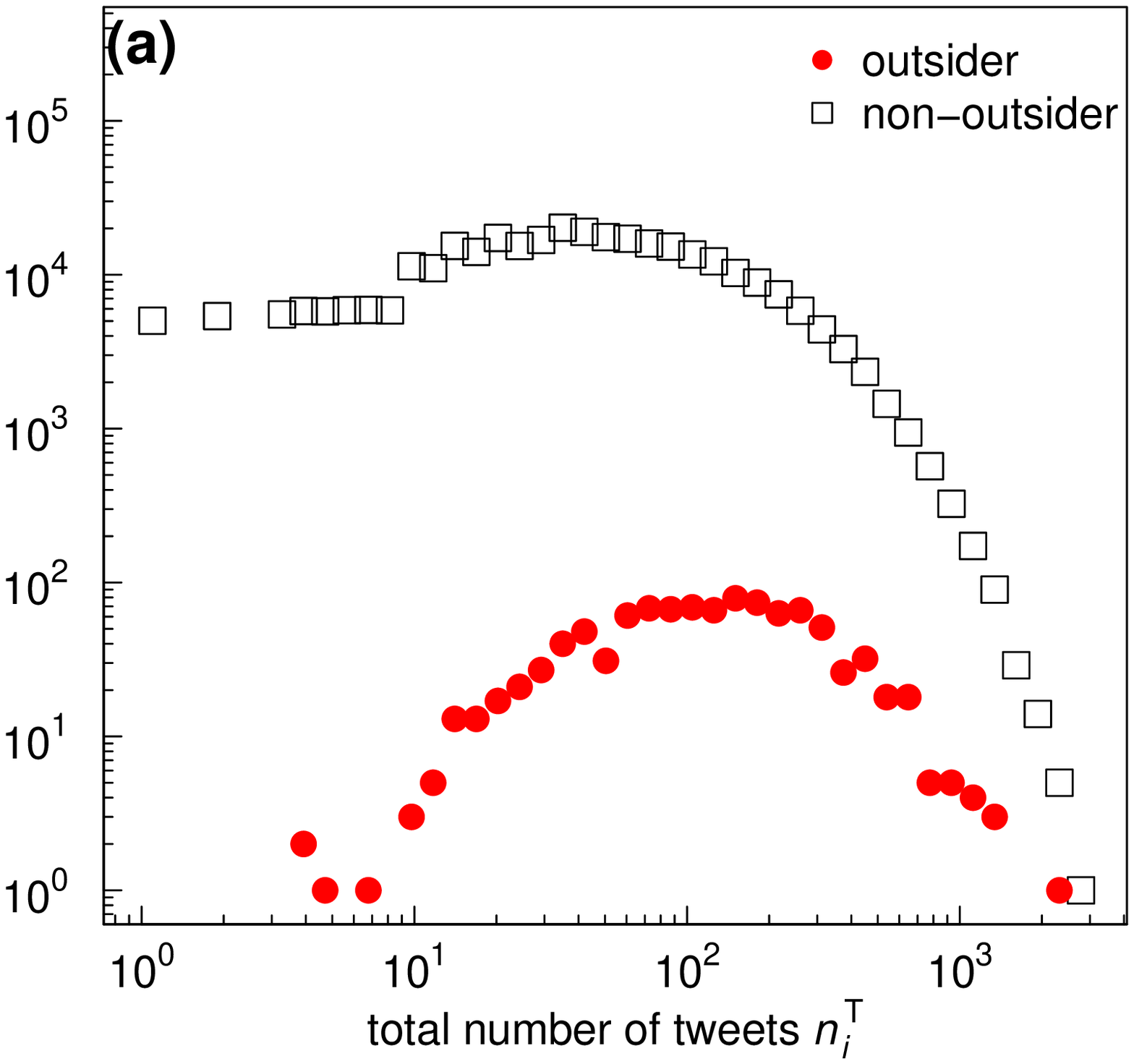}
\includegraphics[width=0.48\hsize]{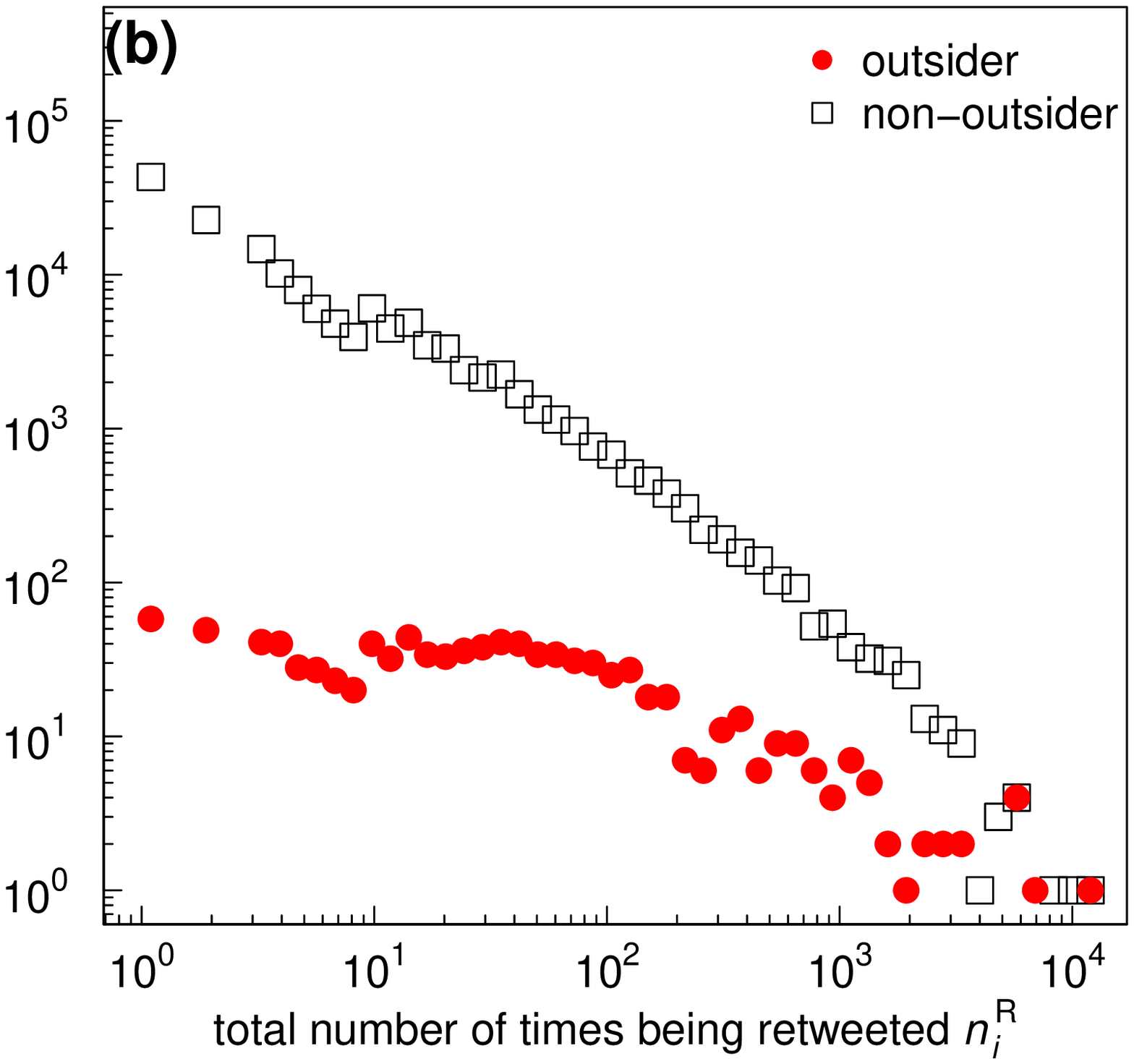}
\caption{
Histograms showing the total number of (a) tweets $n^{\rm T}_i$ and (b) the times being retweeted $n^{\rm R}_i$, for outsiders (circles) and non-outsiders (squares). 
}
\label{fig:other_activity}
\end{figure}

In the previous section, we confirmed that outsiders exhibited different properties in terms of $V_i$ and $C_i$ compared with the remaining nodes (see Fig.~\ref{fig:local_structure}).
In this section, we examine the other activities of outsiders on Twitter in addition to Mentions, in order to further understand the roles of outsiders in communication activities on Twitter.
In principle, the activities performed on Twitter are based on posting tweets, where the posted tweets are categorized into three types according to their contents.
The first type comprises Mentions in which a user mentions or directs tweets to other users.
The second type comprises Retweets in which a user replicates a tweet that was originally posted by another user.
The third type comprises simple tweets which is any tweets other than Mention or Retweet.
To quantify the other activities of users, we use the following two measures. 
First, the total number of tweets posted by user~$i$, denoted by $n^{\rm T}_i$, is a proxy of the activity level of user~$i$.
Second, the total number of times that the tweets made by user~$i$ are retweeted, which is denoted by $n^{\rm R}_i$, is a proxy of the popularity of user~$i$.
Figures.~\ref{fig:other_activity}(a) and (b) show the histograms of $n^{\rm T}_i$ and $n^{\rm R}_i$, respectively, for outsiders and the remaining nodes.
On average, $n^{\rm T}_i$ and $n^{\rm R}_i$ are larger for outsiders than those of non-outsiders,
i.e., average of $n_i^{\rm T} = 173.64$ $(71.19)$ and the average of $n_i^{\rm R} = 47.47$ $(13.38)$ for outsiders (non-outsiders).
These results suggest that outsiders are likely to be more active and popular on Twitter than other users.

\subsection*{Other data sets}
In the previous sections, we showed that the existence of outsiders is a characteristic of the Mention network.
Thus, we might ask whether this is a unique property of communication on Twitter or if other online communication networks also exhibit this property.
To answer this question, we perform the same analyses to five other network data sets that are publicly available online: the Enron~\cite{Klimt2004}, EU-email~\cite{Leskovec2007}, Facebook~\cite{Viswanath2009}, Slashdot~\cite{Gomez2008}, and Wikipedia~\cite{Leskovec2010} networks.
All of these network data sets are obtained from online communication logs and made available online by the Koblenz network collection~\cite{Kunegis2013}.
These networks are originally directed, and we draw an undirected link between node pairs if the two nodes have reciprocal contacts,
in the same manner as the definition of the Mention network (see Methods).  

The basic statistics for these network data sets and the figures are shown in Supplementary Table S1 and Supplementary Figures S1, S2, S3, S4, and S5,
but we provide a brief summary of the results here.
The outsider model does not appear to be applicable to the Enron and EU-email networks, because both networks exhibit disassortative mixing patterns.
This may be logical if we consider that the communication activities in these two networks are constrained by their formal organization structure, i.e., a focal company and institution.
Furthermore, the outsider model is not applicable to the Slashdot and Wikipedia networks, because these two networks exhibit no obvious degree correlation.
Only the Facebook network exhibits a similar pattern to that we observed in the Mention network, i.e., assortative mixing between nodes with small $k$ values, low connectivity between outsiders, and larger $V_i$ values and smaller $C_i$ values for outsiders.
These results suggest that the existence of outsiders might be a characteristic of online communication networks that are defined by rather private and casual communication.
 
\section*{Discussion}
In this study, we showed that the structure of the Mention network on Twitter is characterized by the existence of outsiders, i.e., a small number of distinctive users who are not involved in any particular user groups but who are connected to different groups.
When we removed outsiders from the network, the remaining part of the network exhibited an assortative mixing pattern which agrees with our conventional understanding of social networks.
In addition to Mention activity, we also confirmed that outsiders are more active and popular than other users on Twitter.
The existence of outsiders should be considered carefully when studying processes taking place in Twitter, such as information spreading processes and group organizations.

The presence of outsiders is related to the structural hole theory~\cite{Burt1995}, which has been proposed in previous studies of social network analysis.
When two node groups in a network are separated from each other, there is a structural hole between the two groups.
If a node brokers the two groups by connecting the links, this node may benefit by capturing non-redundant information from the two groups and controlling the flow of information between them.
Thus, outsiders in the Mention network hold the brokerage of structural holes.
The unique feature of outsiders is that they are separate from any node groups, although this is not necessary for nodes that bridge structural holes.

Our results are consistent with the previous study on a sampled Twitter network~\cite{Grabowicz2012},
which showed that the tweets posted by the intermediate users located in different groups (or communities) were more likely to be retweeted than those of users within a single community.
In terms of the definition of the intermediate users,
our analysis can be regarded as a complementary approach to that described in Ref.~\cite{Grabowicz2012}.
In particular, we identified outsiders initially by node removal and we then investigated their connections to the remainder of the network,
whereas all of the nodes were clustered into communities (although a node might belong to no community) and the nodes that belonged to multiple communities were then identified in Ref.~\cite{Grabowicz2012}. 

The classification of Twitter users was also proposed in the previous study based on ego network structure~\cite{Arnaboldi2013}.
In Ref.~\cite{Arnaboldi2013}, the users were categorized as occasional users, regular users, or aficionados, based on the length of their active periods.
The majority of users were occasional users who tended to make frequent contacts with a number of other users initially but they stopped posting tweets after a short period, unlike regular users and aficionados.
The number of aficionados was rather small, but they tended to maintain stable ego networks for longer periods.
These results appear to contradict our finding that a small number of outsiders actively communicate with users in different groups, whereas the majority of users have conversations within their own groups.
This disparity may be attributable to the difference in the length of observation period, i.e., seven days for our data set and up to seven years (the time of the oldest available tweet depended on each specific user) for the data set considered in Ref.~\cite{Arnaboldi2013}.
Because occasional users tend to stop using Twitter after a short period, aficionados are more likely to be observed than occasional users in a given short observation period, such as seven days.
Therefore, our data set might be biased toward aficionados.
Further our analysis should be extended using a data set acquired over a longer period in future research. 

The reason why outsiders are present on Twitter is still unclear from our analysis,
although we suggest the following two causes.
First, they may be users who want to promote their popularity online~\cite{Naaman2010}.
Some users, such as bloggers or lesser-known TV personalities, may try to engage with a number of their followers at the same time by chatting with the use of this low effort communication tool.
Second, some of outsiders may represent the user accounts of so-called bots, which behave automatically on Twitter according to computer programs~\cite{Chu2010,Tavares2013}.
Some of the bot accounts are programmed to search for specific tweets posted by other users (e.g., those containing predefined keywords) and to reply to the users by sending Mentions.

Our analysis was based on the Mention network that mainly comprised Japanese users,
and the same analysis of networks composed of users with other languages would provide more general anthropological insights~\cite{Takhteyev2012,Mocanu2013,Saito2014}.
We discarded the weights of links (e.g., the interaction frequency between node pairs) in the present study,
but structural analysis that considers the link weights would be useful in future work
because the link weights may play a crucial role in organizing social networks, e.g., in the ``strength of weak ties'' phenomenon~\
\cite{Granovetter1973,Onnela2007,Takaguchi2011,Grabowicz2012}.

We should also discuss the relationship between outsiders in the Mention network and a similar phenomenon that has been observed in networks in a completely different context, that is, the metabolic systems of living organisms.
In Ref.~\cite{Huss2007}, the substrates of metabolic systems were considered in the form of a graph in which each node represented a metabolite, i.e., a substance involved in biochemical reactions, and an undirected link was drawn between two nodes if one could produce the other.
The so-called currency metabolites that appeared in widely different reaction processes and that played an important role in the system were identified purely on the basis of the network structure~\cite{Huss2007}.
The resulting currency metabolites overlapped greatly with those conventionally suggested in the previous biological studies, such as water, oxygen, and hydrogen ion.
In addition, these currency metabolites share some characteristics with outsiders, i.e., very few are present in networks and they are apart from any modules (i.e., groups), they have a large degree, and the resultant network becomes more module-like after they are removed.
Thus, although more careful investigations should be performed to extend this simple analogy,  
this similarity between two networks in different contexts might suggest that these special classes of nodes have important effects in general problems,
e.g., robustness of networks against damages and efficiency of information flows on networks.

\section*{Methods}

\subsection*{Twitter data set and construction of the Mention network}
We crawled public tweets posted by a set of designated users via Twitter API.
We determined these users by a snowball sampling in the following manner.
On 15th March, 2011, we selected about $30$~users who had the highest numbers of followers and we collected all the public tweets posted by them after that date.
Next, we tracked back and collected their tweets until we reached the posts on 11th March, 2011.
When we found new user IDs in the collected tweets (i.e., in Mentions and Retweets), we added the new users to the set of designated users and started collecting their subsequent tweets.   
Because all of the initial designated users used Twitter in Japanese, most of the subsequent users also communicated in Japanese. 
In total, we had designated $499,733$~users by 1st December, 2011.
We collected all of the tweets posted by these users from 1st to 7th December, 2011, which comprised $48,949,334$~tweets including $19,667,443$~Mentions and $4,123,662$~Retweets.
Before analyzing the data set, we anonymized the user IDs and discarded the text contents of the tweets.

Using the tweets collected, we constructed the Mention network $G = (V, E)$ as follows.
We considered a network $G^\prime = (V^\prime, E^\prime)$,
where $V^\prime$ is the set of designated users,
and there is link $(i,j) \in E^\prime$ if two users $i$ and $j$ send Mentions to each other in both directions within the observation period.
We defined the Mention network $G = (V, E)$ 
as the largest connected component of network $G^\prime$. 

\subsection*{Implementation of node removal based on assortativity-preference scheme}

The naive computation of the degree assortativity coefficient $r$ (defined by Eq.~\eqref{eq:assortativity}) requires $O(M)$ time,
and recalculating $r$ after removing all the nodes one by one requires $O(N M)$ time, which is infeasible for large networks such as the Mention network.
Therefore, we used an efficient implementation of node removal, which avoids fully recomputing the terms in $r$, as described below.

Let us define
$A_i \equiv \sum_{j \in {\cal N}_i} k_i k_j$,  $B_i = \sum_{j \in {\cal N}_i} (k_i^2 + k_j^2)/2$,  and $C_i \equiv \sum_{j \in {\cal N}_i} ((k_i + k_j)/2)^2$. 
Then, we have
\begin{align}
 r = \frac{\sum_{i} A_i - \sum_{i} C_i}{\sum_{i} B_i - \sum_{i} C_i}.
\end{align}
When node $i$ is removed,
only the degrees of $i$'s neighbors $j \in {\cal N}_{i}$ are changed,
and thus only $A_\ell$, $B_\ell$, and $C_\ell$ need to be recomputed for the second neighboring nodes $\ell$ of node~$i$, i.e., $\ell \in \bigcup_{j \in {\cal N}_{i}} {\cal N}_{j}$.
Therefore, recomputing $r$ after removing node~$i$ requires $O( k_{i}^{(2)} )$ time, where $k_{i}^{(2)}$ is the
number of nodes that are the second neighbors of node~$i$.
In general, $k_{i}^{(2)} \ll M$ holds true and this implementation facilitates assortativity-preference node removal in a manageable amount of computational time.


\section*{Acknowledgments}
The network data sets we used in Other data sets were downloaded from the Koblenz Network Collection (http://konect.uni-koblenz.de/).

\section*{Author contributions}
M.T. collected and organized the data set. T.T. and T.M. conceived and designed the research. 
T.T. and T.M. analyzed the data. T.T., T.M., M.T., and K.K. discussed the results and wrote the manuscript. 

\section*{Competing financial interests}
The authors declare no conflict of interest associated with this manuscript.

\clearpage

\clearpage

\clearpage

\clearpage

\clearpage

\clearpage
\setcounter{page}{1}
\begin{center}
{\bf {\Large Supplementary Information}}\\
\vspace*{3mm}
{\large for}\\
\vspace*{3mm}
\textit{{\large Taro Takaguchi, Takanori Maehara, Masashi Toyoda, \\ and Ken-ichi Kawarabayashi}}\\
\vspace*{3mm}
{\bf {\large Existence of outsiders as a characteristic of online communication networks}}\\
\end{center}

\newpage

\renewcommand{\thesection}{S\arabic{section}}
\renewcommand{\thefigure}{S\arabic{figure}}
\setcounter{figure}{0}
\renewcommand{\thetable}{S\arabic{table}}
\renewcommand{\theequation}{S.\arabic{equation}}
\section{Rewiring of links connected to outsiders}
We first remove all the links connected to outsiders and then rewire the links while keeping the degree of all the nodes.
In total, $46,465$ links associated with outsiders are rewired ($\sim 5\%$ of $M = 927,352$ links in the original mention network). 
This rewiring process keeps the total number of links and the degree distribution (it possibly create self-loops but this effect is negligible since the number of the rewired links is sufficiently large).
By contrast, the rewiring process may change $C_i$, $r$, and the connectedness of the network.
If it holds true that outsiders cast their links to other nodes in a random manner,
the following properties should be observed for the rewired network:
(1) an increase in $r$ after the removal of the same set of outsiders,
(2) low connectivity between outsiders,
(3) similar $V_i$ and $C_i$ values for outsiders as those in the original Mention network. 

\begin{figure}
\centering
\includegraphics[width=0.48\hsize]{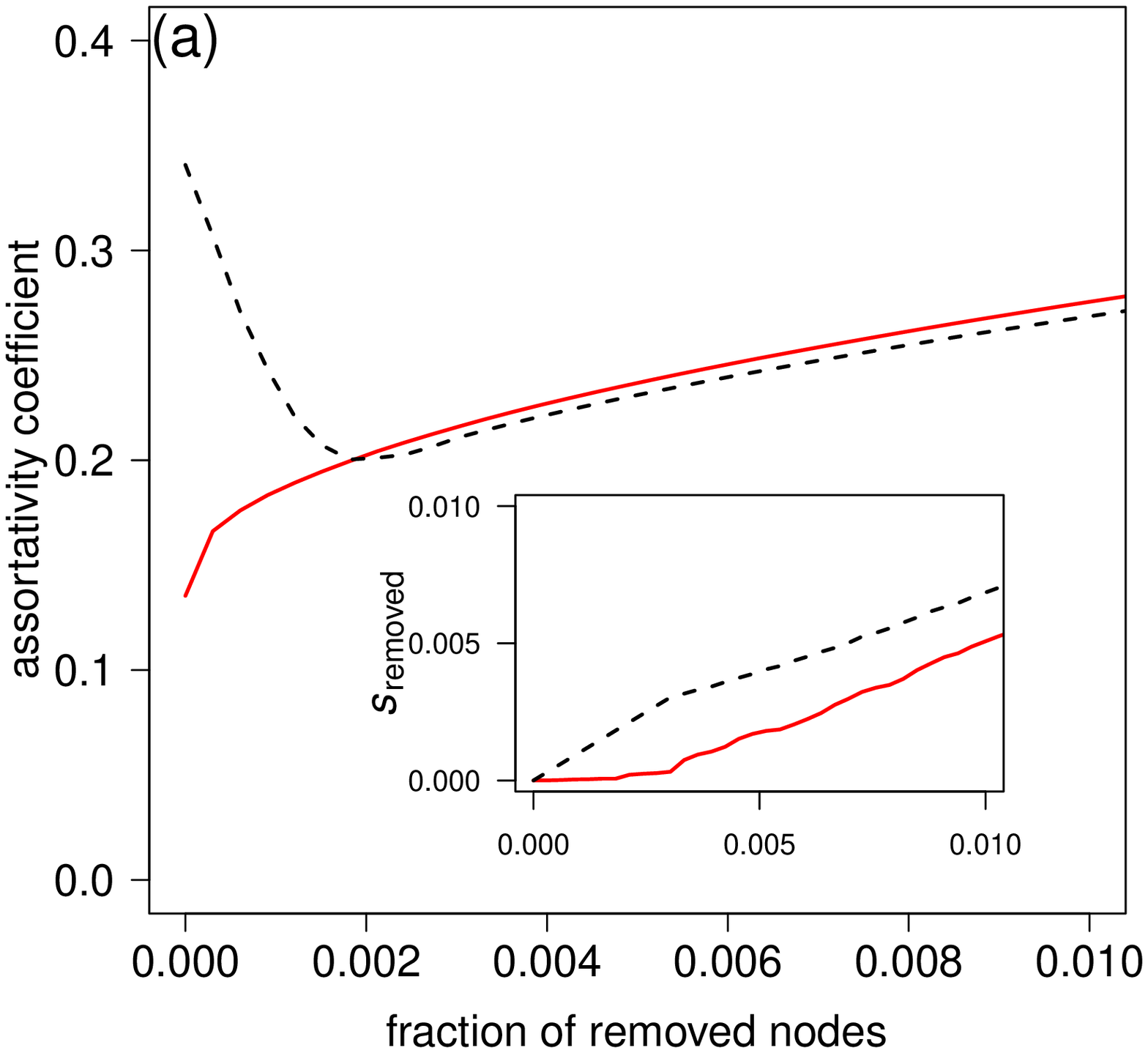}
\includegraphics[width=0.48\hsize]{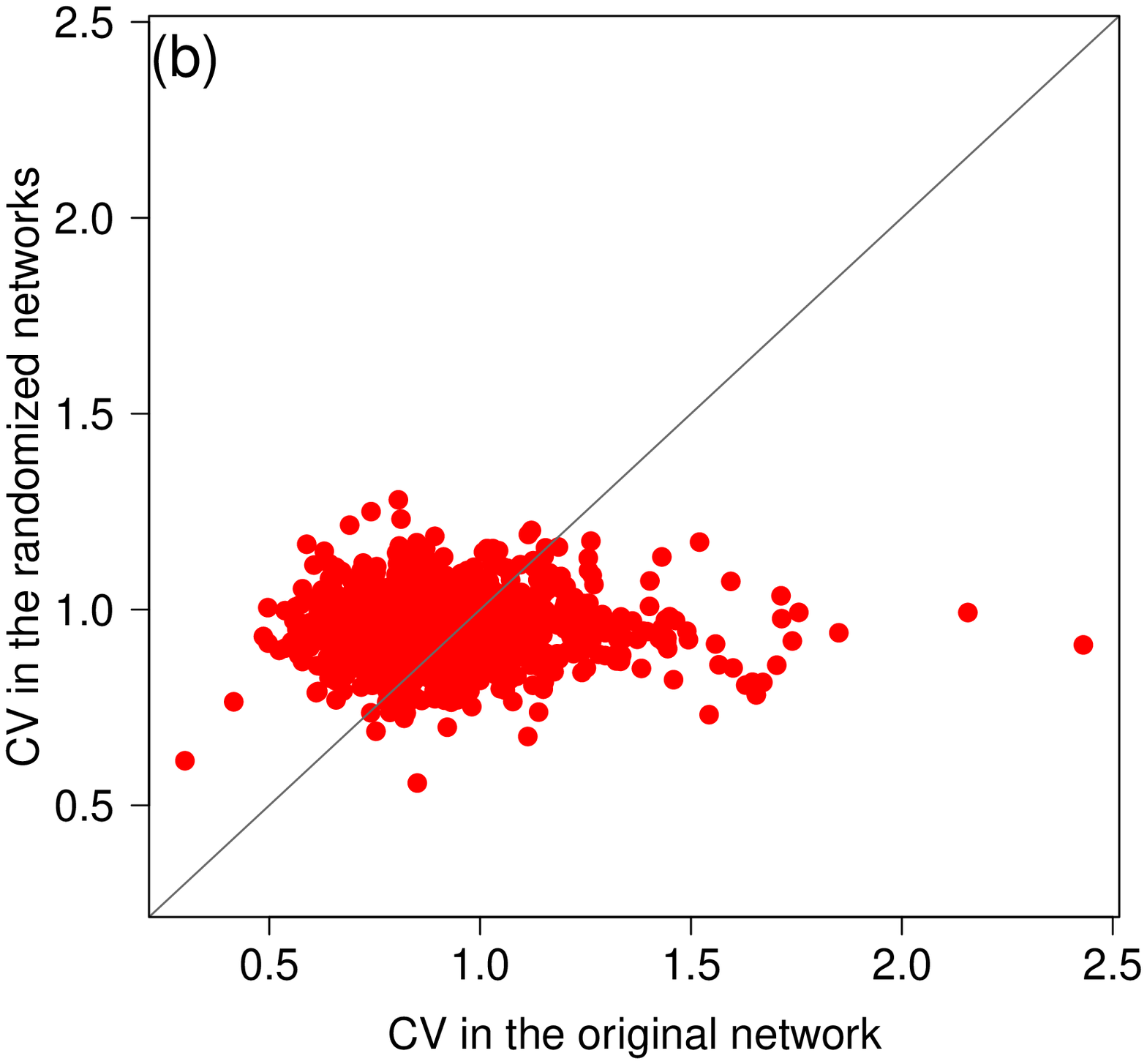}
\includegraphics[width=0.48\hsize]{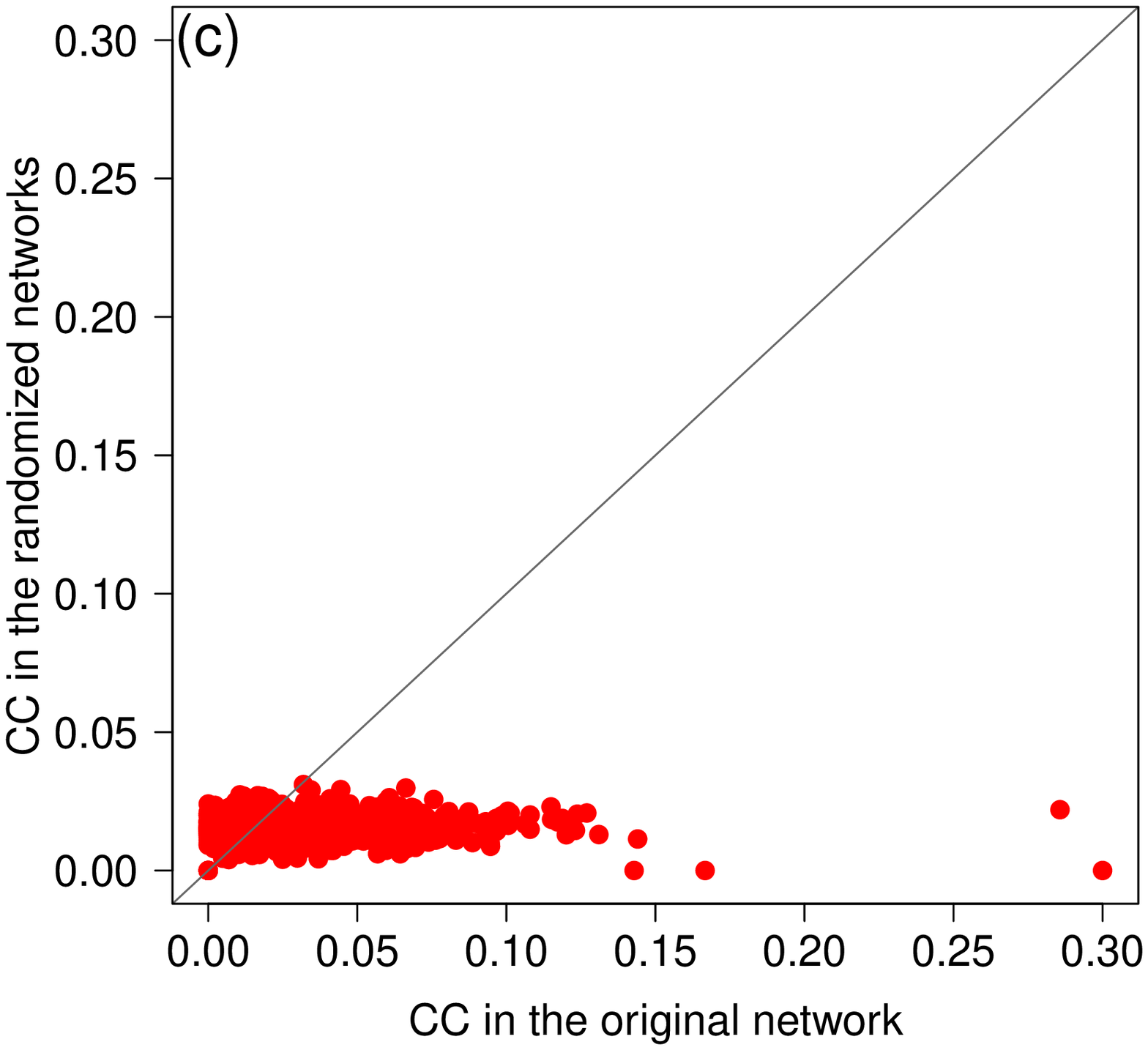}
\caption{Results for a rewired mention network.
(a) Degree assortativity coefficient $r$ of the resultant networks (main panel) and the sizes of the largest connected component of the removed nodes $s_{\rm removed}$ (inset),
as a function of the proportion of nodes removed $f_{\rm removed}$.
The dashed lines are the results for the rewired network and the solid lines the original network as a reference. 
(b) Scatter plot of the diversity of the neighbors' degree $V_i$ of outsiders in the original network (horizontal axis) against $V_i$ in the rewired network (vertical axis).
Each dot corresponds to an outsider.
(c) Scatter plot of the local clustering coefficient $C_i$ of outsiders in the original network (horizontal axis) against $C_i$ in the rewired network (vertical axis).
}
\label{fig:rewired}
\end{figure}

In Fig.~\ref{fig:rewired}, the results for a rewired network are shown, which indicate that the links between outsiders and other nodes are not completely random.
First, as shown in Fig.~\ref{fig:rewired}(a), the $r$ value of the rewired network without node removal is equal to $0.341$ and larger than $0.135$ for the original Mention network.
This implies that outsiders help the rewired network be assortative, which is also supported by the fact that $r$ value decreases in $f_{\rm removed} \lesssim 0.002$.
The $s_{\rm removed}$ value in the rewired network is larger than the original network, and outsiders are more connected to each other.
The rewiring process reduces the variety of $V_i$ and $C_i$ among outsiders as shown in Figs.~\ref{fig:rewired}(b) and (c).
While $V_i$ of outsiders take values within $[0.3, 2.5]$ in the original network, $V_i$ in the rewired network concentrate around unity.
In a similar way, $V_i$ of outsiders take values within $[0, 0.3]$, $C_i$ in the rewired network take value close to zero, regardless the $V_i$ values in the original network.
These results suggest that the links between outsiders and other nodes play an important role in determining the local structure around outsiders
and that the links are not created in a random manner.

\section{Results for other data sets}
\begin{figure}
\centering
\includegraphics[width=0.35\hsize]{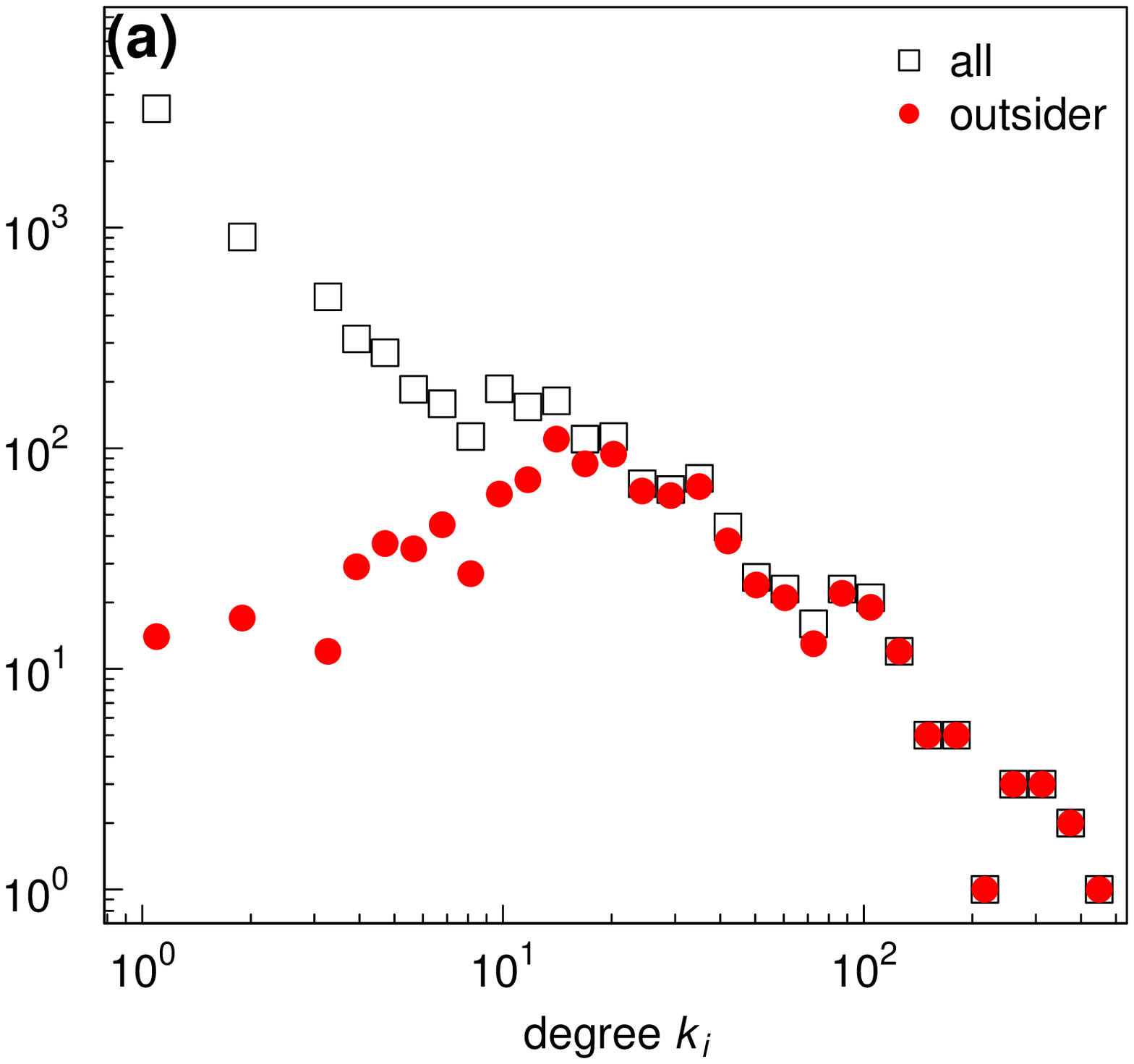}
\includegraphics[width=0.35\hsize]{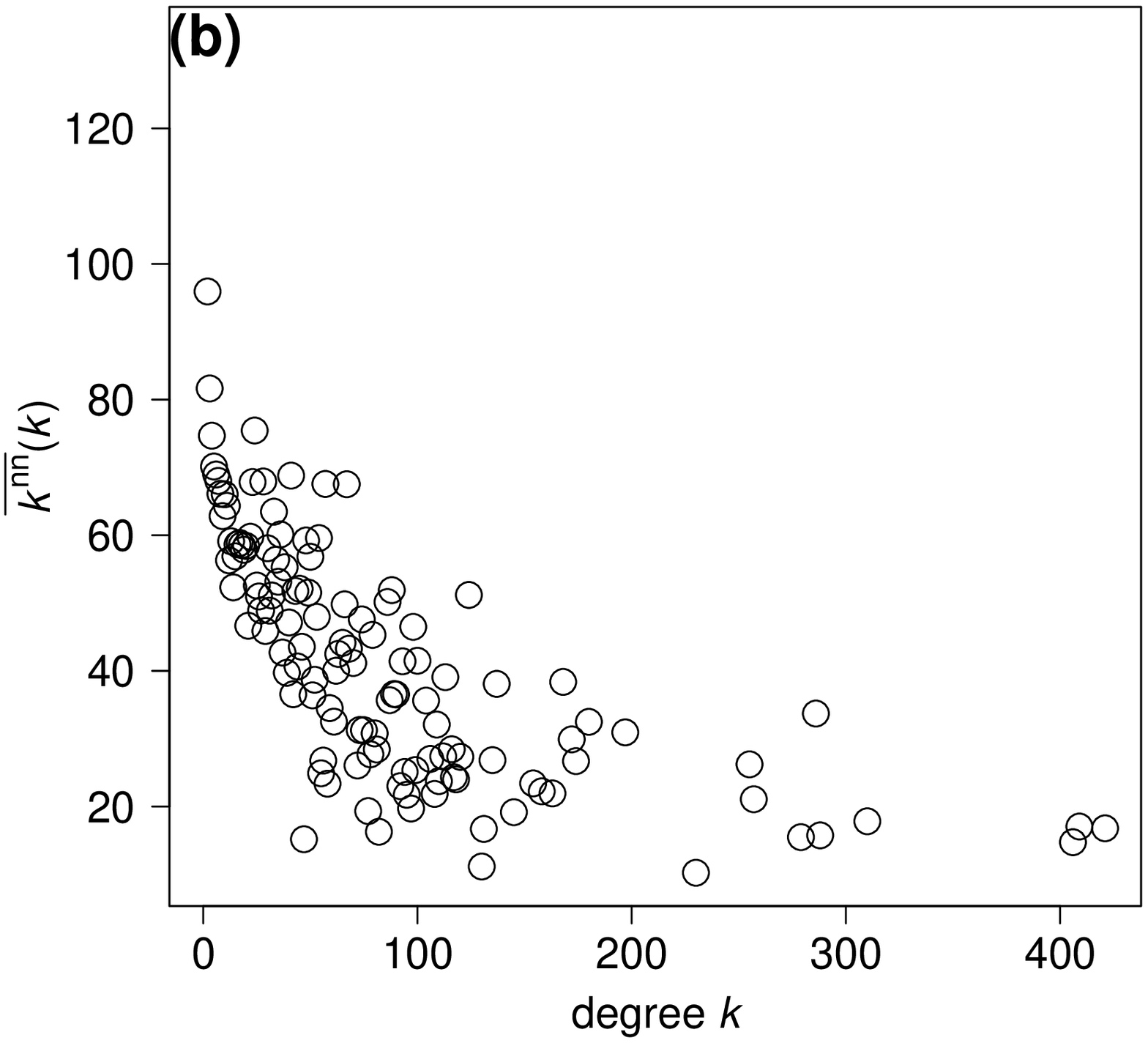}
\includegraphics[width=0.35\hsize]{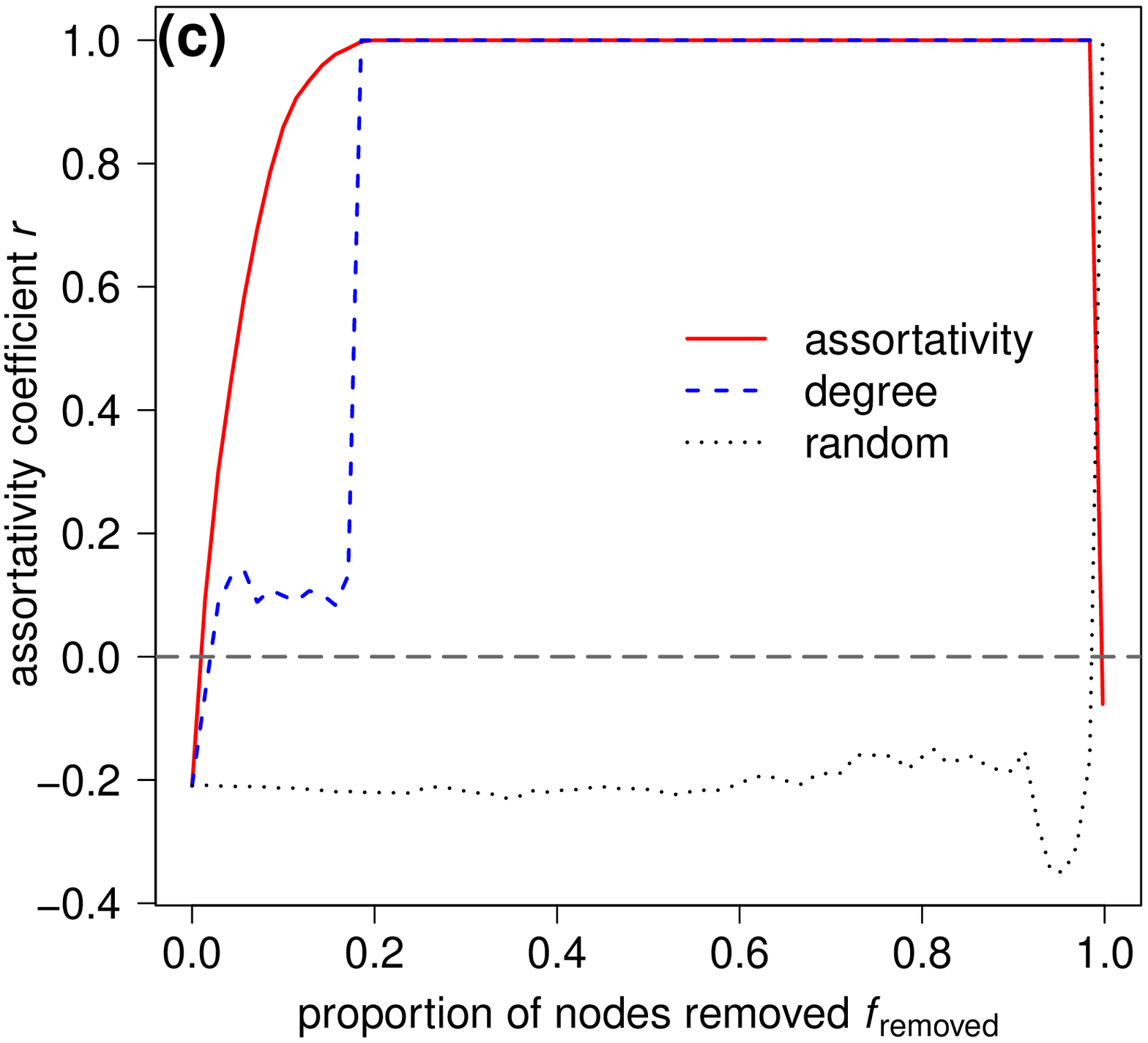}
\includegraphics[width=0.35\hsize]{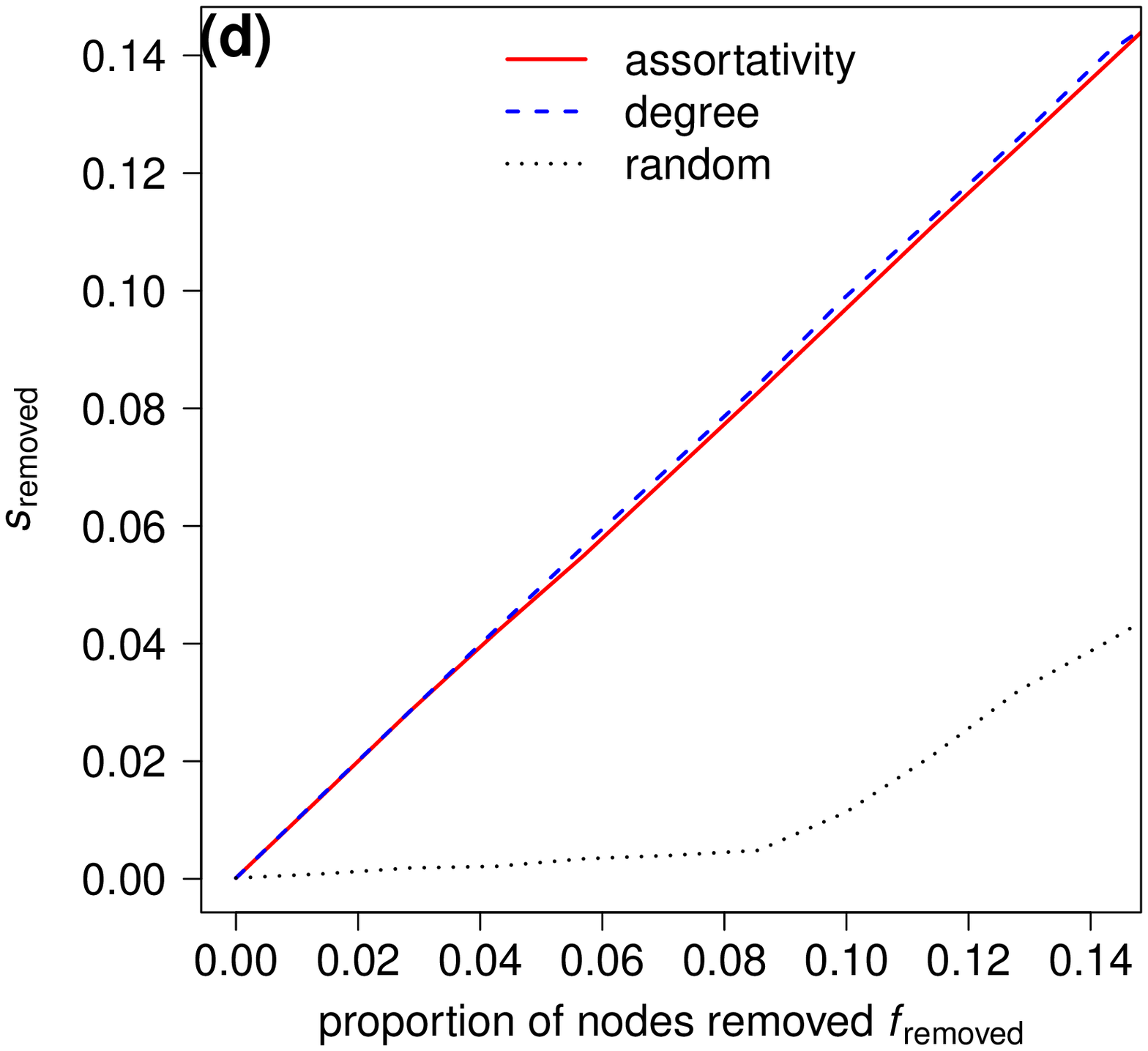}
\includegraphics[width=0.35\hsize]{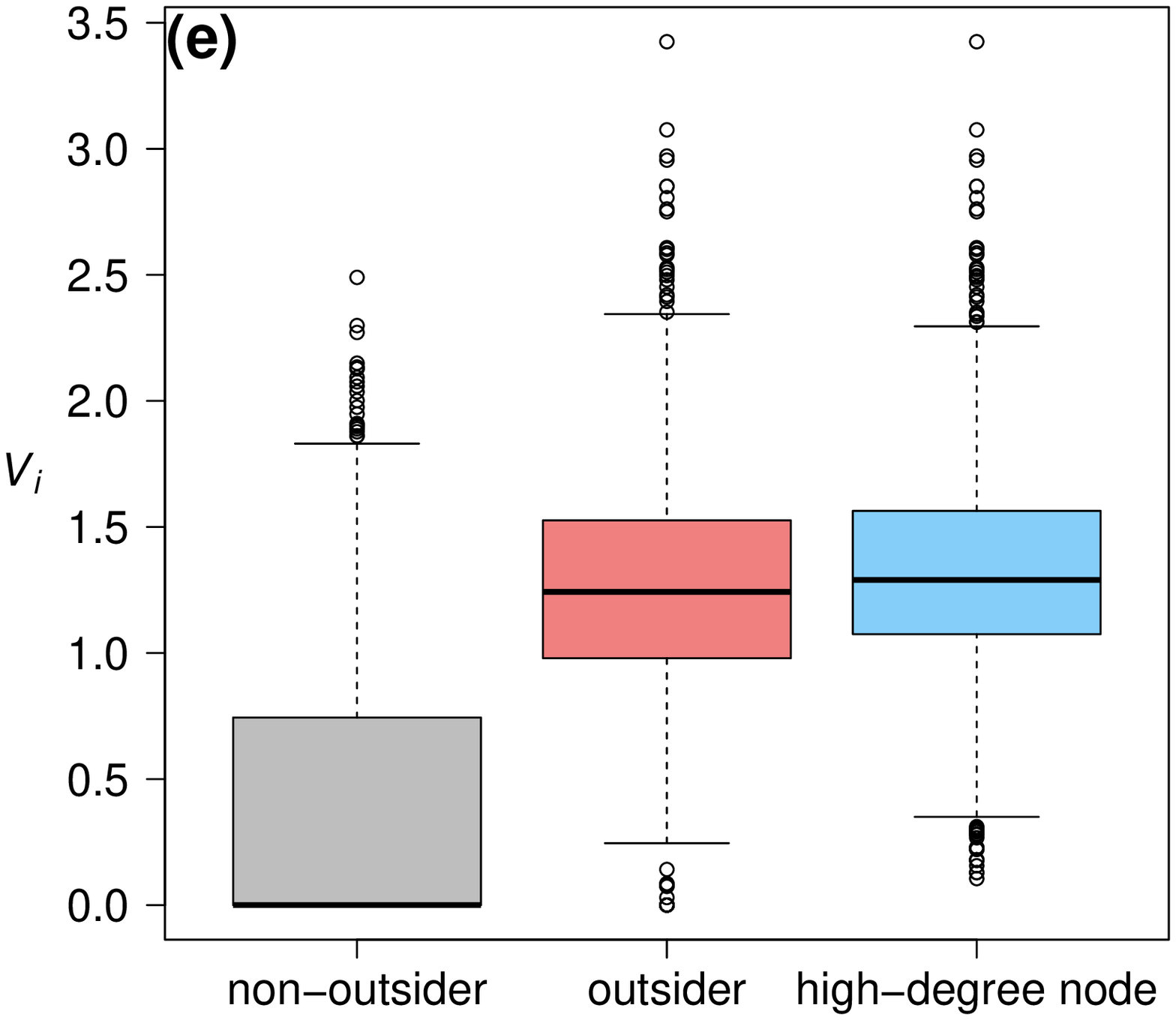}
\includegraphics[width=0.35\hsize]{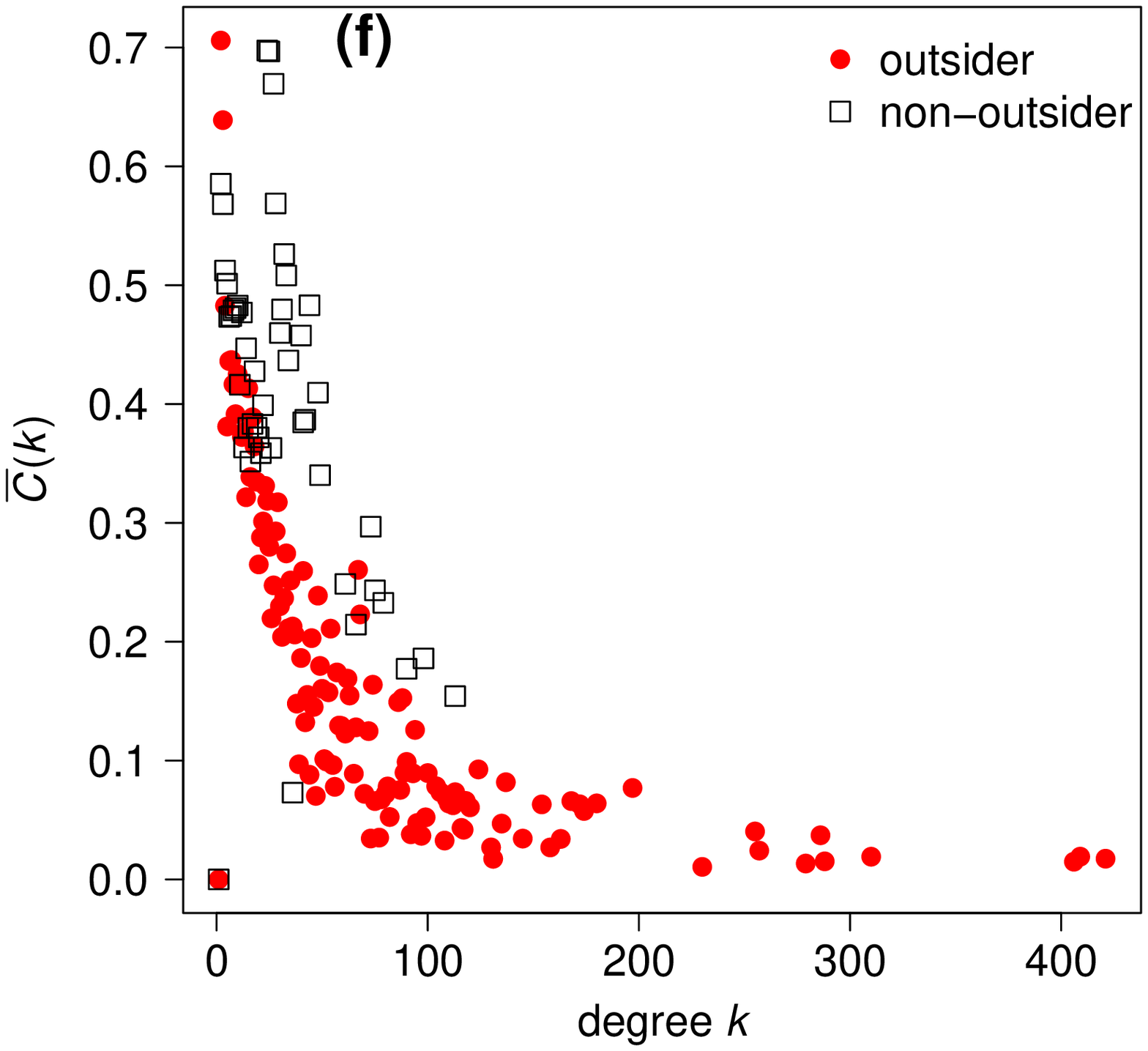}
\caption{Results for the Enron network.
(a) Histogram showing the node degree for outsiders and non-outsiders.
(b) The average degree of nodes adjacent to the nodes with degree $k$.
(c) Assortativity coefficient $r$ (solid lines) and (d) the sizes of the largest connected component of removed nodes $s_{\rm removed}$, as a function of the proportion of  nodes removed $f_{\rm removed}$.
(e) Diversity of neighbors' degree $V_i$. (f) Average local clustering coefficient $\overline{C}(k)$ as a function of node degree $k$.}
\label{fig:enron}
\end{figure}

\begin{figure}
\centering
\includegraphics[width=0.35\hsize]{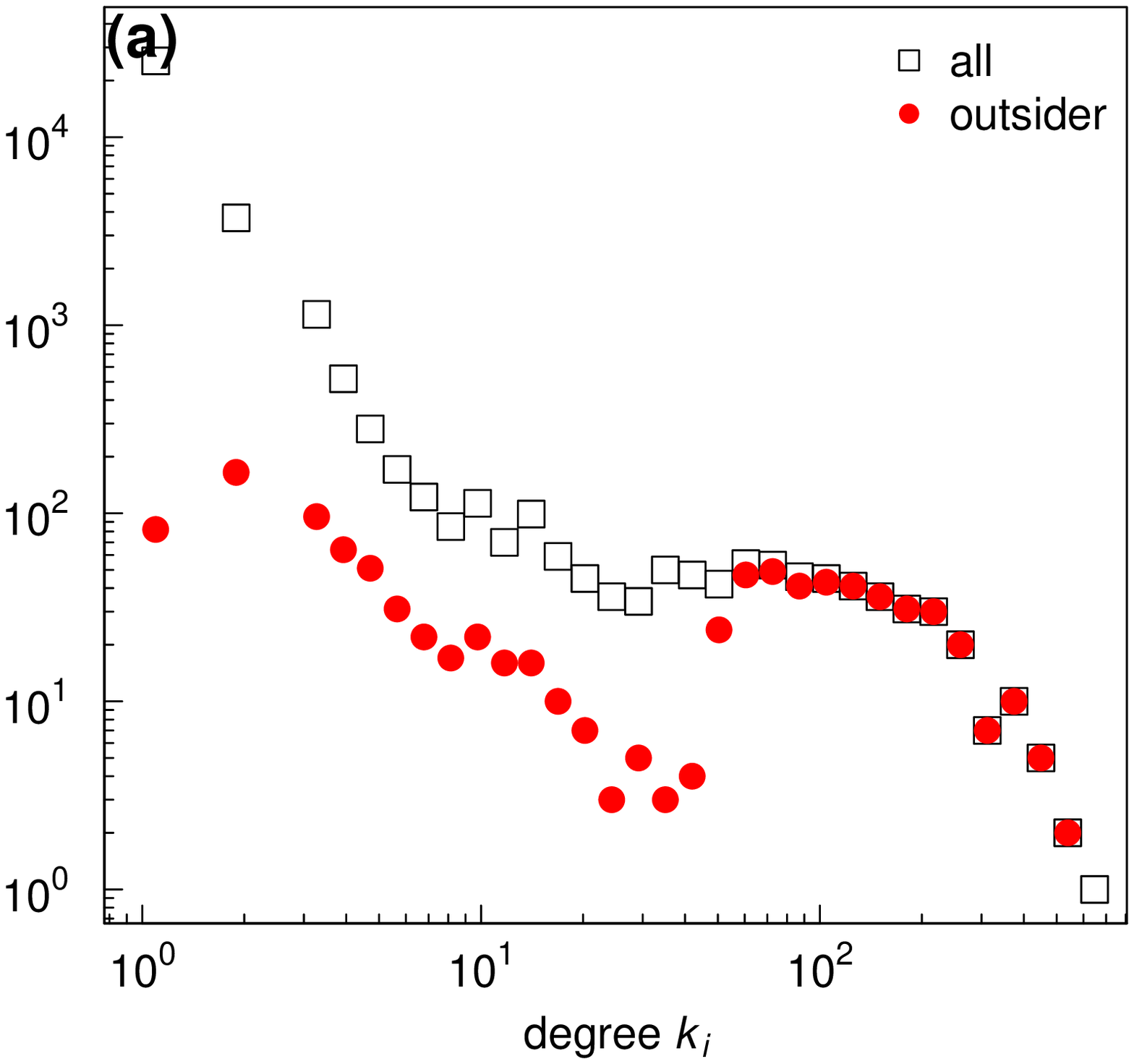}
\includegraphics[width=0.35\hsize]{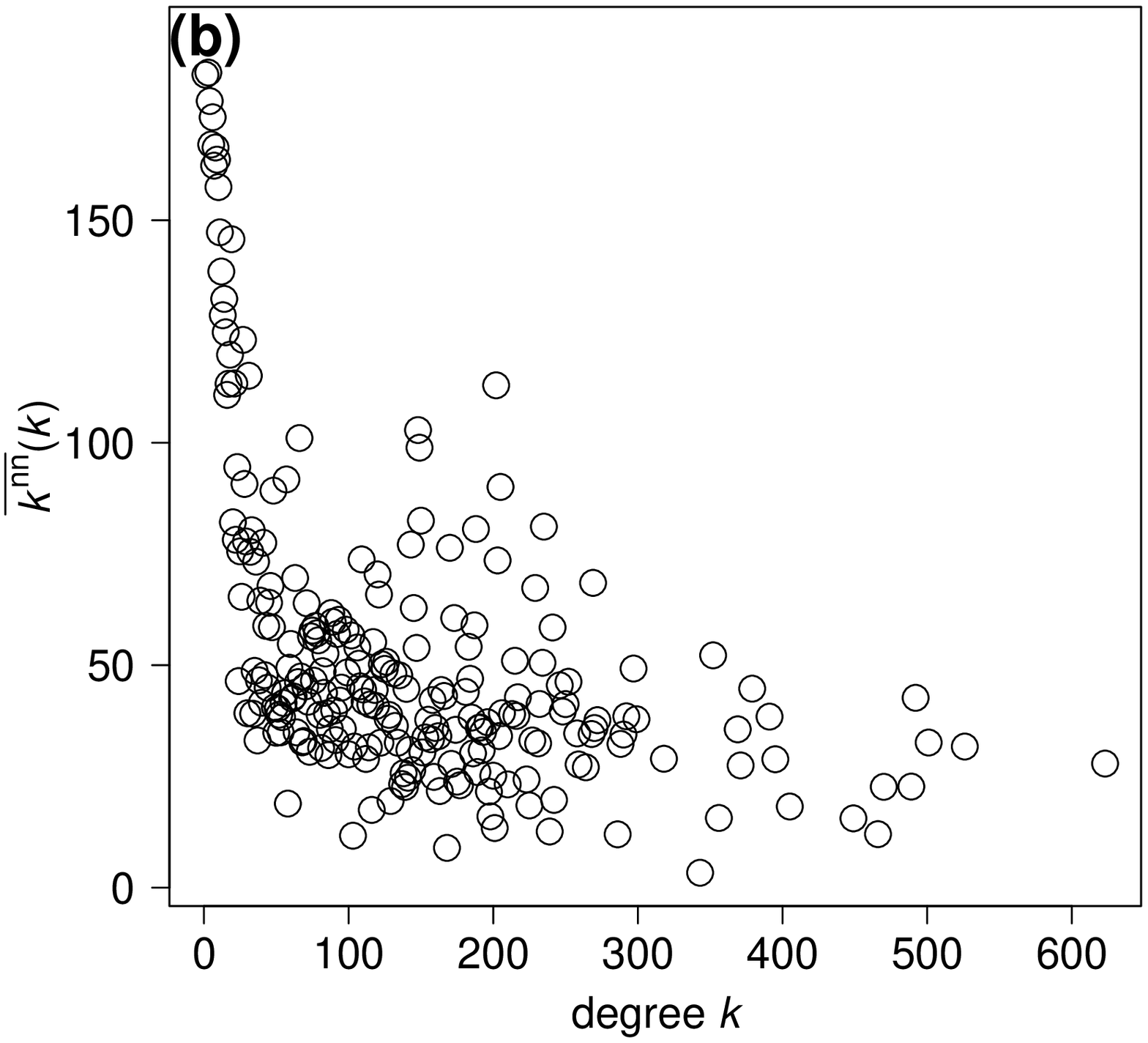}
\includegraphics[width=0.35\hsize]{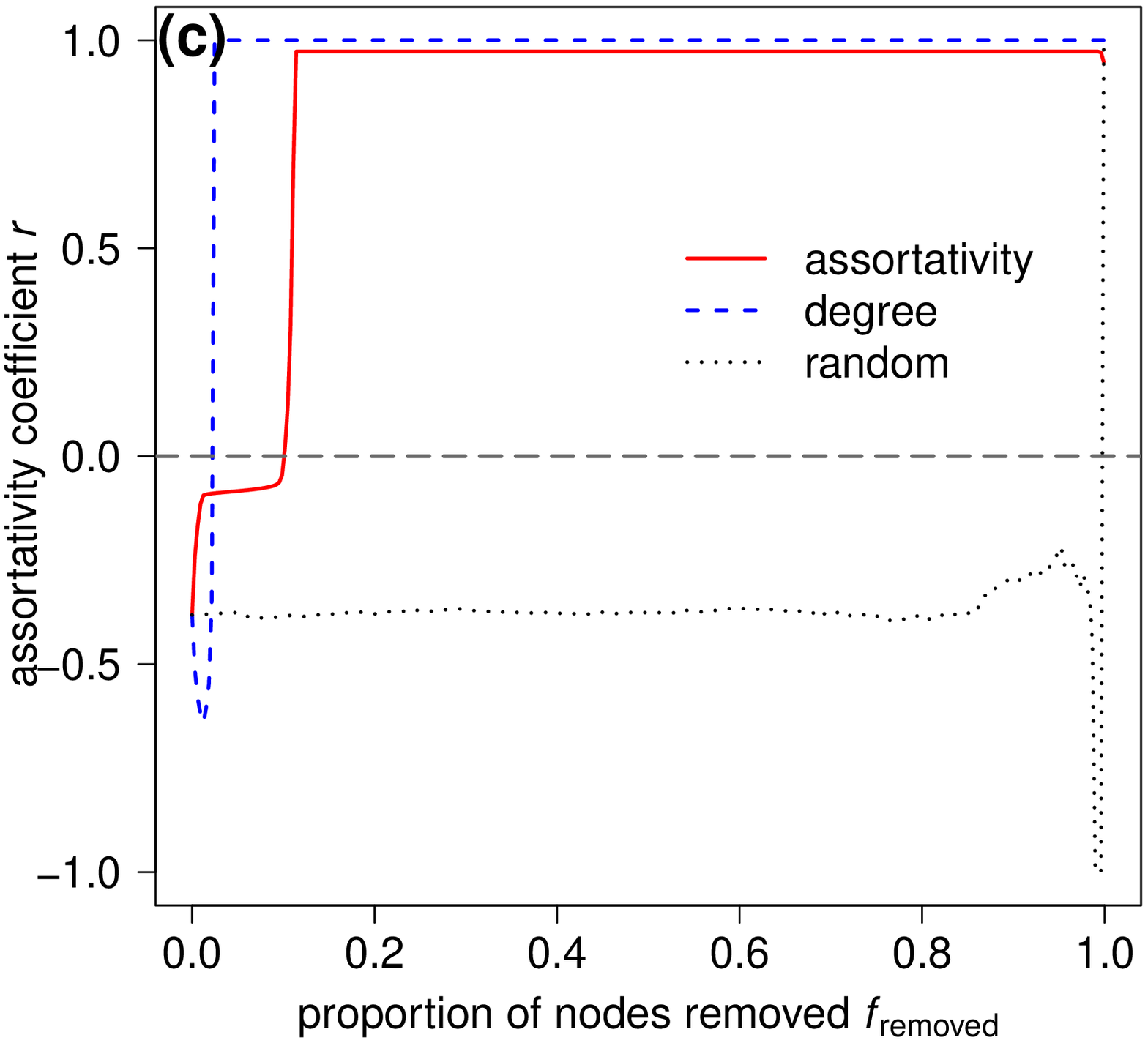}
\includegraphics[width=0.35\hsize]{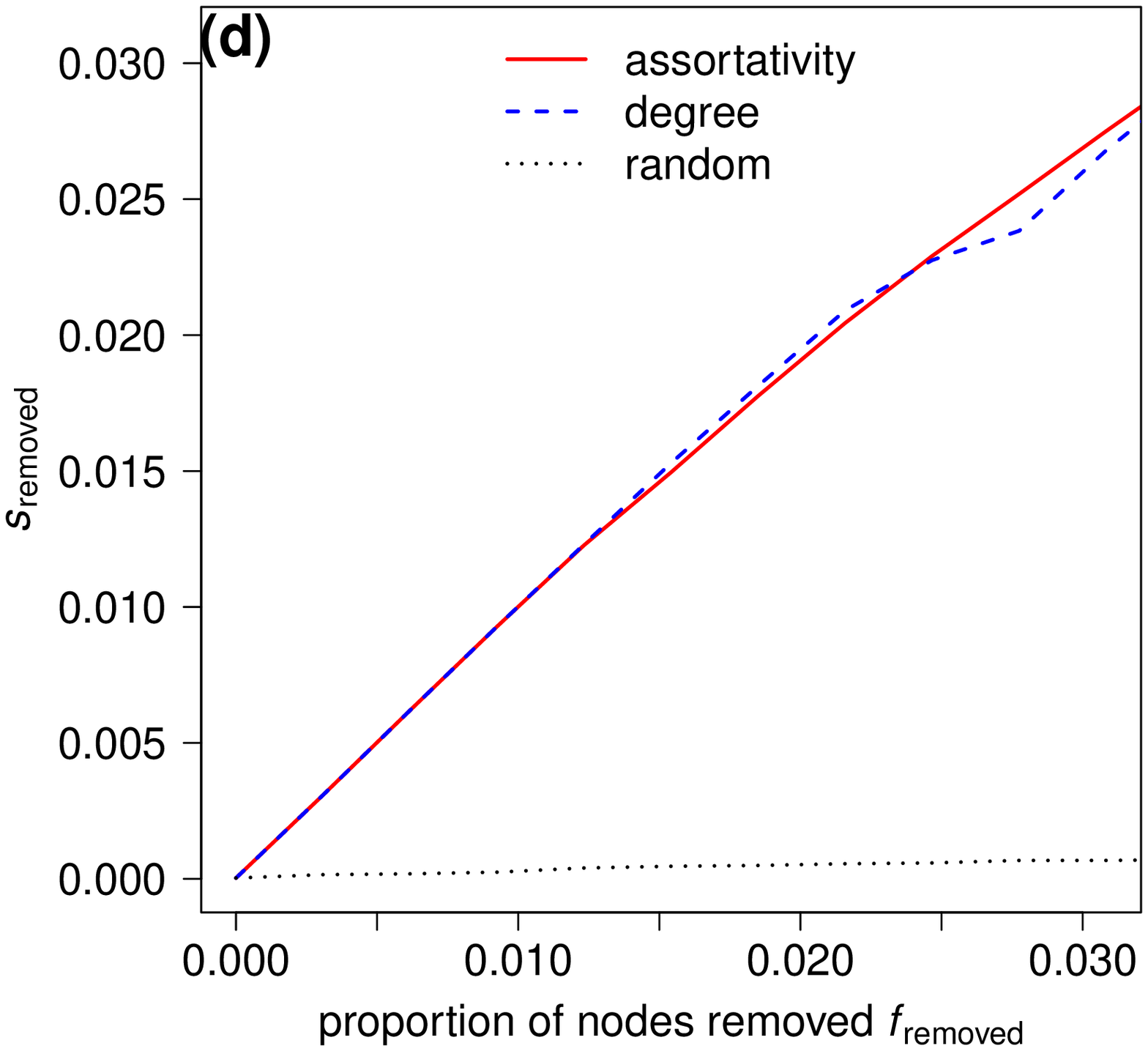}
\includegraphics[width=0.35\hsize]{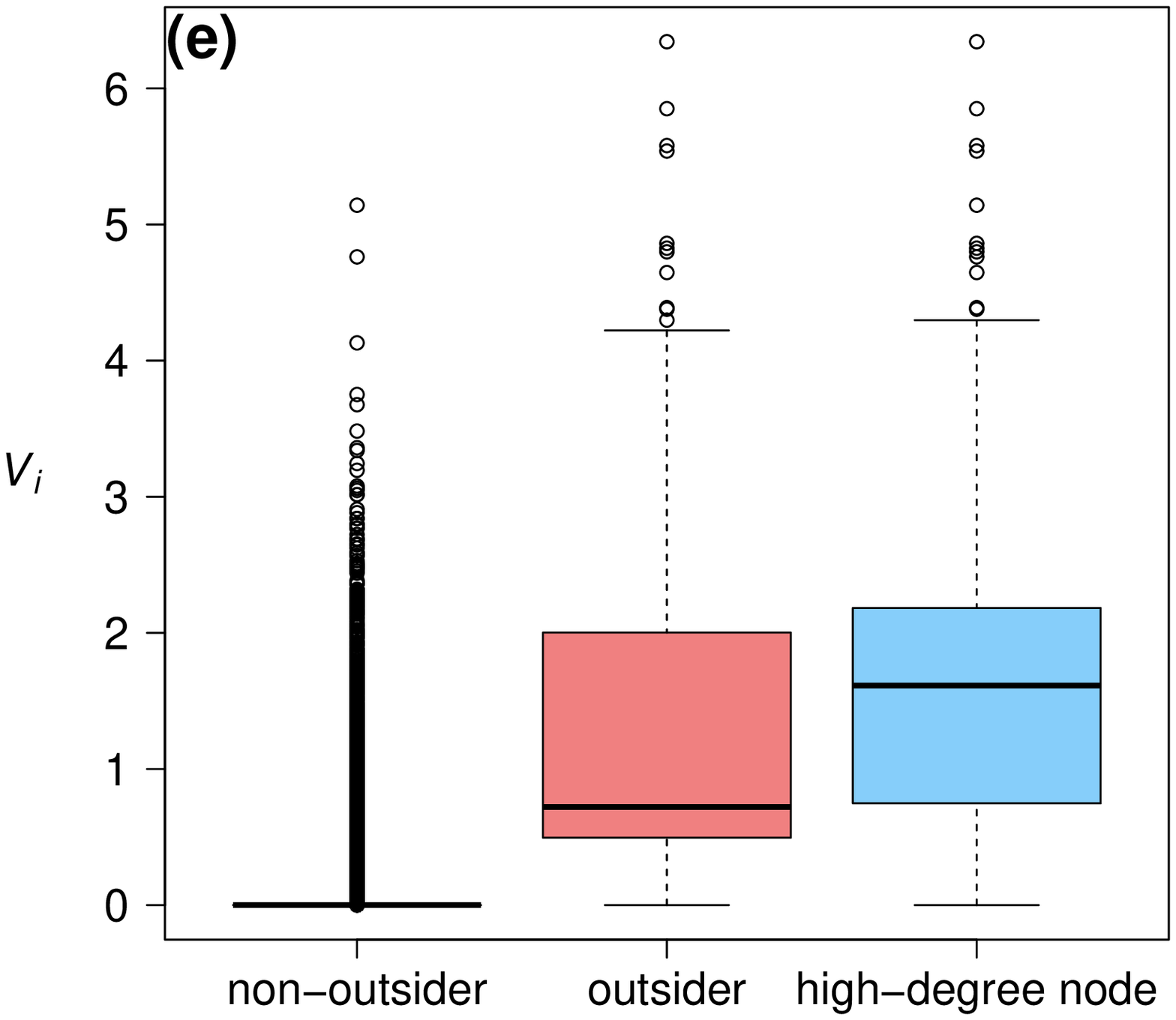}
\includegraphics[width=0.35\hsize]{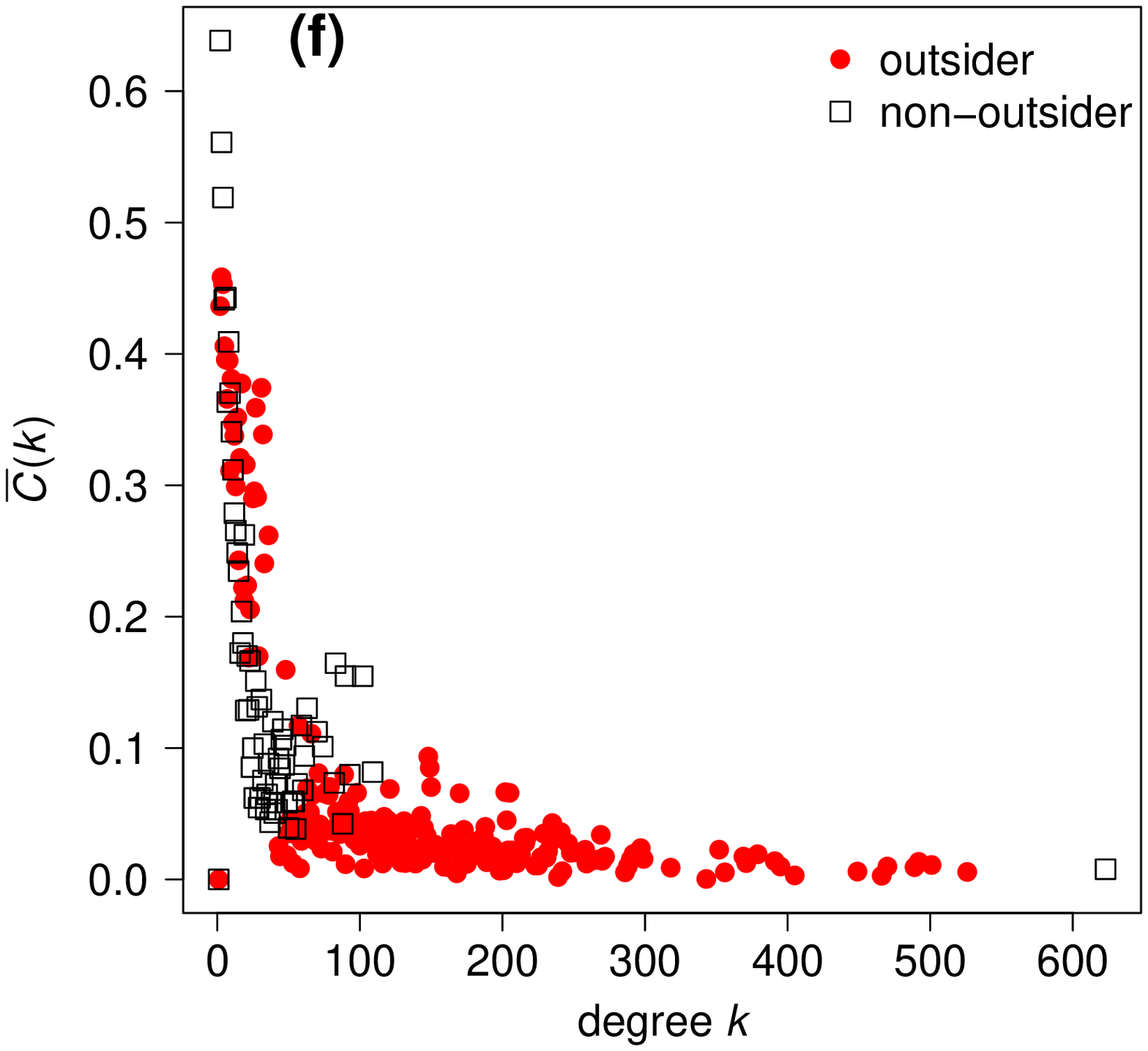}
\caption{Results for the EU-email network.
(a) Histogram showing the node degree for outsiders and non-outsiders.
(b) The average degree of nodes adjacent to the nodes with degree $k$.
(c) Assortativity coefficient $r$ (solid lines) and (d) the sizes of the largest connected component of removed nodes $s_{\rm removed}$, as a function of the proportion of  nodes removed $f_{\rm removed}$.
(e) Diversity of neighbors' degree $V_i$. (f) Average local clustering coefficient $\overline{C}(k)$ as a function of node degree $k$.
}
\label{fig:eu_email}
\end{figure}

\begin{figure}
\centering
\includegraphics[width=0.35\hsize]{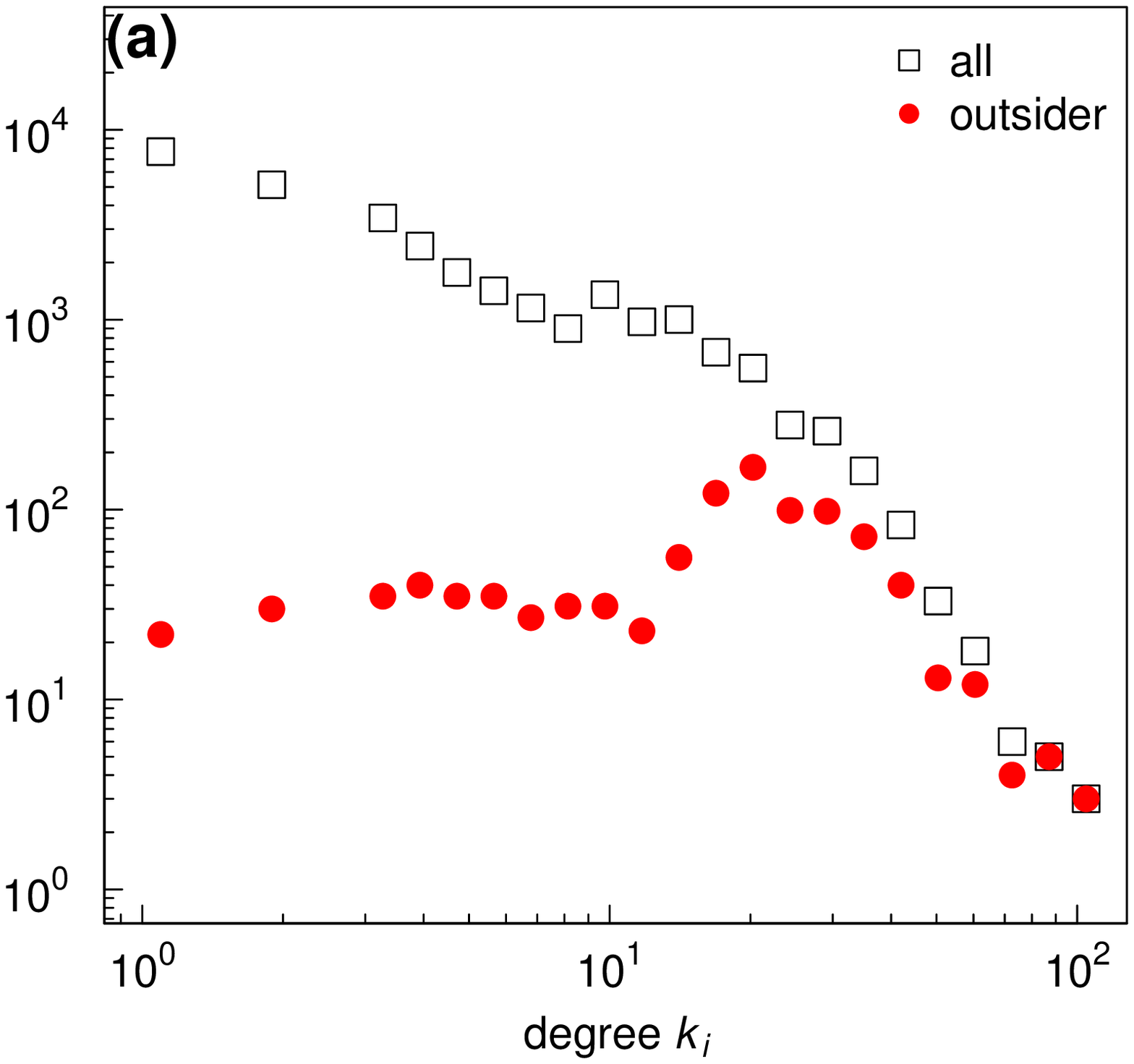}
\includegraphics[width=0.35\hsize]{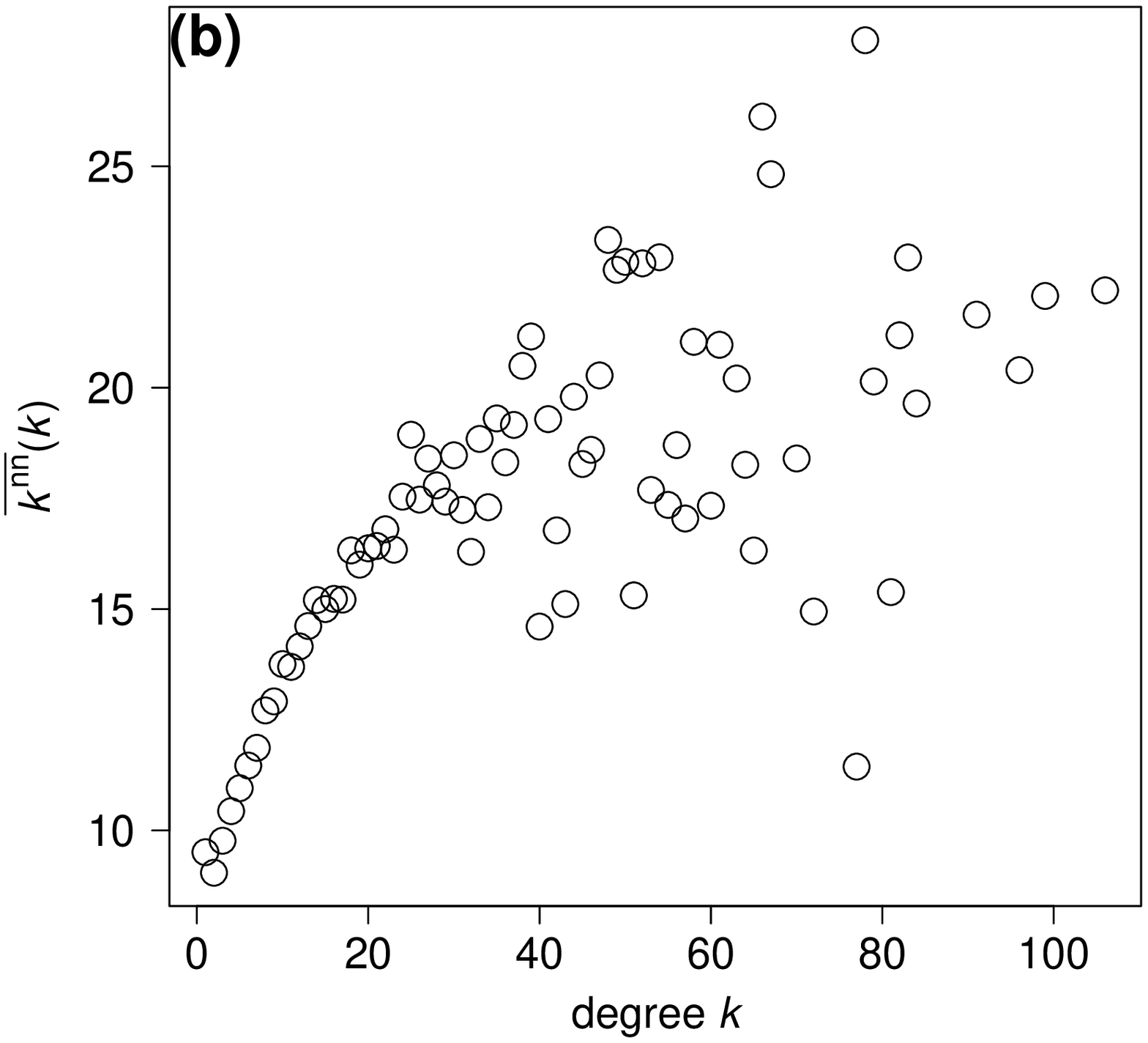}
\includegraphics[width=0.35\hsize]{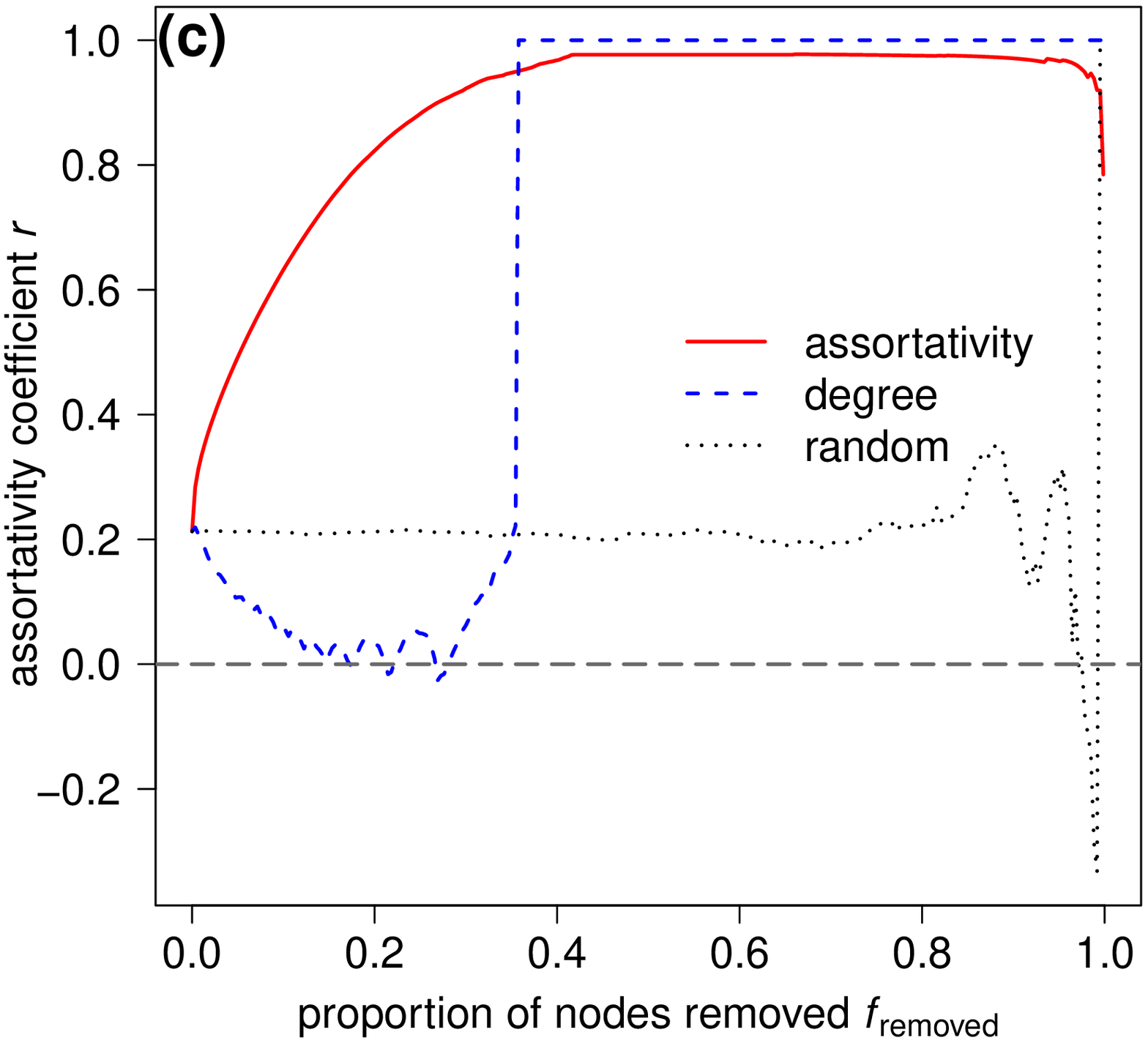}
\includegraphics[width=0.35\hsize]{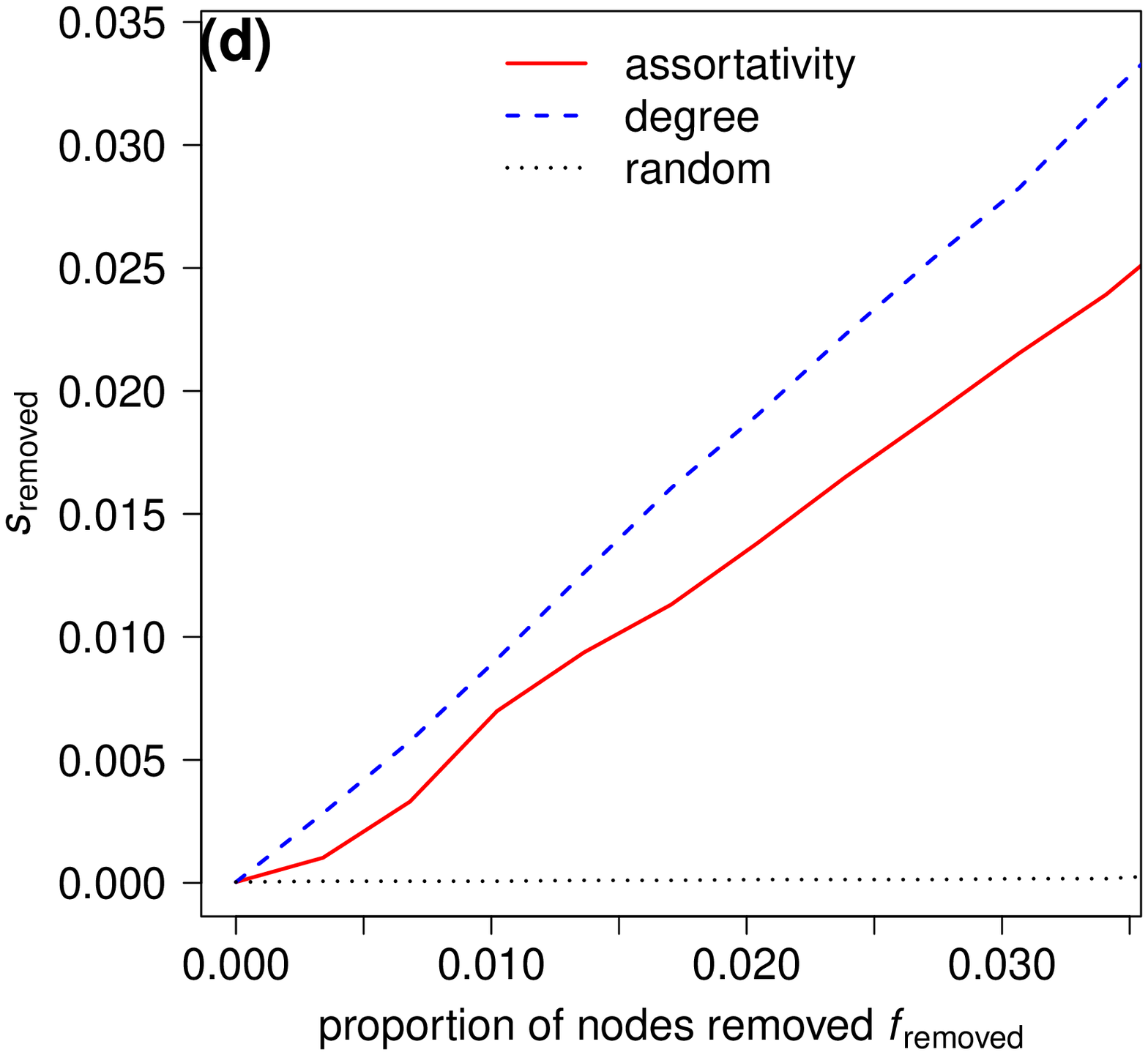}
\includegraphics[width=0.35\hsize]{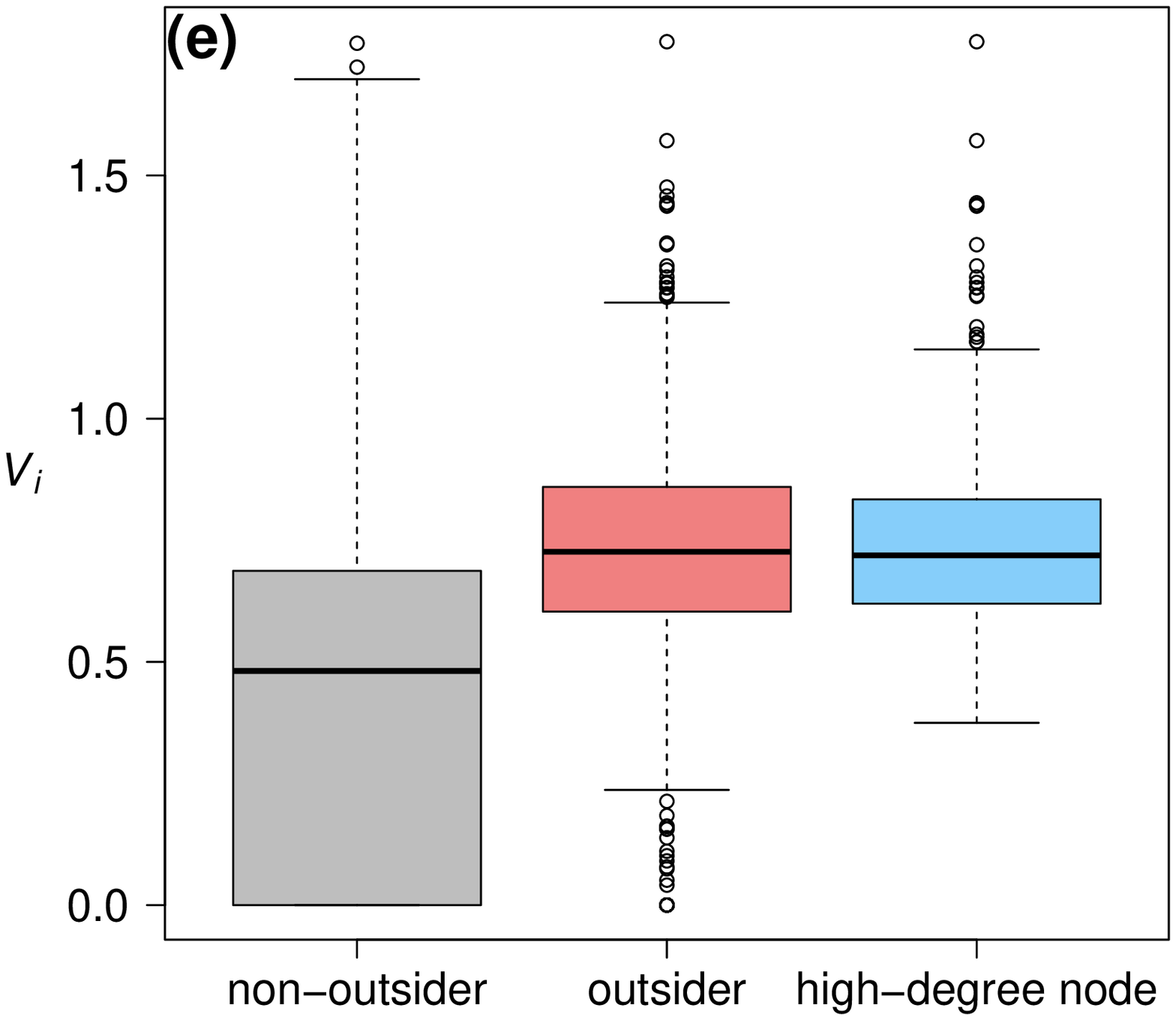}
\includegraphics[width=0.35\hsize]{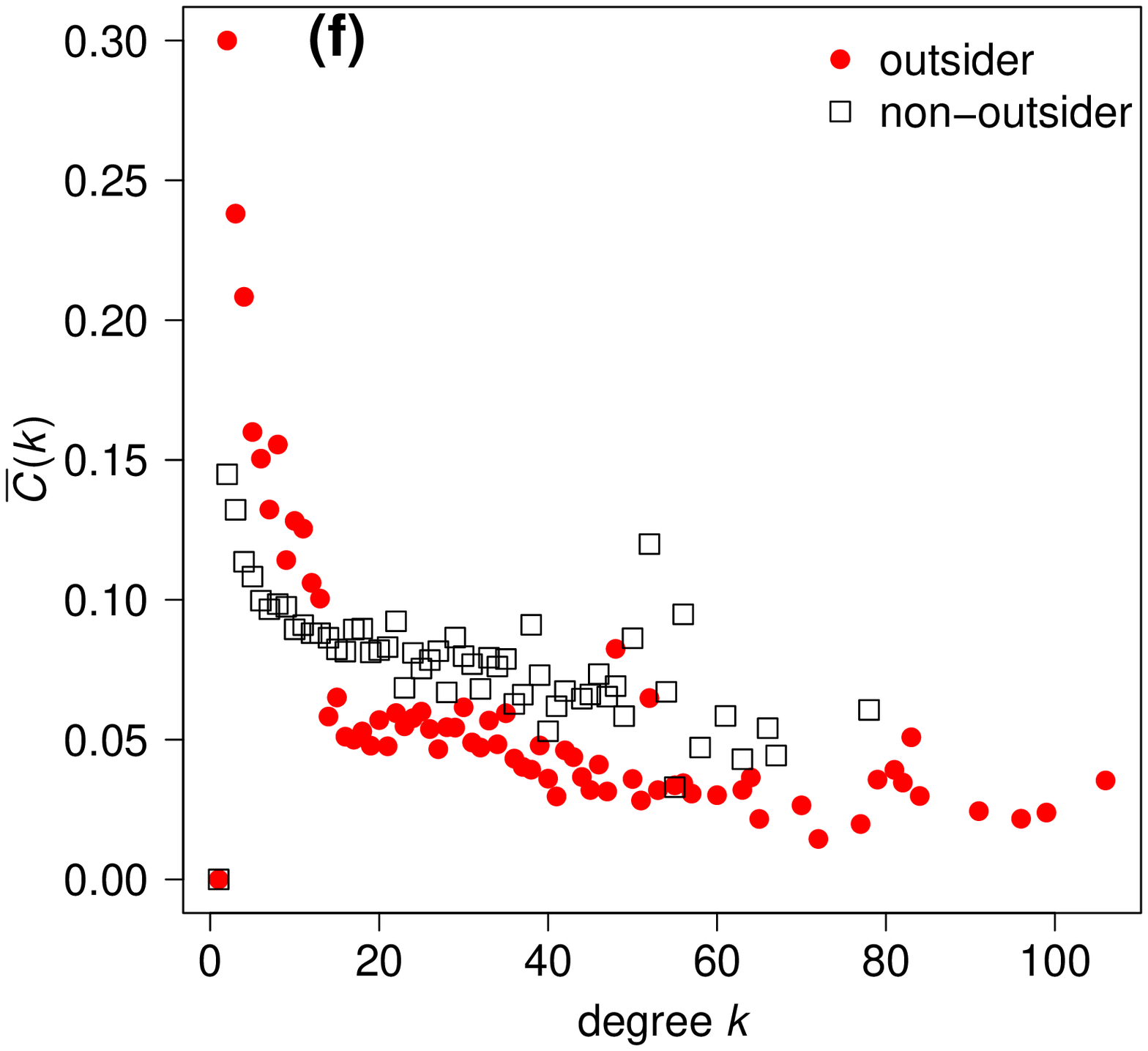}
\caption{Results for the Facebook network.
(a) Histogram showing the node degree for outsiders and non-outsiders.
(b) The average degree of nodes adjacent to the nodes with degree $k$.
(c) Assortativity coefficient $r$ (solid lines) and (d) the sizes of the largest connected component of removed nodes $s_{\rm removed}$, as a function of the proportion of  nodes removed $f_{\rm removed}$.
(e) Diversity of neighbors' degree $V_i$. (f) Average local clustering coefficient $\overline{C}(k)$ as a function of node degree $k$.
}
\label{fig:facebook}
\end{figure}

\begin{figure}
\centering
\includegraphics[width=0.35\hsize]{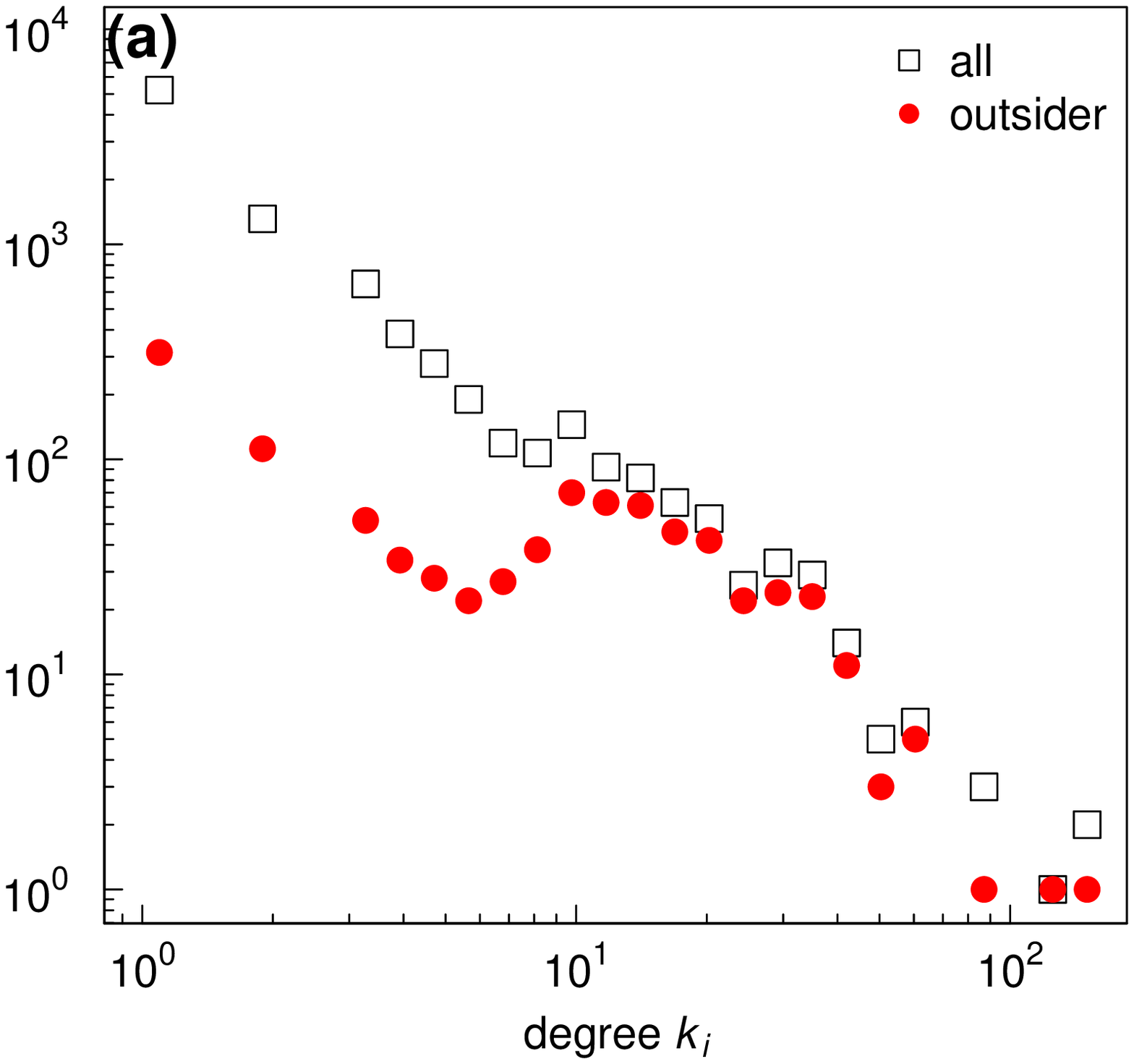}
\includegraphics[width=0.35\hsize]{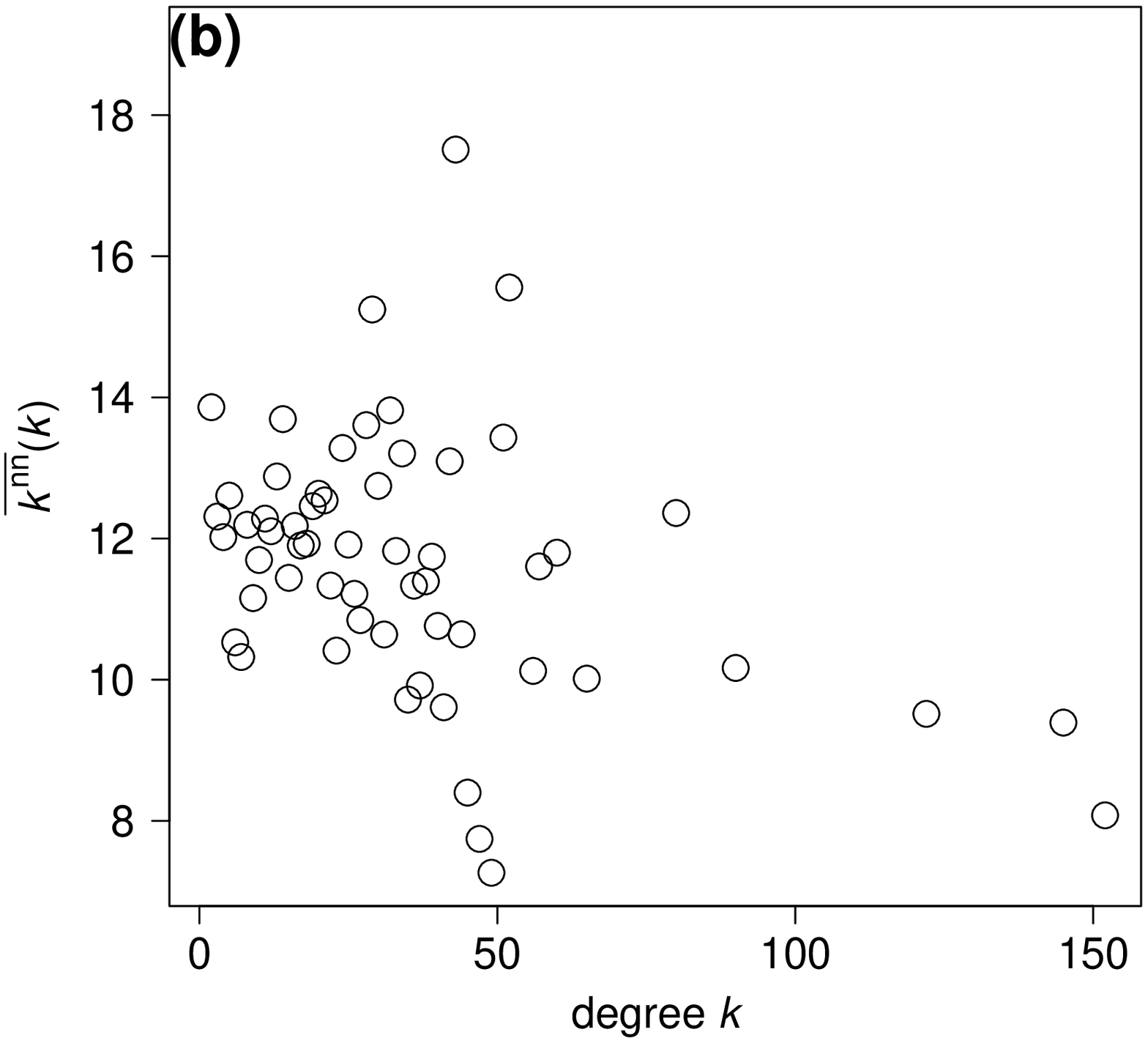}
\includegraphics[width=0.35\hsize]{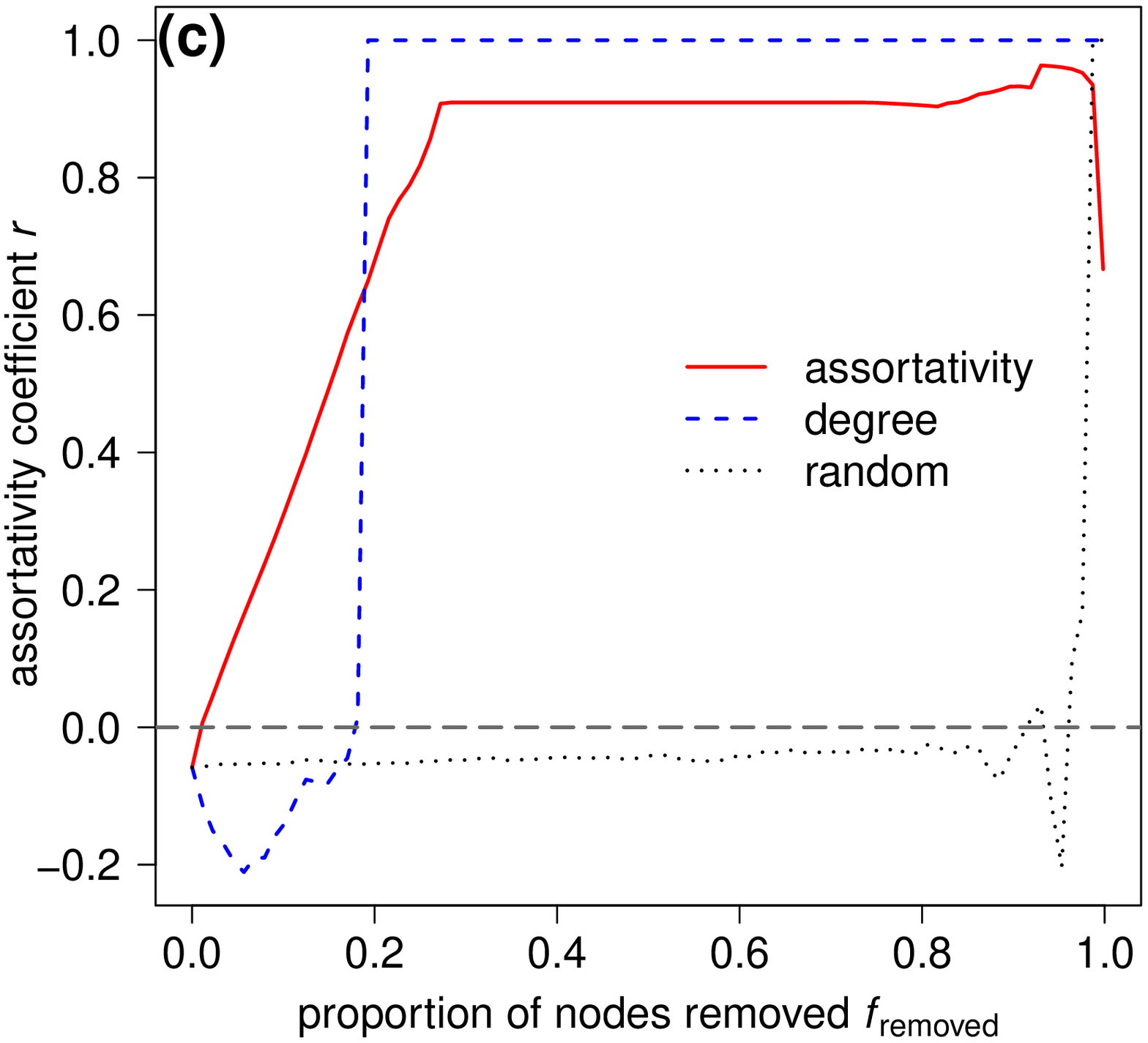}
\includegraphics[width=0.35\hsize]{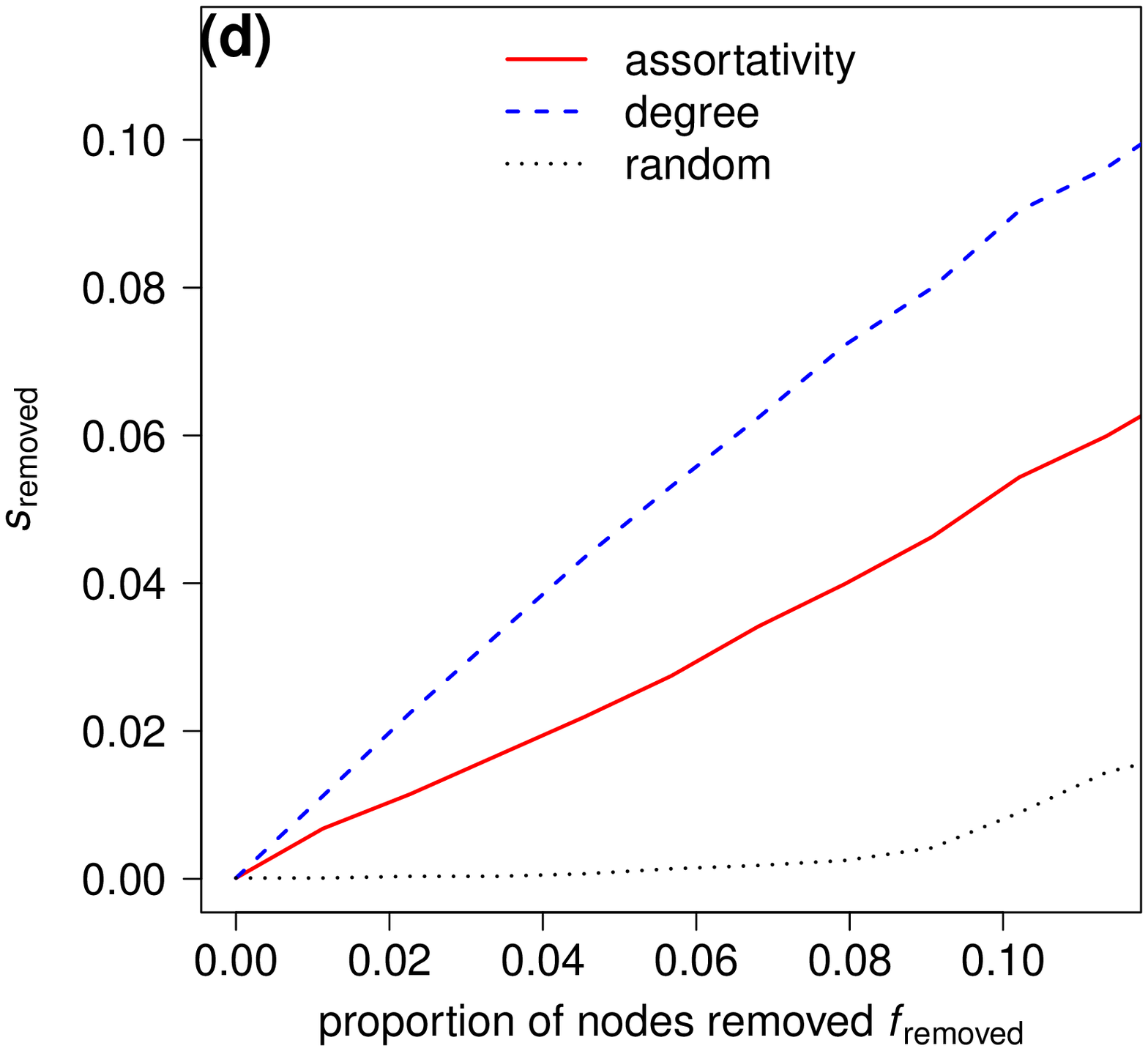}
\includegraphics[width=0.35\hsize]{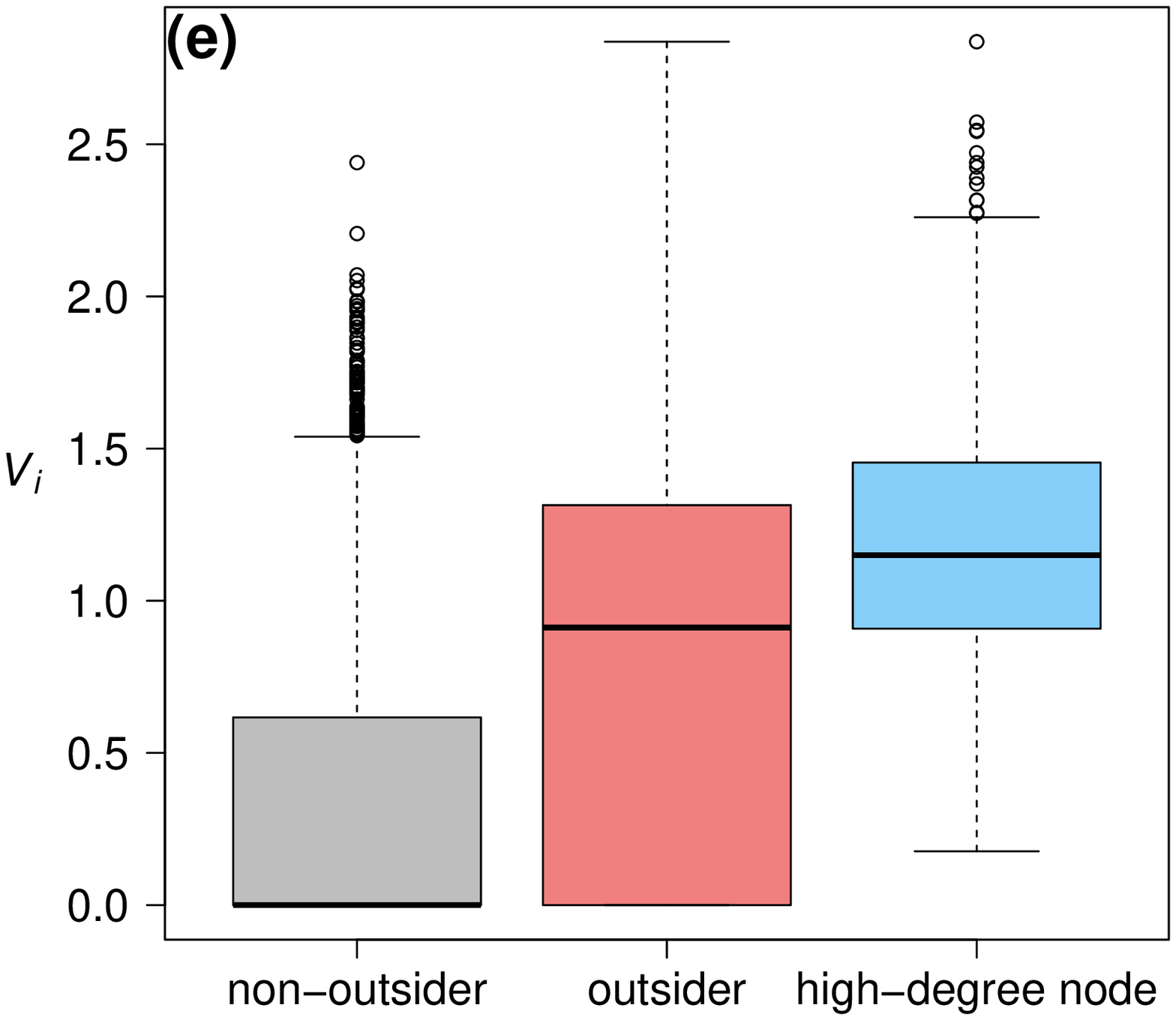}
\includegraphics[width=0.35\hsize]{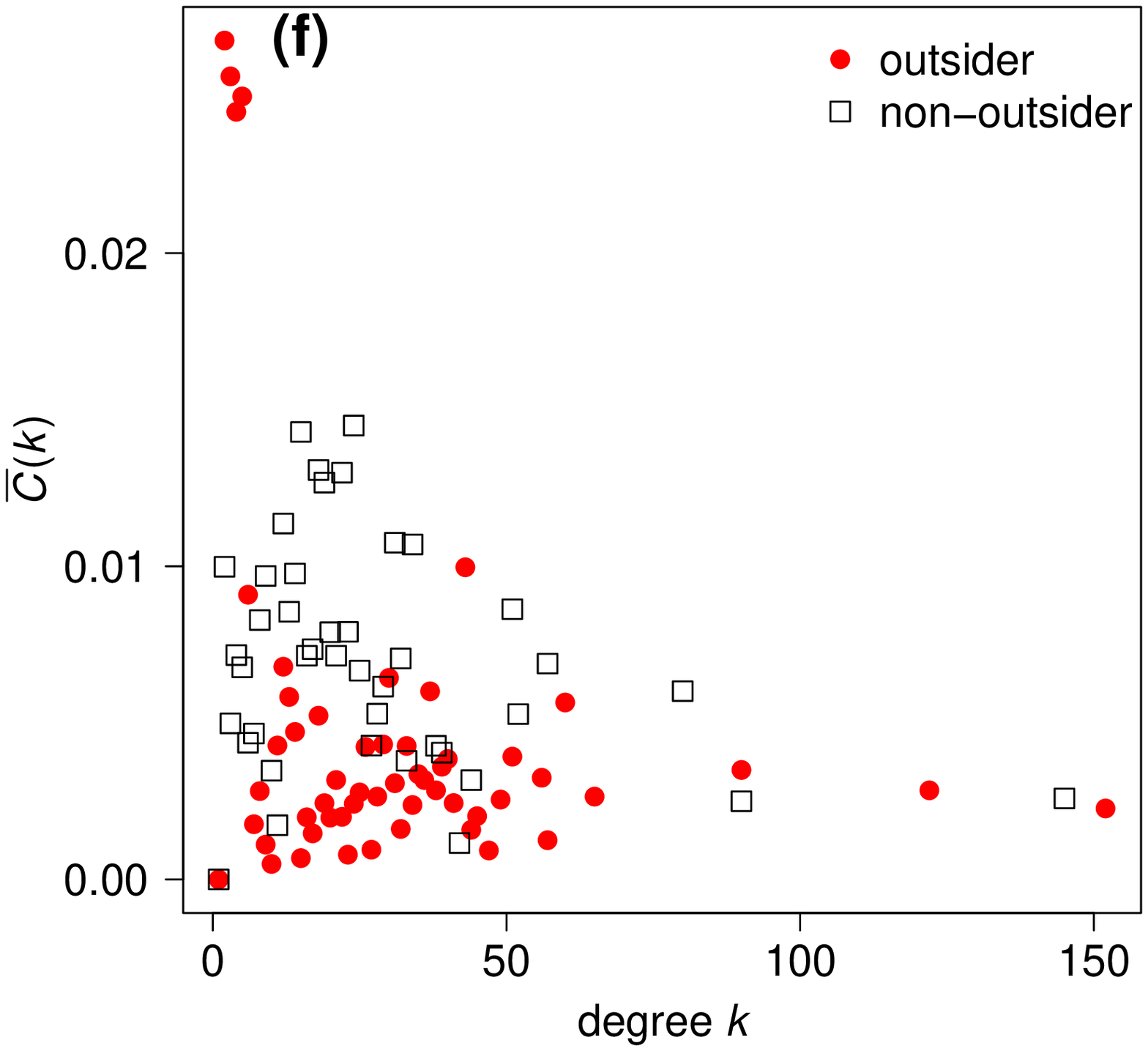}
\caption{Results for the Slashdot network.
(a) Histogram showing the node degree for outsiders and non-outsiders.
(b) The average degree of nodes adjacent to the nodes with degree $k$.
(c) Assortativity coefficient $r$ (solid lines) and (d) the sizes of the largest connected component of removed nodes $s_{\rm removed}$, as a function of the proportion of  nodes removed $f_{\rm removed}$.
(e) Diversity of neighbors' degree $V_i$. (f) Average local clustering coefficient $\overline{C}(k)$ as a function of node degree $k$.
}
\label{fig:slashdot}
\end{figure}

\begin{figure}
\centering
\includegraphics[width=0.35\hsize]{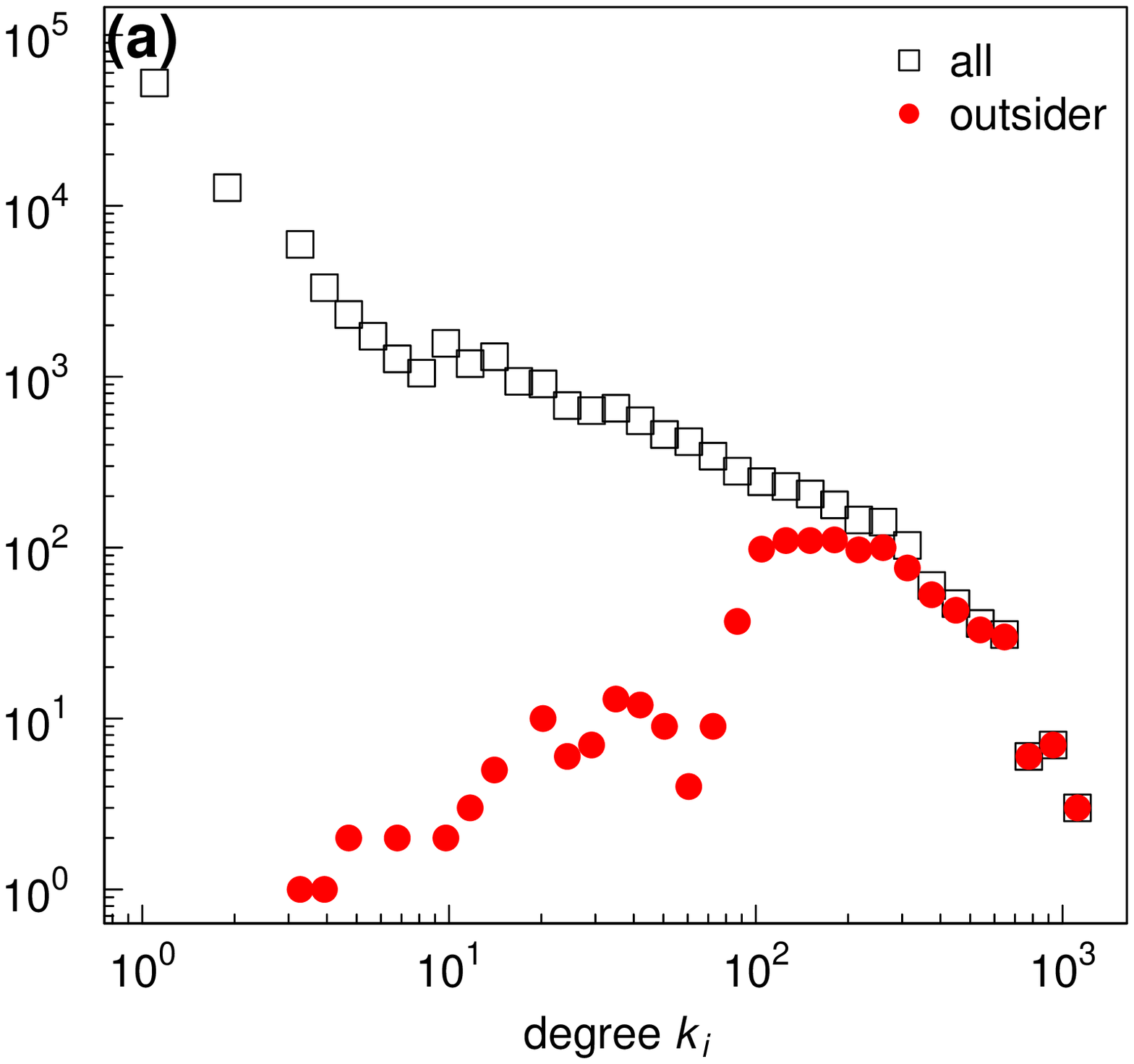}
\includegraphics[width=0.35\hsize]{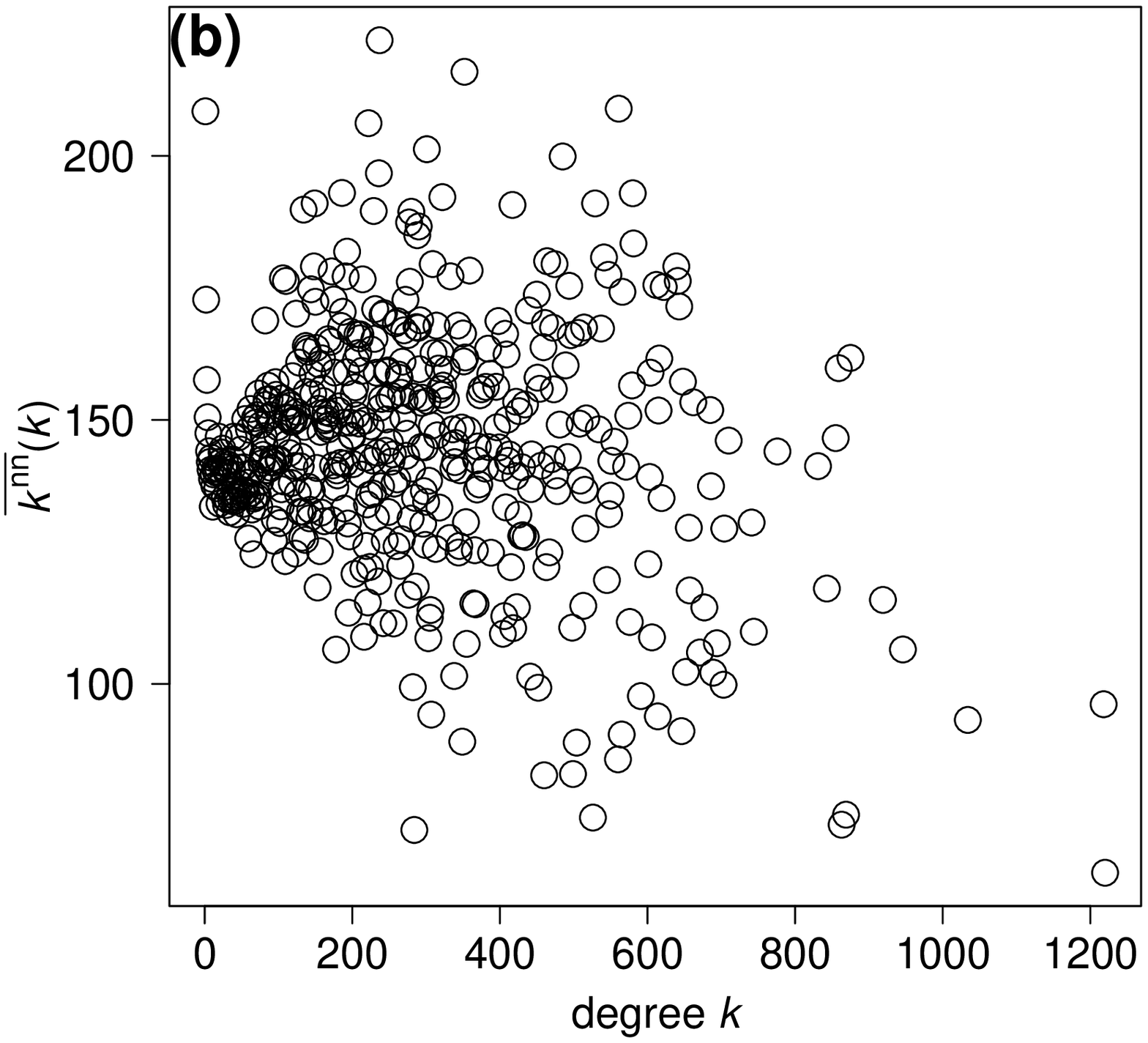}
\includegraphics[width=0.35\hsize]{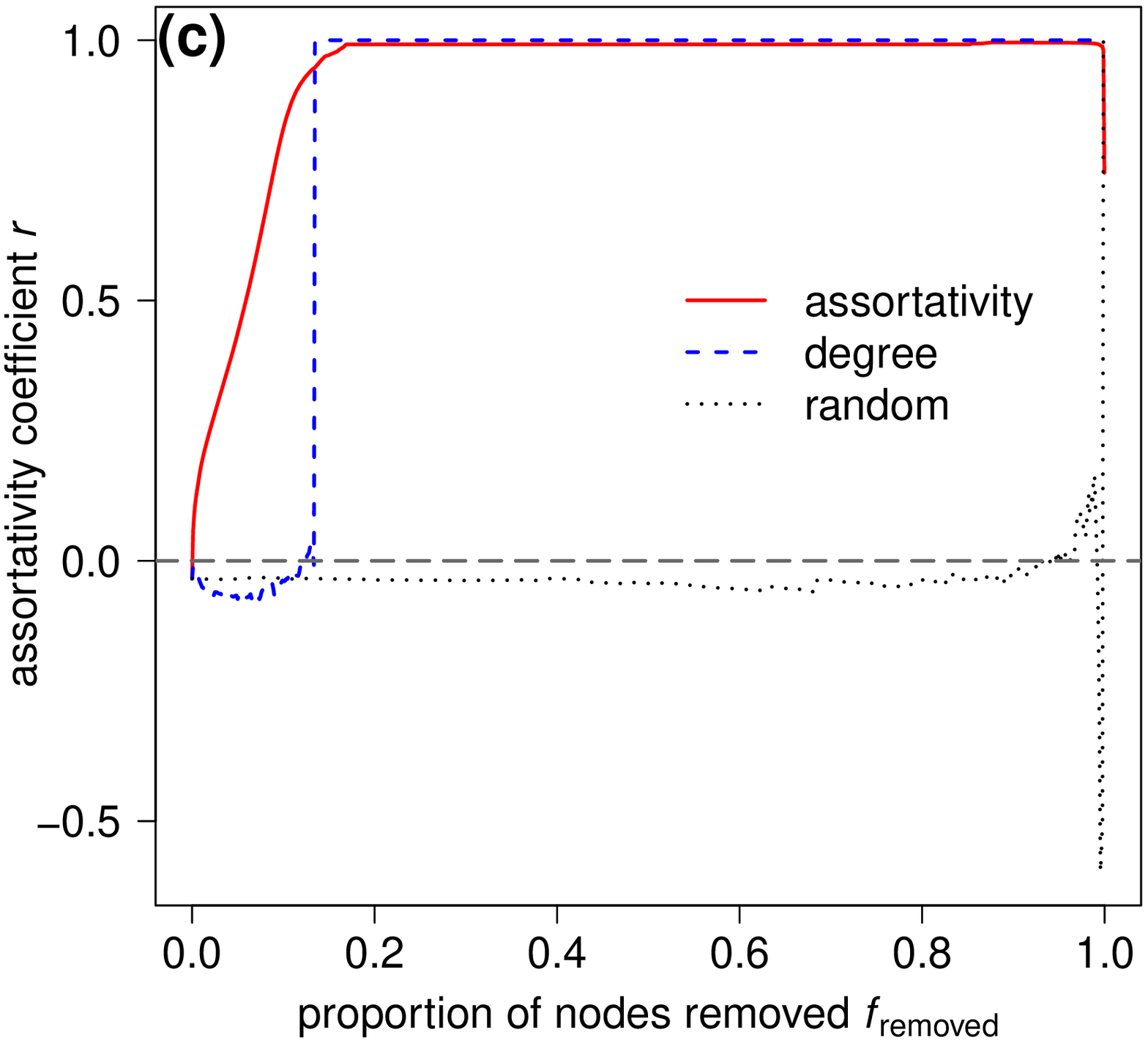}
\includegraphics[width=0.35\hsize]{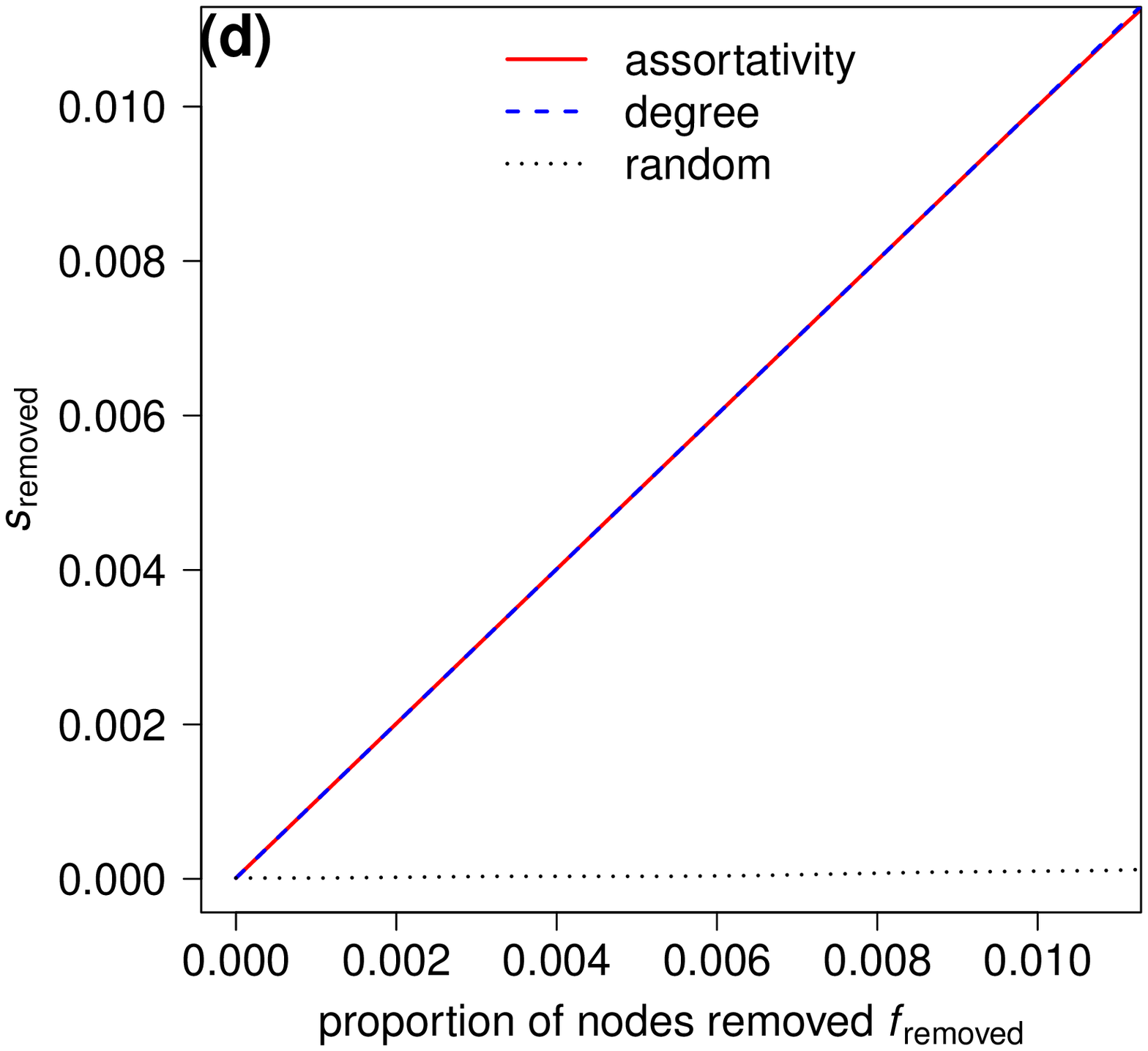}
\includegraphics[width=0.35\hsize]{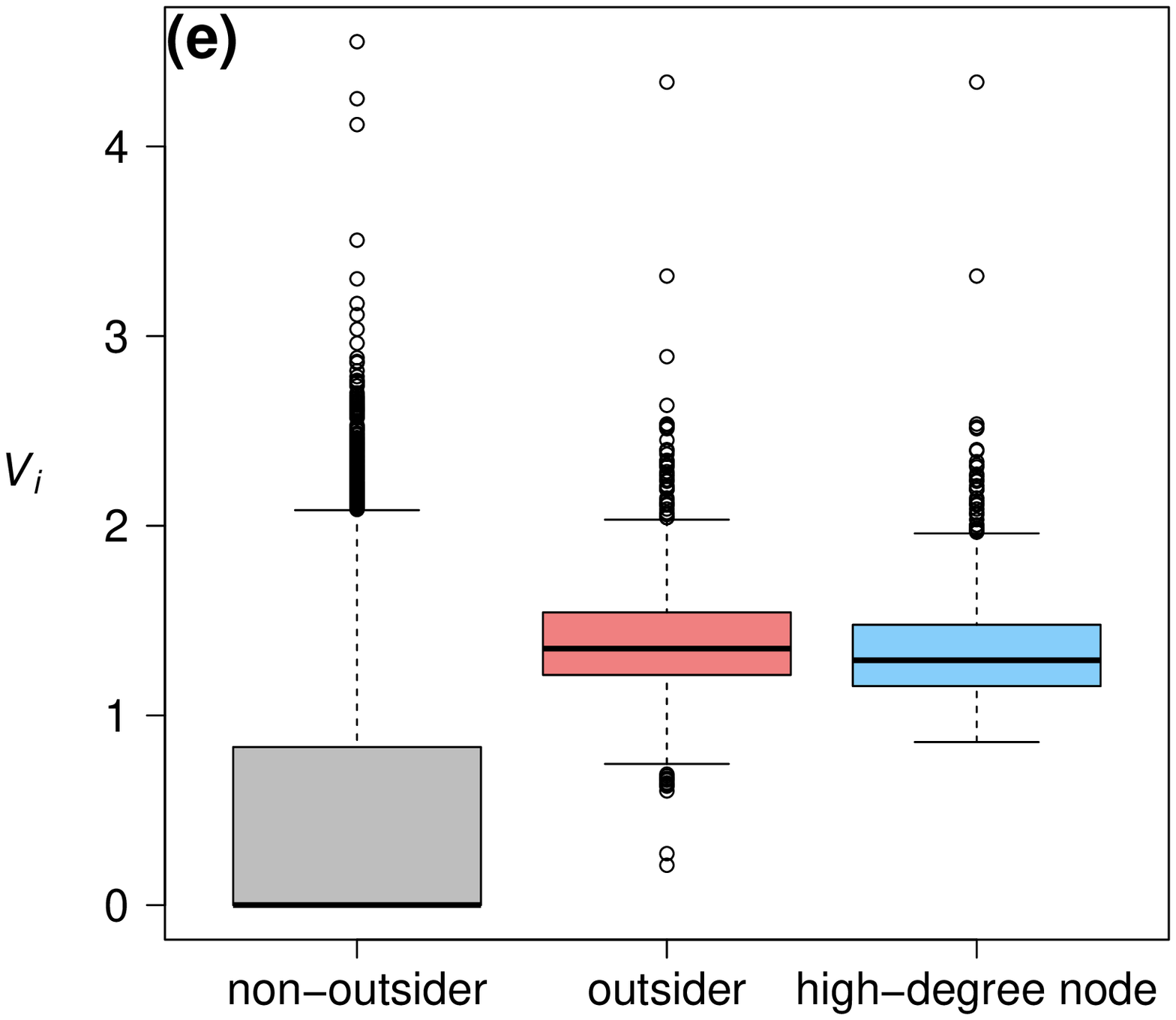}
\includegraphics[width=0.35\hsize]{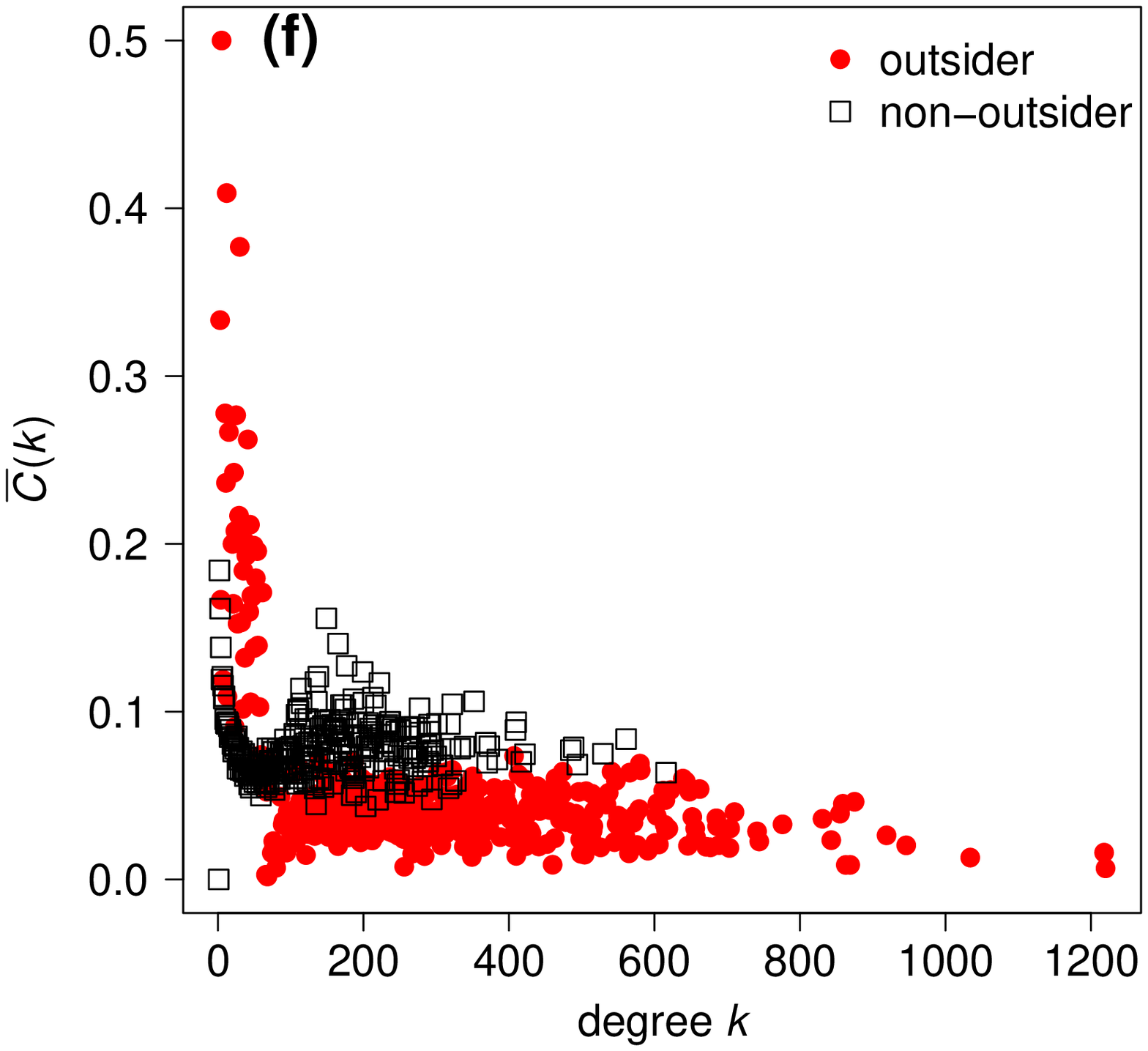}
\caption{Results for the Wiki-talk network.
(a) Histogram showing the node degree for outsiders and non-outsiders.
(b) The average degree of nodes adjacent to the nodes with degree $k$.
(c) Assortativity coefficient $r$ (solid lines) and (d) the sizes of the largest connected component of removed nodes $s_{\rm removed}$, as a function of the proportion of  nodes removed $f_{\rm removed}$.
(e) Diversity of neighbors' degree $V_i$. (f) Average local clustering coefficient $\overline{C}(k)$ as a function of node degree $k$.
}
\label{fig:wiki}
\end{figure}

We examine the outsider model in other network data sets to investigate the generality of the existence of outsiders that we confirmed in the Mention network.
The five network data sets we use are the Enron~\cite{Klimt2004}, EU-email~\cite{Leskovec2007}, Facebook~\cite{Viswanath2009}, Slashdot~\cite{Gomez2008}, and Wiki-talk~\cite{Leskovec2010} networks that were all originally made available online by the Koblenz Network Collection (http://konect.uni-koblenz.de/)~\cite{Kunegis2013}.
In all of these networks, we define the nodes by the users of the online communication tools and the links by the reciprocal interactions between pairs of users.
The Enron and EU-email are based on the record of email exchanges within a company and an European institution.
The Facebook network is based on the the post of messages by a user on another user's personal page in Facebook.
The Slashdot and Wiki-talk networks are based on the message exchanges between the users of these websites.
The basic statistics of the five networks are summarized in Tab.~\ref{tab:stats}
The results for the five networks are shown in Figs.~\ref{fig:enron}, \ref{fig:eu_email}, \ref{fig:facebook}, \ref{fig:slashdot}, and \ref{fig:wiki}.
We fix the number of outsiders $N_{\rm outsider} = 1,000$ for all the networks.


\begin{thebibliography}{10}

\bibitem{Kwak2010}
Haewoon Kwak, Changhyun Lee, Hosung Park, and Sue Moon.
\newblock {What is Twitter, a Social Network or a News Media?}
\newblock In {\em Proceedings of the 19th international conference on World
  Wide Web}, pages 591--600, Raleigh, NC, 2010. ACM.

\bibitem{Golder2011}
S.~A. Golder and M.~W. Macy.
\newblock {Diurnal and Seasonal Mood Vary with Work, Sleep, and Daylength
  Across Diverse Cultures}.
\newblock {\em Science}, 333(6051):1878--1881, September 2011.

\bibitem{Sasahara2013}
Kazutoshi Sasahara, Yoshito Hirata, Masashi Toyoda, Masaru Kitsuregawa, and
  Kazuyuki Aihara.
\newblock {Quantifying collective attention from tweet stream.}
\newblock {\em PLOS ONE}, 8(4):e61823, January 2013.

\bibitem{Sakaki2010}
Takeshi Sakaki, M~Okazaki, and Y~Matsuo.
\newblock {Earthquake Shakes Twitter Users : Real-time Event Detection by
  Social Sensors}.
\newblock In {\em Proceedings of the 19th international conference on World
  Wide Web}, pages 851--860, Raleigh, NC, 2010. ACM.

\bibitem{Sano2013}
Yukie Sano, Kenta Yamada, Hayafumi Watanabe, Hideki Takayasu, and Misako
  Takayasu.
\newblock {Empirical analysis of collective human behavior for extraordinary
  events in the blogosphere}.
\newblock {\em Physical Review E}, 87(1):012805, January 2013.

\bibitem{Szell2014}
Michael Szell, S\'{e}bastian Grauwin, and Carlo Ratti.
\newblock {Contraction of online response to major events.}
\newblock {\em PLOS ONE}, 9(2):e89052, January 2014.

\bibitem{Gonzalez-Bailon2011}
Sandra Gonz\'{a}lez-Bail\'{o}n, Javier Borge-Holthoefer, Alejandro Rivero, and
  Yamir Moreno.
\newblock {The Dynamics of Protest Recruitment through an Online Network}.
\newblock {\em Scientific reports}, 1:197, January 2011.

\bibitem{Ahn2007}
Yong-Yeol Ahn, Seungyeop Han, Haewoon Kwak, Sue Moon, and Hawoong Jeong.
\newblock {Analysis of topological characteristics of huge online social
  networking services}.
\newblock In {\em Proceedings of the 16th international conference on World
  Wide Web - WWW '07}, pages 835--844, Banff, Alberta, Canada, 2007. ACM.

\bibitem{Grabowicz2012}
Przemyslaw~a Grabowicz, Jos\'{e}~J Ramasco, Esteban Moro, Josep~M Pujol, and
  Victor~M Eguiluz.
\newblock {Social features of online networks: the strength of intermediary
  ties in online social media.}
\newblock {\em PLOS ONE}, 7(1):e29358, January 2012.

\bibitem{Arnaboldi2013}
Valerio Arnaboldi, Marco Conti, Andrea Passarella, and Robin Dunbar.
\newblock {Dynamics of personal social relationships in online social networks:
  a study on twitter}.
\newblock In {\em Proceedings of the first ACM conference on Online social
  networks}, pages 15--26, Boston, MA, 2013. ACM.

\bibitem{twitterURL}
{Twitter}.

\bibitem{Newman2002}
M.E.J. Newman.
\newblock {Assortative Mixing in Networks}.
\newblock {\em Physical Review Letters}, 89(20):208701, October 2002.

\bibitem{Newman2003}
M.E.J. Newman.
\newblock {Mixing patterns in networks}.
\newblock {\em Physical Review E}, 67(2):026126, February 2003.

\bibitem{Newman2003a}
M.~Newman and Juyong Park.
\newblock {Why social networks are different from other types of networks}.
\newblock {\em Physical Review E}, 68(3):036122, September 2003.

\bibitem{Java2007}
A~Java, X~Song, Tim Finin, and Belle Tseng.
\newblock {Why We Twitter : An Analysis of a Microblogging Community}.
\newblock In {\em Proceedings of the 9th WebKDD and 1st SNA-KDD 2007 workshop
  on Web Mining and Social Network Analysis}, pages 56--65, San Jose, CA, 2007.
  ACM.

\bibitem{Boyd2010}
D.~Boyd, S.~Golder, and G.~Lotan.
\newblock {Tweet, Tweet, Retweet: Conversational Aspects of Retweeting on
  Twitter}.
\newblock In {\em Proceedings of the 43rd Hawaii International Conference on
  System Sciences}, pages 1--10, Honolulu, HI, January 2010. IEEE.

\bibitem{Cha2010}
Meeyoung Cha, H~Haddadi, F~Benevenuto, and PK~Gummadi.
\newblock {Measuring User Influence in Twitter: The Million Follower Fallacy}.
\newblock In {\em Proceedings of Fourth International Conference on Weblogs and
  Social Media}, pages 10--17, Washington, D.C., 2010. AAAI.

\bibitem{Sousa2010}
Daniel Sousa, Lu¥'is Sarmento, and Eduarda~Mendes Rodrigues.
\newblock {Characterization of the twitter@ replies network: are user ties
  social or topical?}
\newblock In {\em Proceedings of the 2nd International Workshop on Search and
  Mining User-Generated Contents}, pages 63--70, Tronto, Ontario, Canada, 2010.
  ACM.

\bibitem{Goncalves2011}
Bruno Gon¥clves, Nicola Perra, and Alessandro Vespignani.
\newblock {Modeling Users' Activity on Twitter Networks: Validation of Dunbar's
  Number}.
\newblock {\em PLOS ONE}, 6(8):e22656, 2011.

\bibitem{Bliss2012}
Catherine~a. Bliss, Isabel~M. Kloumann, Kameron~Decker Harris, Christopher~M.
  Danforth, and Peter~Sheridan Dodds.
\newblock {Twitter reciprocal reply networks exhibit assortativity with respect
  to happiness}.
\newblock {\em Journal of Computational Science}, 3(5):388--397, September
  2012.

\bibitem{Watts1998}
D~J Watts and S~H Strogatz.
\newblock {Collective dynamics of `small-world' networks.}
\newblock {\em Nature}, 393(6684):440--442, June 1998.

\bibitem{Molloy1995}
Michael Molloy and Bruce Reed.
\newblock {A critical point for random graphs with a given degree sequence}.
\newblock {\em Random Structures and Algorithms}, 6(1995):161--179, 1995.

\bibitem{Newman2010}
M~E~J Newman.
\newblock {\em {Networks: an Introduction}}.
\newblock Oxford: Oxford University Press, 2010.

\bibitem{Hu2009}
HB~Hu and XF~Wang.
\newblock {Disassortative mixing in online social networks}.
\newblock {\em EPL (Europhysics Letters)}, 86(1):18003, 2009.

\bibitem{Serrano2007}
MA~Serrano, M~Bogun\'{a}, R.~Pastor-Satorras, and A~Vespignani.
\newblock {Correlations in Complex Networks}.
\newblock In G~Caldarelli and A.~Vespignani, editors, {\em Large Scale
  Structure and Dynamics of Complex Net- works: From Information Technology to
  Finance and Natural Science}, chapter~3, pages 35--65. World Scientific,
  Singapore, 2007.

\bibitem{Alderson2007}
David Alderson and Lun Li.
\newblock {Diversity of graphs with highly variable connectivity}.
\newblock {\em Physical Review E}, 75(4):046102, April 2007.

\bibitem{Whitney2008}
DE~Whitney and David Alderson.
\newblock {Are technological and social networks really different ?}
\newblock In A~Minai, D.~Braha, and Y.~Bar-Yam, editors, {\em Unifying Themes
  in Complex Systems}, chapter~10, pages 74--81. Springer, Berlin, 2008.

\bibitem{Menche2010}
J\"{o}rg Menche, Angelo Valleriani, and Reinhard Lipowsky.
\newblock {Asymptotic properties of degree-correlated scale-free networks}.
\newblock {\em Physical Review E}, 81(4):046103, April 2010.

\bibitem{Pastor-Satorras2001}
Romualdo Pastor-Satorras, Alexei V\'{a}zquez, and Alessandro Vespignani.
\newblock {Dynamical and Correlation Properties of the Internet}.
\newblock {\em Physical Review Letters}, 87(25):258701, November 2001.

\bibitem{Vazquez2002}
Alexei V\'{a}zquez, Romualdo Pastor-Satorras, and Alessandro Vespignani.
\newblock {Large-scale topological and dynamical properties of the Internet}.
\newblock {\em Physical Review E}, 65(6):066130, June 2002.

\bibitem{Mislove2007}
Alan Mislove, Massimiliano Marcon, Krishna~P. Gummadi, Peter Druschel, and
  Biplab Bhattacherjee.
\newblock {Measurement and analysis of online social networks}.
\newblock In {\em Proceedings of the 7th ACM SIGCOMM conference on Internet
  Measurement}, pages 29--42, San Diego, CA, 2007. ACM.

\bibitem{Chun2008}
Hyunwoo Chun, H~Kwak, Y.-H. Eom, Y.-Y. Ahn, Sue Moon, and Hawoong Jeong.
\newblock {Comparison of online social relations in volume vs. interaction: a
  case study of cyworld}.
\newblock In {\em Proceedings of the Eighth ACM SIGCOMM Conference on Internet
  Measurement}, volume~v, pages 57--69, Vouliagmini, Greece, 2008. ACM.

\bibitem{Zhou2004}
Shi Zhou and RJ~Mondrag\'{o}n.
\newblock {The Rich-Club Phenomenon in the Internet Topology}.
\newblock {\em IEEE Communications Letters}, 8(3):180--182, 2004.

\bibitem{Colizza2006}
V.~Colizza, a.~Flammini, M.~a. Serrano, and a.~Vespignani.
\newblock {Detecting rich-club ordering in complex networks}.
\newblock {\em Nature Physics}, 2(2):110--115, January 2006.

\bibitem{Freeman1977}
LC~Freeman.
\newblock {A set of measures of centrality based upon betweenness}.
\newblock {\em Sociometry}, 40:35--41, 1977.

\bibitem{Klimt2004}
Bryan Klimt and Yiming Yang.
\newblock {The Enron Corpus: A New Dataset for Email Classification Research}.
\newblock In {\em Machine Learning: ECML 2004 Lecture Notes in Computer Science
  Volume 3201}, pages 217--226. Springer Berlin Heidelberg, 2004.

\bibitem{Leskovec2007}
Jure Leskovec, Jon Kleinberg, and Christos Faloutsos.
\newblock {Graph evolution: Desification and Shrinking diameters}.
\newblock {\em ACM Transactions on Knowledge Discovery from Data}, 1(1):2,
  March 2007.

\bibitem{Viswanath2009}
Bimal Viswanath, Alan Mislove, Meeyoung Cha, and Krishna~P. Gummadi.
\newblock {On the evolution of user interaction in Facebook}.
\newblock In {\em Proceedings of the 2nd ACM Workshop on Online Social
  Networks}, page~37, Barcelona, Spain, 2009. ACM.

\bibitem{Gomez2008}
Vicen\c{c} G\'{o}mez, Andreas Kaltenbrunner, and Vicente L\'{o}pez.
\newblock {Statistical analysis of the social network and discussion threads in
  slashdot}.
\newblock In {\em Proceedings of the 17th international conference on World
  Wide Web}, pages 645--654, Beijing, China, 2008. ACM.

\bibitem{Leskovec2010}
Jure Leskovec, DP~Huttenlocher, and JM~Kleinberg.
\newblock {Governance in Social Media: A case study of the Wikipedia promotion
  process}.
\newblock In {\em Proceedings of 4th Int'l AAAI Conference on Weblogs and
  Social Media}, pages 98--105, Washington, D.C., 2010. AAAl.

\bibitem{Burt1995}
R.S. Burt.
\newblock {\em {Structural Holes: The Social Structure of Competition}}.
\newblock Harvard University Press, 1995.

\bibitem{Naaman2010}
Mor Naaman, Jeffrey Boase, and Chih-hui Lai.
\newblock {Is it Really About Me ? Message Content in Social Awareness
  Streams}.
\newblock In {\em Proceedings of the 2010 ACM Conference on Computer Supported
  Cooperative Work}, pages 189--192, Savannah, Georgia, USA, 2010. ACM.

\bibitem{Chu2010}
Zi~Chu, Steven Gianvecchio, Haining Wang, and S~Jajodia.
\newblock {Who is tweeting on Twitter: human, bot, or cyborg?}
\newblock In {\em Proceedings of the 26th Annual Computer Security Applications
  Conference}, pages 21--30, Austin, Texus, USA, 2010. ACM.

\bibitem{Tavares2013}
Gabriela Tavares and Aldo Faisal.
\newblock {Scaling-laws of human broadcast communication enable distinction
  between human, corporate and robot Twitter users.}
\newblock {\em PLOS ONE}, 8(7):e65774, January 2013.

\bibitem{Takhteyev2012}
Yuri Takhteyev, Anatoliy Gruzd, and Barry Wellman.
\newblock {Geography of Twitter networks}.
\newblock {\em Social Networks}, 34(1):73--81, January 2012.

\bibitem{Mocanu2013}
Delia Mocanu, Andrea Baronchelli, Nicola Perra, Bruno Gon\c{c}alves, Qian
  Zhang, and Alessandro Vespignani.
\newblock {The Twitter of Babel: mapping world languages through microblogging
  platforms.}
\newblock {\em PLOS ONE}, 8(4):e61981, January 2013.

\bibitem{Saito2014}
Kodai Saito and Naoki Masuda.
\newblock {Two types of well followed users in the followership networks of
  Twitter}.
\newblock {\em PLOS ONE}, 9(1):e84265, January 2014.

\bibitem{Granovetter1973}
Mark Granovetter.
\newblock {The strength of weak ties}.
\newblock {\em American Journal of Sociology}, 78(6):1360--1380, May 1973.

\bibitem{Onnela2007}
J-P Onnela, J~Saram\"{a}ki, J~Hyv\"{o}nen, G~Szab\'{o}, D~Lazer, K~Kaski,
  J~Kert\'{e}sz, and A-L Barab\'{a}si.
\newblock {Structure and tie strengths in mobile communication networks.}
\newblock {\em Proceedings of the National Academy of Sciences of the United
  States of America}, 104(18):7332--7336, May 2007.

\bibitem{Takaguchi2011}
Taro Takaguchi, Mitsuhiro Nakamura, Nobuo Sato, Kazuo Yano, and Naoki Masuda.
\newblock {Predictability of Conversation Partners}.
\newblock {\em Physical Review X}, 1(1):011008, September 2011.

\bibitem{Huss2007}
Mikael Huss and Petter Holme.
\newblock {Currency and commodity metabolites: their identification and
  relation to the modularity of metabolic networks}.
\newblock {\em IET Systems Biology}, 1(5):280--285, September 2007.

\end{thebibliography}

\begin{thebibliography}{6}

\bibitem{Klimt2004}
Klimt,~B. and Yang,~Y.
\newblock {The Enron corpus: a new dataset for email classification research}.
\newblock In {\em Machine learning: ECML 2004}, pp. 217-226 (Springer Berlin Heidelberg, 2004).

\bibitem{Leskovec2007}
Leskovec,~J., Kleinberg,~J. \& Faloutsos,~C.
\newblock {Graph evolution: desification and shrinking diameters}.
\newblock {\em ACM Trans.  Knowl. Discov. Data} {\bf 1}, 2 (2007).

\bibitem{Viswanath2009}
Viswanath,~B., Mislove,~A., Cha,~M. \& Gummadi,~K.~P.
\newblock {On the evolution of user interaction in Facebook}.
\newblock In {\em Proceedings of the Second ACM Workshop on Online Social Networks}, pp. 37--42, Barcelona, Spain, ACM (2009).

\bibitem{Gomez2008}
G\'{o}mez,~V., Kaltenbrunner,~A. \& L\'{o}pez,~V.
\newblock {Statistical analysis of the social network and discussion threads in slashdot}.
\newblock In {\em Proceedings of the 17th International Conference on World Wide Web}, pp. 645--654, Beijing, China, ACM (2008).

\bibitem{Leskovec2010}
Leskovec,~J., Huttenlocher,~D.~P. and Kleinberg,~J.~M.
\newblock {Governance in social media: a case study of the Wikipedia promotion process}.
\newblock In {\em Proceedings of Fourth International AAAI Conference on Weblogs and Social Media}, pp. 98--105, Washington, D.C., USA, AAAl (2010).

\bibitem{Kunegis2013}
Kunegis,~J. 
\newblock {KONECT: the Koblenz network collection.}
\newblock In {\em Proceedings of the 22nd International Conference on World Wide Web Companion.}, pp. 1343--1350, Rio de Janeiro, Brazil, ACM (2013).

\end{thebibliography}

\clearpage

\pagestyle{empty}
\clearpage

\clearpage

\clearpage

\clearpage

\clearpage

\clearpage
\begin{table}
\centering
\caption{Summary of basic statistics of the network data sets used: the total number of nodes $N$ and links $M$, the average clustering coefficient $C$, and the degree assortativity coefficient $r$.}
\label{tab:stats}
\begin{tabular}{|c|r|r|r|r|}\hline
Name & $N$ & $M$ & $C$ & $r$\\ \hline
Enron & $7,015$ & $22,474$ & $0.240$ & $-0.209$\\\hline
EU-email & $32,430$ & $54,397$ & $0.113$ & $-0.382$\\\hline
Facebook & $29,342$ & $79,230$ & $0.084$ & $0.213$\\\hline
Slashdot & $8,815$ & $12,859$ & $0.003$ & $-0.058$\\\hline
Wiki-talk & $92,117$ & $360,767$ & $0.059$ & $-0.034$\\\hline
\end{tabular}
\end{table}

\end{document}